\documentclass[useAMS,usenatbib,usegraphicx]{mn2e}

\newcommand{\BV}{Brunt-V\"ais\"al\"a }
\newcommand{\teff}{\mbox{${T}_{\rm eff}$}}

\newcommand{\msol}{\mbox{${\rm M}_{\odot}\;$}}
\newcommand{\msun}{\mbox{${\rm M}_{\odot}$}}

\newcommand{\simgt}{\lower.5ex\hbox{$\; \buildrel > \over \sim \;$}}
\newcommand{\simlt}{\lower.5ex\hbox{$\; \buildrel < \over \sim \;$}}

\title[Probing the properties of convective cores through g modes]{Probing the properties of convective cores through g modes: high-order g modes in SPB and $\gamma$ Doradus stars}
\author[A. Miglio, J. Montalb\'an, A. Noels and P. Eggenberger]{Andrea Miglio, Josefina Montalb\'an, Arlette Noels and Patrick Eggenberger\\
Institut d'Astrophysique et de G\'eophysique de l'Universit\'e de Li\`ege,
All\'ee du 6 Ao\^ut, 17 B-4000 Li\`ege, Belgium}

\begin{document}
\date{Accepted 2008}

\pagerange{\pageref{firstpage}--\pageref{lastpage}} \pubyear{2002}

\maketitle

\label{firstpage}

\begin{abstract}
In main sequence stars the periods of high-order gravity modes are sensitive probes of stellar cores and, in particular, of the chemical composition gradient that develops near the outer edge of the convective core.
We present an analytical approximation of high-order g modes that takes into account the effect of the $\mu$ gradient near the core. We show that in main-sequence models, similarly to the case of white dwarfs, the periods of high-order gravity modes are accurately described by a uniform period spacing superposed to an oscillatory component. The periodicity and amplitude of such component are related, respectively, to the location and sharpness of the $\mu$ gradient.

We investigate the properties of  high-order gravity modes for
stellar models in a mass domain range between 1 and 10 \msun, and
the effects of the stellar mass, evolutionary state, and
extra-mixing processes on period spacing features. In particular, we
show that for models of a typical SPB star, a chemical mixing that
could likely be induced by the slow rotation observed in these
stars, is able to significantly change the g-mode spectra of the
equilibrium model. Prospects and challenges for the asteroseismology
of $\gamma$ Doradus and SPB stars are also discussed.
\end{abstract}


\section{Introduction}
It is well known that a stratification in the chemical composition
of stellar models directly influences the properties of gravity
modes. The signatures of chemical stratifications have been
extensively investigated theoretically, and observed, in pulsating
white dwarfs \citep[see e.g.][for a review]{Kawaler95}. The
influence of chemical composition gradients on g-modes in main
sequence stars has been partly addressed and suggested in the works
by \citet{Berthomieu88} and \cite{Dziembowski93}. Inspired by these
works and following the approach of \citet{Berthomieu88} and
\citet{Brassard92} we study the properties of high-order, low-degree
gravity modes in  main sequence stellar models.

High-order gravity modes are observed in two classes of main-sequence stars: $\gamma$ Doradus and Slowly Pulsating B stars.
 The former are main-sequence stars with masses around 1.5 \msol \citep[see e.g.][]{Guzik00} that show both photometric and line profile variations. Their spectral class is A7-F5 and their effective temperature is between 7200-7700 K on the ZAMS and 6900-7500 K above it \citep{Handler99}. $\gamma$ Dor stars are multiperiodic oscillators with  periods between 8 hours and 3 days (high-order g-modes). The modulation of the radiative flux by convection at the base of the convective envelope was proposed as the excitation mechanism for such stars \citep[see e.g.][]{Guzik00, Dupret04a}.

 Slowly Pulsating B  stars (SPB) are multiperiodic main-sequence stars with masses from about 3 to 8 \msol and spectral type B3-B8 \citep{Waelkens91}. High order g-modes of periods typically between 1 and 3 days are found to be excited by the $\kappa$-mechanism acting in the region of the metal opacity bump located at $T\sim 2\times 10^5$ K in the stellar interior \citep[see e.g.][]{Dziembowski93}. Recent observations \citep[see][]{Jerzykiewicz05,Handler06,Chapellier06} and theoretical instability analysis \citep{Pamyatnykh99, Miglio07} also suggest high-order g-modes being excited in a large fraction of the more massive $\beta$ Cephei pulsators: the seismic modelling of these hybrid pulsators looks very promising as it would benefit from the information on the internal structure carried by both low order p and g modes ($\beta$ Cephei type oscillation modes) and high order g modes (SPB-type pulsation).

The seismic modelling of $\gamma$ Doradus and
SPB stars is a formidable task to undertake. The frequencies of high-order g modes are in fact closely spaced  and  scan be severely perturbed by the effects of rotation \citep[see e.g.][]{Dintrans00,Suarez05}. Nonetheless, the high scientific interest of these classes of pulsators has driven efforts in both the observational and theoretical domain.
Besides systematic photometric and spectroscopic ground-based surveys carried out on $\gamma$ Dor \citep[see][]{Mathias04} and SPB stars \citep[see][]{DeCat02}, the long and uninterrupted photometric observations planned with COROT \citep{Baglin06,Mathias06} will allow to significantly increase the number and accuracy of the observed frequencies.

On the theoretical side, as suggested by \citet{Suarez05} in the case of $\gamma$ Dor stars, a seismic analysis becomes feasible for slowly rotating targets. In these favorable cases  the first-order asymptotic approximation \citep{Tassoul80} can be used as a tool to derive the buoyancy radius of the star \citep[see][]{Moya05} from the observed frequencies.
Nevertheless, the g-mode spectra of these stars contain much more information on the internal structure of the star. In this paper we describe in detail the information content carried by the periods of high-order g modes,
 and show that the effect of chemical composition gradients can be easily included as a refinement of the
asymptotic approximation of \citet{Tassoul80}.

After an introduction to the properties of gravity modes in main-sequence stars (Sec.\ref{sec:modetrap}), we present in Sec. \ref{sec:approx} the analytical approximation of high-order g-mode frequencies that will be used in the subsequent sections. In Sec. \ref{sec:numeric} we describe the properties of numerically computed g-mode frequencies in main-sequence stars in the  mass domain 1-10 \msun.
The effect of adding extra-mixing at the outer edge of the convective core (rotationally induced turbulence, overshooting, diffusion) is investigated in Sec. \ref{sec:extra}. In Sec. \ref{sec:realworld} we estimate how of the effects of rotation and of current observational limitations affect asteroseismology of main sequence high-order g modes pulsators. A summary is finally given in Sec. \ref{sec:conclusions}.


\section{The properties of trapped \lowercase{g}-modes}
\label{sec:modetrap}
As it is well known, the period spectrum of gravity modes is determined by the spatial distribution of the \BV frequency ($N$) which is defined as:

\begin{equation}
N^2=g\left(\frac{1}{\Gamma_1 p}\frac{{\rm d}p}{{\rm d}r}-\frac{1}{\rho}\frac{{\rm d}\rho}{{\rm d}r}\right)\;{\rm .}
\end{equation}

\noindent
$N$ can be approximated, assuming the ideal gas law for a fully-ionized gas, as:

\begin{equation}
N^2\simeq \frac{g^2\rho}{p}\left(\nabla_{\rm ad}-\nabla+\nabla_\mu \right)\;{\rm ,}
\label{eq:bvgr}
\end{equation}

\noindent
where

\begin{equation}
\nabla=\frac{{\rm d}\ln{T}}{{\rm d} \ln{p}},\;\;\nabla_{\rm ad}=\left(\frac{\partial\ln{T}}{\partial\ln{p}}\right)_{\rm ad}\;\;{\rm and}\;\; \nabla_\mu=\frac{{\rm d}\ln{\mu}}{{\rm d}\ln{p}} \;\;{\rm .}
\end{equation}

\noindent
The term $\nabla_\mu$ gives the explicit contribution of a change of chemical composition to $N$.
The first order asymptotic approximation developed by \citet{Tassoul80} shows that, in the case of a model that consists of an inner convective core and an outer radiative envelope  \citep[we refer to the work by][ for a complete analysis of other possible cases]{Tassoul80}, the periods of low-degree, high-order g modes 
are given by:

\begin{equation}
P_k=\frac{\pi^2}{L\,\int_{x_0}^{1}{\frac{|N|}{x} {\rm d}x}} \left(2k+n_e\right)
\label{eq:asy}
\end{equation}

\noindent
where $L=[\ell(\ell+1)]^{1/2}$ (with $\ell$ the mode degree), $n_e$ the effective polytropic index of the surface layer, $x$ the normalized radius and $x_0$ corresponds to the boundary of the convective core. In order to avoid confusion with $n_e$, the radial order of g modes is represented by $k$.

Following Eq. \ref{eq:asy}, the periods are asymptotically equally spaced in $k$ and the spacing decreases with increasing $L$. It is therefore natural to introduce, in analogy to the large frequency separation of p modes, the {\it period spacing} of gravity modes, defined as:
\begin{equation}
\Delta P=P_{k+1}-P_{k}\;{\rm .}
\end{equation}
In the following sections we will show that deviations from a constant $\Delta P$ contain information on the chemical composition gradient left by a convective core evolving on the main sequence.

\begin{figure}
\begin{center}
\resizebox{0.8\hsize}{!}{\includegraphics[angle=0]{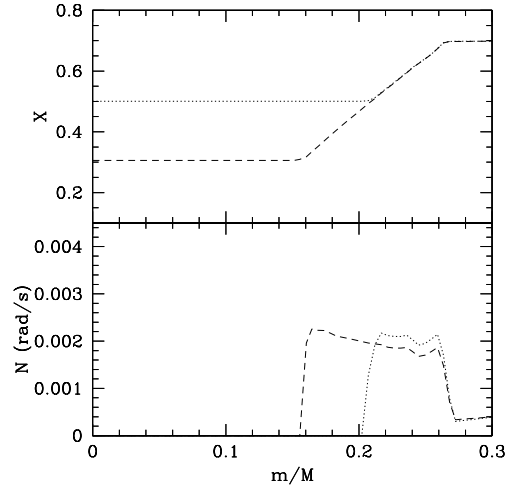}}
\caption{\small {\it Upper panel:} Hydrogen abundance in the core of 6 \msol models on the main sequence. $X_c \simeq 0.5$ (dotted line) and at $X_c \simeq 0.3$ (dashed line). The convective core recedes during the evolution and leaves behind a chemical composition gradient.
{\it Lower panel:} Buoyancy frequency $N$ as a function of the normalized mass.}\label{fig:6trapbvx}
\end{center}
\end{figure}
\begin{figure}
\begin{center}
\resizebox{0.8\hsize}{!}{\includegraphics[angle=0]{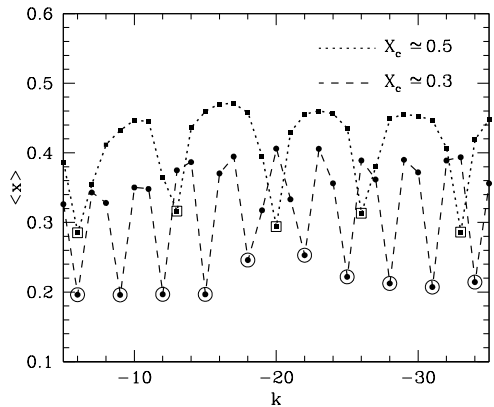}}
\caption{\small $\langle x \rangle$ for 6 \msol models on the main sequence at two different points in the evolution: $X_c \simeq 0.5$ (dotted lines) and at $X_c \simeq 0.3$ (dashed line). Modes of different radial order are periodically trapped in the region of chemical composition gradient. Trapped modes are marked with open symbols.}\label{fig:6trapx}
\end{center}
\end{figure}
\begin{figure}
\begin{center}
\resizebox{0.9\hsize}{!}{\includegraphics[angle=0]{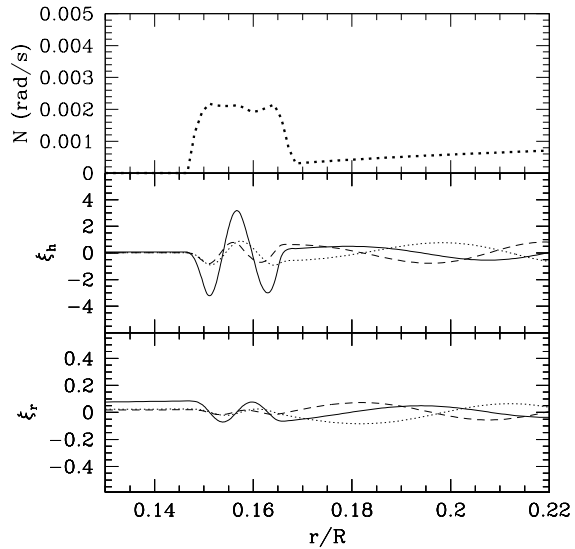}}
\caption{\small  We consider a $6$ \msol model with $X_c \simeq 0.5$.  {\it Upper panel:} $N$ against fractional radius. {\it Central panel:}  Horizontal component of the displacement for three $\ell=1$ gravity modes of radial order 18 (dotted line), 20 (continuous line)  and 22 (dashed line). The eigenfunction corresponding to $k=20$ is partly trapped in the region of mean molecular weight where the sharp variation of $N$ is located. {\it Lower panel:} As in the central panel, this time the radial displacement is shown. The eigenfunctions corresponding to different modes are normalized to have the same total pulsation energy.}\label{fig:6trap}
\end{center}
\end{figure}
\begin{figure}
\begin{center}
\resizebox{0.85\hsize}{!}{\includegraphics[angle=0]{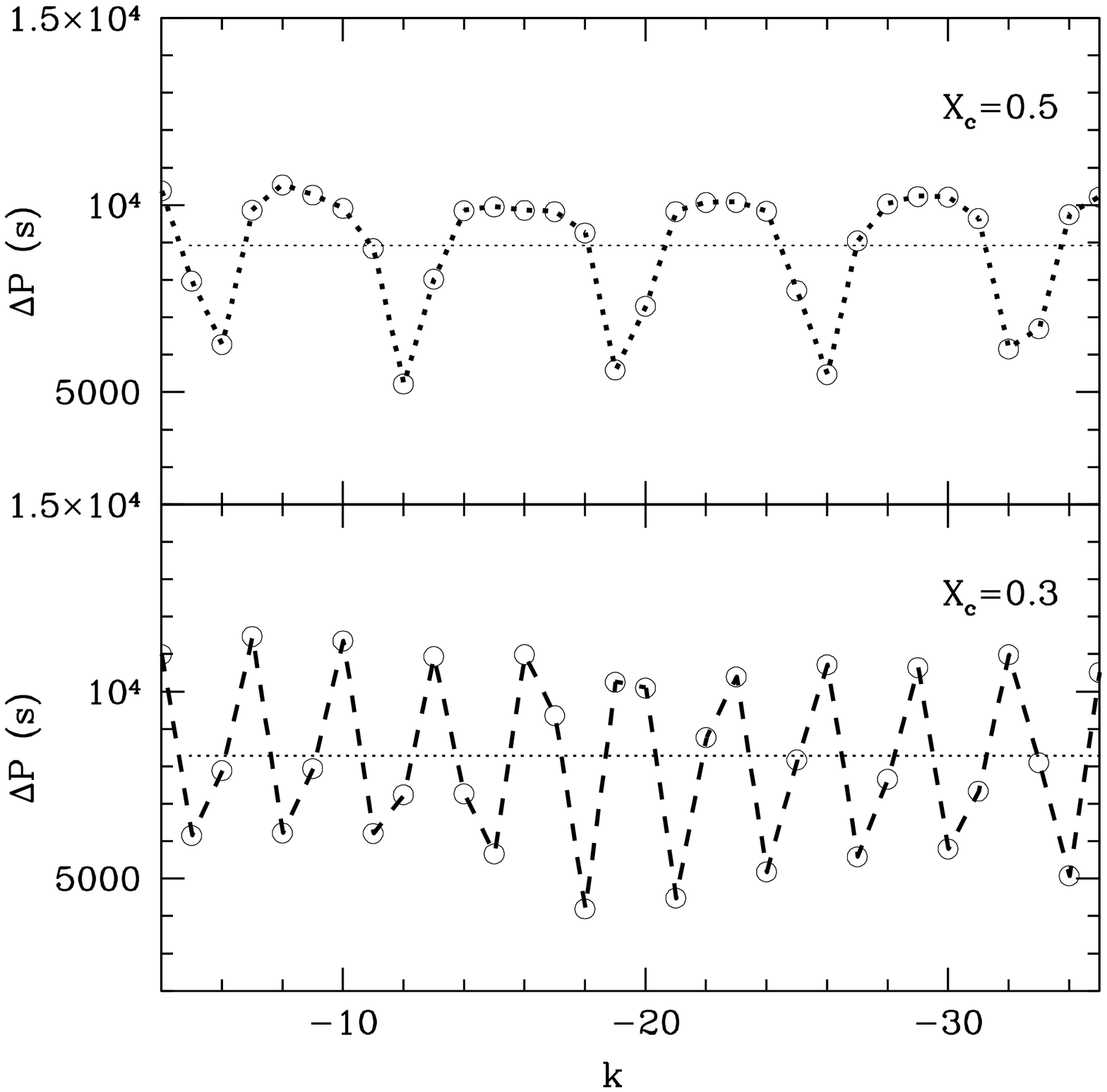}}
\caption{\small Period spacing for the same 6 \msol models as in Fig. \ref{fig:6trapx}. The periods of the components (in terms of $k$) are approximately 7 and 3. Horizontal dotted lines represent constant period spacing predicted by the asymptotic approximation (Eq. \ref{eq:asy}).}\label{fig:6trapg}
\end{center}
\end{figure}

We consider as a first example two models of a 6 \msol star evolving on the main-sequence.
The behaviour of $N$ and of the chemical composition profile is represented in Fig. \ref{fig:6trapbvx}. The convective core is fully mixed and, therefore, the composition is uniform ($\nabla_\mu=0$). However, in stars in this mass range, the convective core shrinks during the evolution, leaving behind a steep gradient in the hydrogen abundance X. This causes a sharp peak in $\nabla_\mu$ and in $N$: does this feature leave a clear signature in the properties of g-modes?

This question was addressed by \citet{Brassard91} while studying the seismic effects of compositional layering in white-dwarfs. The authors found that a sharp feature in the buoyancy frequency could lead to a resonance condition that may trap modes in different regions of the model.

A first indicator of such a trapping is the behaviour of $\langle x\rangle$ defined by:
\begin{equation}
\langle x \rangle = \frac{\int_0^1{x |\bmath{\delta r}|^2 dm}}{\int_0^1{|\bmath{\delta r}|^2dm}}
\end{equation}
where $x=r/R$ and {\boldmath$\delta r$} is the total displacement vector. As shown in Fig. \ref{fig:6trapx} modes of different radial order $k$ are {\it periodically confined} closer to the center of the star.

In Fig. \ref{fig:6trap} we show the behaviour of the eigenfunctions for modes of radial orders around a trapped mode: the partly trapped mode has, compared to ``neighbour'' modes, a larger amplitude in the region of mean molecular weight gradient.

In white dwarfs it has been theoretically predicted and observed
\citep[see, for instance, the recent work of][]{Metcalfe03} that the
period spacing $\Delta P(k)=P_{k+1}-P_k$ is not constant, contrary
to what is predicted by the first order asymptotic approximation of
gravity modes. This has been interpreted as the signature of
chemical composition gradients in the envelope and in the core of
the star. In analogy with the case of white dwarfs, in models with a
convective core, we expect the formation of a {\it nonuniform period
distribution}; this is in fact the case as is presented in Fig.
\ref{fig:6trapg}. In that figure we plot the period spacing derived
by using the adiabatic oscillation code LOSC \citep{Scuflaire07b}
for models of 6 \msol\ at two different stages in the main sequence
evolution.  The period spacing presents clear deviations from the
uniformity that would be expected in a model without sharp
variations in $N$. How these deviations are related to the
characteristics of the chemical composition gradients will be
studied in the following sections.

\section[Approximate expression of \lowercase{g}-modes periods]{Approximate analytical expression of \lowercase{g}-modes period spacing}
\label{sec:approx} In this section we derive two approximate
expressions that relate deviations from a uniform period
distribution to the characteristics of the $\mu$-gradient region.
These simplified expressions could also represent a useful tool to
give a direct interpretation of an observed period spectrum. Though
a first description of these approximated expressions was outlined
in \citet{Miglio06a} and \citet{Miglio06}, we here present a more
detailed analysis.

We recall that deviations from the asymptotic expressions of the
frequencies of high-order pressure modes have been widely studied in
the context of helioseismology. The oscillatory features in the
oscillation spectrum of solar oscillation modes allowed modes to
infer the properties of localized variations of the solar structure,
e.g. at the base of the convective envelope and in the second helium
ionization region \citep[see e.g.][]{Gough90, jcd91b, Monteiro94,
Basu95, Monteiro05, Houdek07a}.
\subsection{Variational principle}
\label{sec:varia}
A first and simple approach to the problem is to make use of the variational principle for adiabatic stellar oscillations \citep[see e.g.][]{Unno89}. The effect of a sharp feature in the model (a chemical composition gradient, for instance) can be estimated
 from the periodic signature in $\delta P$, defined as the difference between the periods of the star showing such a sharp variation and the periods of an otherwise fictitious smooth model.

We consider a model with a radiative envelope and a convective core whose boundary is located at a normalized radius $x_0$. $N_-$ and $N_+$ are the values of the \BV frequency at the outer and inner border of the $\mu$-gradient region. We define $\alpha=\left(\frac{N_+}{N_-}\right)^{1/2}$ with $N_+\leq N_-$. Then $\alpha=1$ describes the smooth model and $\alpha \to 0$ a sharp discontinuity in $N$.

To obtain a first estimate of $\delta P$, we adopt \citep[following the approach by][]{Montgomery03} the Cowling approximation, that reduces the differential equations of stellar adiabatic oscillations to a system of the second order. Furthermore, since we deal with high-order gravity modes, the eigenfunctions are well described by their JWKB approximation \citep[see e.g. ][]{Gough93}.
We can therefore express $\delta P$ as:
\begin{equation}
\frac{\delta P_k}{P}=2\Pi_0 \int_{0}^{\Pi_0^{-1}}{\left(\frac{\delta N}{N} \right)\cos{\left(\frac{L\,P_k}{\pi\,\Pi_x}+\frac{\pi}{2} \right)}\;{\rm d}\Pi_x^{-1}}
\label{eq:montgo}
\end{equation}
where $L=[\ell(\ell+1)]^{1/2}$, the local  buoyancy radius is defined as:
\begin{equation}\Pi_x^{-1}=\int_{x_0}^x{\frac{|N|}{x'}dx'}\;{\rm ,}
\end{equation}
and the  total buoyancy radius as
\begin{equation}
\Pi_0^{-1}=\int_{x_0}^1{\frac{|N|}{x'}dx'}\;.
\end{equation}
The buoyancy radius of the discontinuity is then:
\begin{equation}
\Pi_\mu^{-1}=\int_{x_0}^{x_\mu}{\frac{|N|}{x'}dx'}\;\;.
\label{eq:pimu}
\end{equation}

\begin{figure}
 \begin{center}
\resizebox{0.48\hsize}{!}{\includegraphics[angle=0]{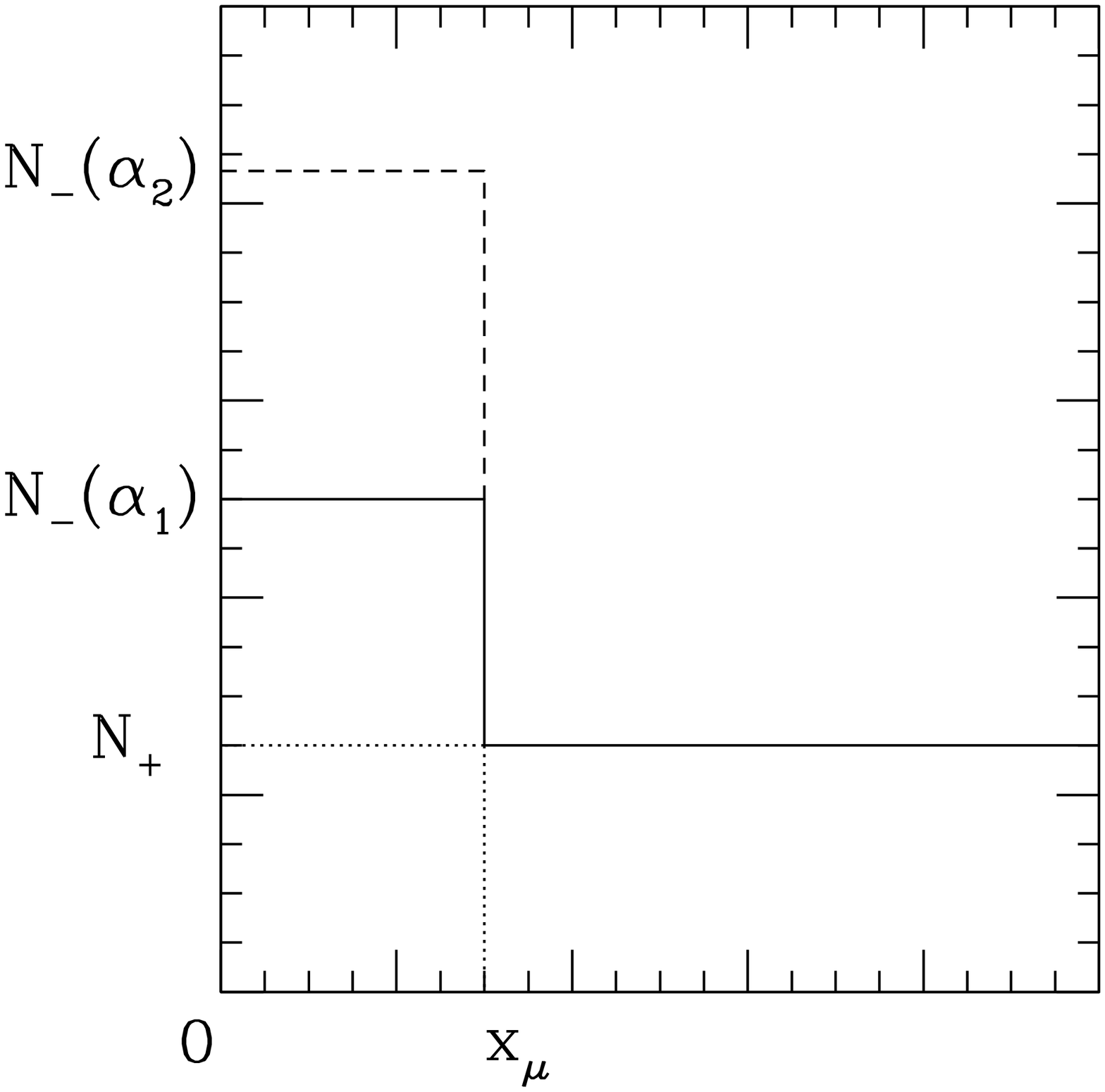}}
\resizebox{0.48\hsize}{!}{\includegraphics[angle=0]{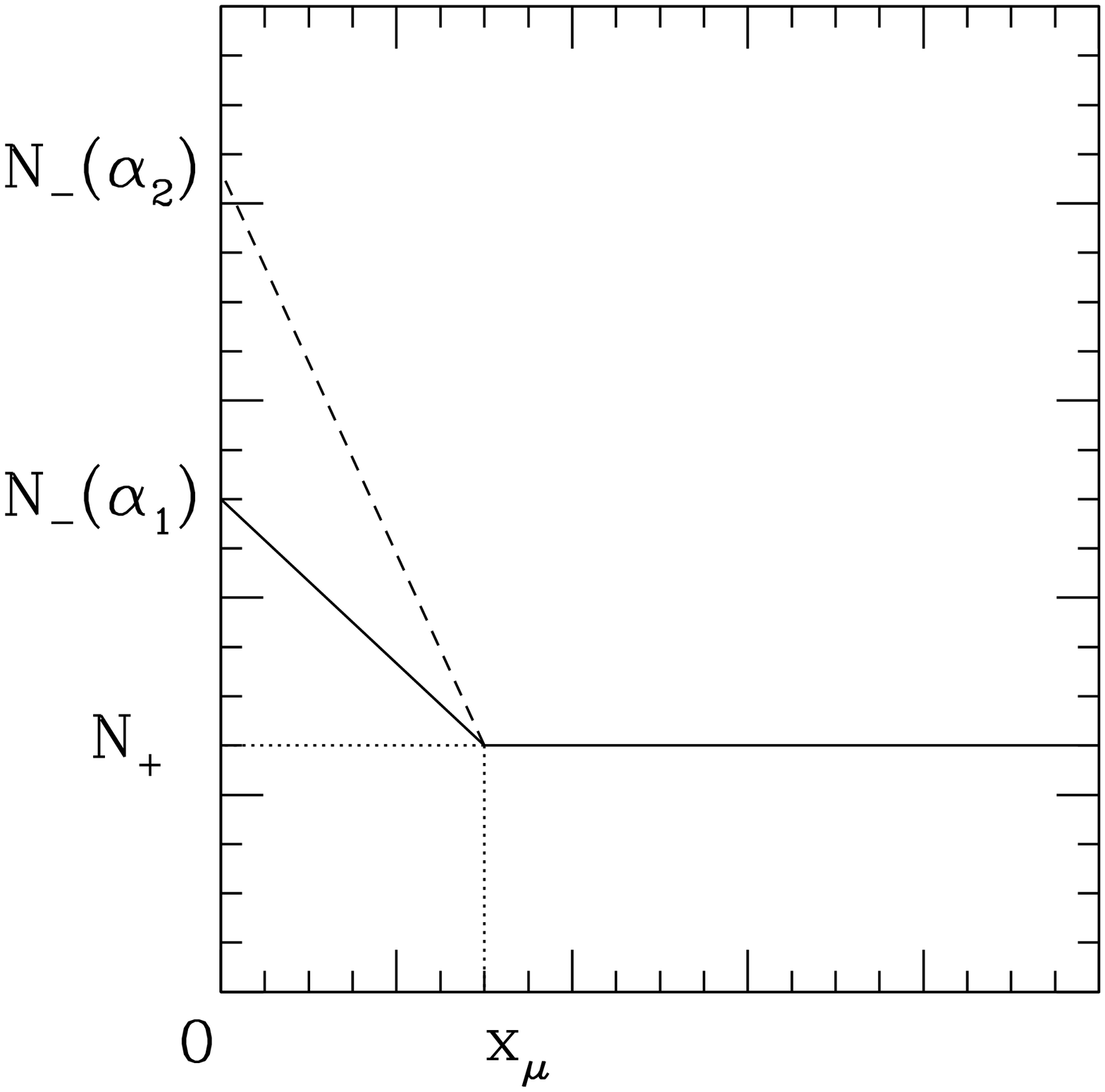}}
\caption{\small {\it Left:} We model the sharp variation of $N$ at $x=x_\mu$ by means of Eq. (\ref{eq:step}) for different values of $\alpha$: $\alpha_1^2=0.5$ (continuous line), $\alpha_2^2=0.3$ (dashed line).  {\it Right:} A smoother variation of $N$ is modeled following Eq. (\ref{eq:ramp}).}
\label{fig:simplen}
 \end{center}
\end{figure}

We model the sharp feature in $\frac{\delta N}{N}$ located at $x=x_\mu$ as:
\begin{equation}
\frac{\delta N}{N}=\frac{1-\alpha^2}{\alpha^2}\, H(x_\mu-x)\;,
\label{eq:step}
\end{equation}
where $H(x)$ is the step function (see left panel of Fig. \ref{fig:simplen}).

Retaining only periodic terms in $\delta P$ and integrating by parts we obtain:
\begin{equation}
\delta P_k\propto\frac{\Pi_0}{L} \frac{1-\alpha^2}{\alpha^2}\cos{\left(\frac{L\,P_k}{\pi\,\Pi_\mu}+\frac{\pi}{2} \right)}\;.
\end{equation}
For small $\delta P$ we can substitute the asymptotic approximation for g-modes periods derived by \citet{Tassoul80} in the expression above:
$$P_k=\pi^2\frac{\Pi_0}{L} \left(2k+\phi'\right)\;,$$
where $\phi'$ is a phase constant that depends on the boundary
conditions of the propagation cavity \citep[see][]{Tassoul80}, and
find
\begin{equation}
\delta P_k\propto\frac{\Pi_0}{L} \frac{1-\alpha^2}{\alpha^2}\cos{\left(2\pi\,\frac{\Pi_0}{\Pi_\mu} k+\pi\frac{\Pi_0}{\Pi_\mu}\phi'+\frac{\pi}{2} \right)}\;.
\label{eq:varia}
\end{equation}
From this simple approach we derive that the signature of a sharp feature in the \BV frequency is a {\it sinusoidal component in the periods of oscillations}, and therefore  in the period spacing, with a periodicity in terms of the radial order $k$ given by
\begin{equation}
\Delta k\simeq\frac{\Pi_\mu}{\Pi_0}\;{.}
\label{eq:variak}
\end{equation}
The amplitude of this sinusoidal component is proportional to the sharpness of the variation in $N$ and does not depend on the order of the mode $k$.

Such a simple approach allows us to easily test the effect of having a less sharp ``glitch'' in the \BV frequency. We model $\delta N$ (Fig. \ref{fig:simplen}, right panel) as a ramp function instead of a step function:
\begin{equation}
\frac{\delta N}{N}=\frac{1-\alpha^2}{\alpha^2}\frac{(x_\mu-x)}{x_\mu-x_0} H(x_\mu-x)\;.
\label{eq:ramp}
\end{equation}

In this case integration by parts leads to a sinusoidal component in $\delta P_k$ whose {\it amplitude is modulated by a factor $1/P_k$} and therefore decreases with increasing $k$, i.e.
\begin{equation}
\delta P_k\propto\frac{1}{P_k}\frac{\Pi_0}{L} \frac{1-\alpha^2}{\alpha^2}\frac{1}{\Pi^{-1}_\mu}\cos{\left(2\pi\,\frac{\Pi_0}{\Pi_\mu} k+\pi\frac{\Pi_0}{\Pi_\mu}\phi'+\frac{\pi}{2} \right)}.
\label{eq:senal}
\end{equation}
  The information contained in the amplitude of the sinusoidal component, as will be presented in Section \ref{sec:extra}, is potentially very interesting. It reflects the different characteristics of the chemical composition gradient resulting, for example, from a different treatment of the mixing process in convective cores, from considering microscopic diffusion or rotationally induced mixing in the models.

Eq. (\ref{eq:varia}) was derived by means of a first order perturbation of the periods neglecting changes in the eigenfunctions. This approximation is valid in the case of small variations relative to a smooth model, therefore it becomes questionable as the change of $N$ at the edge of the convective core becomes large. A more accurate approximation is presented in the following section.

\subsection{Considering the effects of the $\mu$ gradient on the eigenfunctions}
\label{sec:approx2}
We present in this Section a description of mode trapping considering the change in the eigenfunctions due to a sharp feature in $N$. As a second step we derive the effects on the periods of g-modes.

\citet{Brassard92} studied the problem of mode trapping in $\mu$-gradient regions inside white dwarfs. In this section we proceed as \citet{Brassard92}, applying the asymptotic theory as developed in \citet{Tassoul80} to the typical structure of an intermediate mass star on the main sequence.

\citet{Tassoul80}, assuming the Cowling approximation, provided
asymptotic solutions for the propagation of high-order g-modes in
convective and radiative regions, located in different parts of the
star. In order to generalize the expression for these solutions, she
introduced two functions $S_1$ and $S_2$ related to the radial
displacement ($\xi_r$) and pressure perturbation ($p'$) as follows:
\begin{equation}
\sigma^2\xi_r=\rho^{-1/2}x^{-2}|\phi|^{-1/4}S_1
\end{equation}
and
\begin{equation}
\sigma \ell(\ell+1)p'/\rho=\rho^{-1/2}|\phi|^{1/4}S_2\;,
\end{equation}
where $x$ is the normalized radius, $\sigma$ the angular frequency
of the oscillation, and
\begin{equation}
\phi=\frac{\ell(\ell+1)N^2}{x^2}\; .
\end{equation}

The expressions for $S_1$, $S_2$ for different propagation regions
are given in the equations [T79] to [T97]\footnote{Henceforth on
[T$i$] indicates equation number $i$ in the paper by
\citet{Tassoul80}.}.

The only difference from the derivation of \cite{Brassard92}, who assumed an entirely radiative model, is that here we consider a model that consists of a convective core and a radiative envelope. The solutions should then be described by [T80] close to the center, [T96] in the convective core ($x_0$), [T97] in the radiative region with [T82] close to the surface (where the structure of the surface layers of the model is described by an effective polytropic index $n_e$).

\begin{figure}
 \begin{center}
\resizebox{0.85\hsize}{!}{\includegraphics[angle=-90]{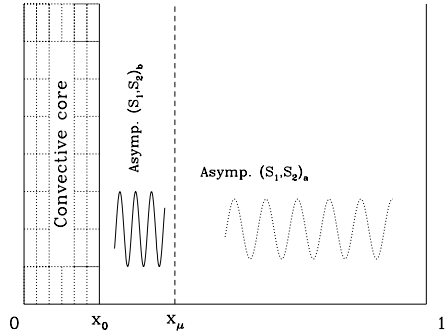}}
\caption{\small A schematic view of the simplified model we consider. The radiative region outside the convective core ($x>x_0$) is divided in two zones: one below and the other above the discontinuity in $N$ located at $x=x_\mu$. In each of these regions we consider the asymptotic expressions for the eigenfunctions that are then joined continuously at $x=x_\mu$.}
\label{fig:simple}
 \end{center}
\end{figure}

Now, as explained in Fig. \ref{fig:simple}, we assume that at $x=x_\mu$ in the radiative zone there is a sharp variation of $N$ due to a $\mu$ gradient and, as in the previous section, we model it as a discontinuity weighted by $\alpha$, where $\alpha=\left(N_+/N_-\right)^{1/2}$ (see Eq. \ref{eq:step}).

We define $\lambda=\sigma^{-1}$,
$$v_0(x)=L\int_{x}^1{\frac{|N|}{x'}dx'}=L\left(\Pi_0^{-1}-\Pi_x^{-1}\right)$$

$$ {\rm and}\;\;v_1(x)=L\int_{x_0}^x{\frac{|N|}{x'}dx'}=L\Pi_x^{-1}\;.$$

For large values of $\lambda v_0$ and $\lambda v_1$, we can write the eigenfunctions in the radiative region above the discontinuity at $x_{\mu}$ as:
\begin{equation}
S_{1\rm a}\propto  k_o\sin{\left(A\right)}\;\; {\rm and} \;\;S_{2\rm a}\propto -k_o\cos{\left(A\right)}\;,
\label{eq:s2+}
\end{equation}
and in the 
$\nabla\mu$ region below the discontinuity we have:
\begin{equation}
S_{1\rm b}\propto k_{1} \cos{\left(B\right)}\;\;{\rm and}\;\;S_{2\rm b}\propto -k_{1} \sin{\left(B\right)}\;,
\label{eq:s2-}
\end{equation}
where
\begin{equation}
A=\lambda v_0(x)-n_e \frac{\pi}{2}-\frac{\pi}{4}\;\;\;\;{\rm ,}\;\;\;\;B=\lambda v_1(x)-\frac{\pi}{4}\;,
\label{eq:AB}
\end{equation}
and $k_0$ and $k_1$ are arbitrary constants. Note that $A$ and $B$ are functions of $x$.

The eigenfrequencies are now obtained by matching continuously the individual solutions in their common domain of
validity.
In particular, imposing the continuity of $p'$ and $\xi_r$ at the location of the discontinuity in $N$ we obtain the following conditions \citep[as in][]{Brassard92}:
\begin{equation}
S_1^+=\alpha S_1^-\;,
\label{eq:s1}
\end{equation}
\begin{equation}
S_2^+\alpha =S_2^-\;,
 \label{eq:s2}
\end{equation}
where $S_{1,2}$ is evaluated above ($S_{1,2}^+$) and below ($S_{1,2}^-$) $x=x_\mu$.

This finally leads to a condition on the eigenfrequencies $1/\lambda$:
\begin{equation}
\cos{\left(A+B\right)}=\frac{1-\alpha^2}{1+\alpha^2}\cos{\left(A-B\right)}\;,
\label{eq:relazsimple}
\end{equation}
With  $A$ and $B$ (Eq.~\ref{eq:AB}) evaluated at $x=x_\mu$, this condition can also be explicitly  written as:
\begin{eqnarray}
\cos{\left(\lambda \frac{L}{\Pi_0}-\frac{n_e \pi}{2}-\frac{\pi}{2}\right)}\nonumber=\\
 =\frac{1-\alpha^2}{1+\alpha^2}\cos{\left(\lambda L\left[\frac{1}{\Pi_0} -\frac{2}{\Pi_\mu}\right]-\frac{n_e \pi}{2}\right)}
{\rm.}
\label{eq:relaz}
\end{eqnarray}

\subsubsection{Further approximations}
A first extreme case for  Eq. (\ref{eq:relaz}) is that corresponding to  $\alpha=1$,
 i.e. no discontinuities in the \BV frequency. In this case Eq. (\ref{eq:relazsimple}) immediately leads to the condition $\cos{\left(A+B\right)}=0$ and therefore to the uniformly spaced period spectrum predicted by Tassoul's first order approximation:
\begin{equation}
P_k=\pi^2\frac{\Pi_0}{L} \left(2k+n_e\right)
\label{eq:notrap}
\end{equation}

Another extreme situation is $\alpha \to 0$, in this case $N$ is so large in the $\mu$-gradient region that all the modes are trapped there; the periods of ``perfectly trapped'' modes are then:
\begin{equation}
P_n=(n+\frac{1}{4})\;2\pi^2\frac{\Pi_\mu}{L} \label{eq:trap}
\end{equation}
where $n=(1,2,3, \ldots)$.

The interval (in terms of radial order $k$) $\Delta k$ between two consecutive trapped modes ($\Delta n = 1$)
can be obtained combining  Eq.~(\ref{eq:notrap}) and Eq.~(\ref{eq:trap}) \citep[see also][]{Brassard92}, and  is roughly given by:
\begin{equation}
\Delta k\simeq\frac{\Pi_\mu}{\Pi_0},
\label{eq:comevaria}
\end{equation}
which corresponds to Eq.~(\ref{eq:variak}).

We choose the 6~\msol models considered in Sec. \ref{sec:modetrap}
to compare the g-mode period spacings predicted by equations
(\ref{eq:variak}) and (\ref{eq:relaz}), with the results obtained
from the frequencies computed with an adiabatic oscillation code.


\begin{figure}
\begin{center}
\resizebox{0.8\hsize}{!}{\includegraphics[angle=0]{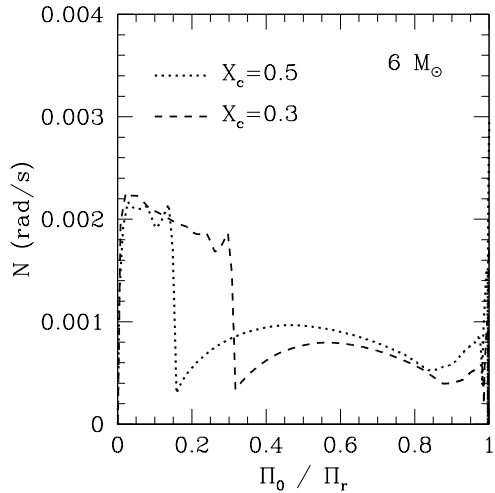}}
\caption{\small The Brunt-V\"ais\"al\"a frequency versus $\Pi_0/\Pi_r$ for $6$ \msol models with $X_c \simeq 0.5$ (dotted line) and $X_c \simeq 0.3 $ (dashed line).}
\label{fig:6trapbv}
\end{center}
\end{figure}

Equation (\ref{eq:variak}) relates the period of the oscillatory
component in the period spacing $\Delta P$ to the location of the
sharp variation in $N$. In Fig. \ref{fig:6trapg} the periods (in
terms of $k$) of the components are approximately 7 and 3 for models
with $X_c=$ 0.5 and 0.3. Following Eq. (\ref{eq:variak}) these
periods should correspond to a location of the discontinuity
(expressed as $\Pi_0/\Pi_\mu \simeq k^{-1}$) of ~0.14 and 0.3: as
shown in Fig. \ref{fig:6trapbv}, these estimates describe very
accurately the locations of the sharp variation of $N$ in the
models.

\begin{figure}
\begin{center}
\resizebox{0.85\hsize}{!}{\includegraphics[angle=-90]{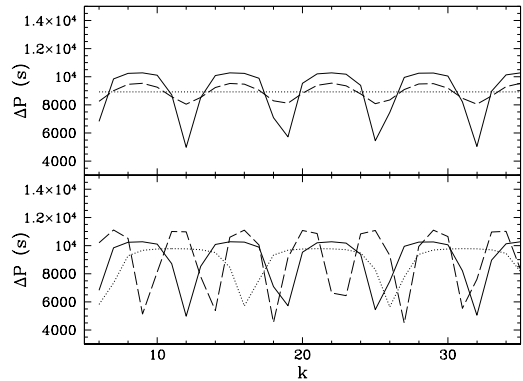}}
\caption{\small Period spacing $\Delta P=P_{k+1}-P_{k}$ calculated
from numerical solutions of Eq. (\ref{eq:relaz}). All solutions are
calculated for  $\ell=1$ modes and $\Pi_0=639$ s. In the {\sl upper
panel} we fix the value of $\Pi_0/\Pi_\mu$  to  0.15 and vary
$\alpha$: $\alpha=0.35$ (continuous line),   $\alpha=0.75$ (dashed
line) and  $\alpha=1$ (dotted line). In the {\sl lower panel} we fix
the value of $\alpha$ to 0.35 and vary  the value of
$\Pi_0/\Pi_\mu$:  $\Pi_0/\Pi_\mu=0.15$ (continuous line), 0.225
(dashed line) and 0.1 (dotted line).} \label{fig:solutions}
\end{center}
\end{figure}

Numerical solutions of Eq. (\ref{eq:relaz}), found using a
bracketing-bisection method \citep[see][]{Press92}, are shown in
Fig. \ref{fig:solutions}. As is clearly visible when comparing Figs.
\ref{fig:solutions} and \ref{fig:6trapg}, we find that the solutions
of Eq. (\ref{eq:relaz}) better match the oscillatory behaviour of
the period spacing than the sinusoids of Eq. (\ref{eq:varia}).

In the following section we extend to a wider range of main-sequence
models the analysis presented for a 6 \msol model.

\section{Application to stellar models}

\label{sec:numeric}
The occurrence of a sharp chemical composition gradient in the central region of a star is determined by the
 appearance of convection in the core and by the displacement of the convective core boundary during the main sequence.
For a given chemical composition, and if no non-standard transport
process is included in the modelling,
 the transition from radiative to convective energy transport, as well as the  shape of the $\mu$ gradients in the
central stellar region, are determined by the mass of the model.
On the other hand, additional mixing processes may  alter the evolution of the convective core
and the detailed properties of the chemical composition profile.

As shown in the previous section, the features of periodic signals
in the period spacing of high order g-modes can provide very
important information on the size of the convective core and on the
mixing-processes able to change the $\mu$ gradients generated during
the evolution.

In this section, we present a survey of the properties of adiabatic
$\ell=1$  high order g-modes in main-sequence stars with masses from
1 to 10~\msun, and for four different evolutionary stages: those
corresponding to a central hydrogen mass fraction $X_{\rm c}$ of
0.7, 0.5, 0.3 and 0.1. All these models were computed with the same
initial chemical composition $(X_0,Z_0)=(0.70,0.02)$. The adiabatic
oscillation frequencies were computed with LOSC
\citep{Scuflaire07b}.

We first study how the properties of high order gravity modes depend
on the mass and the evolutionary stage of the model
(Sec.~\ref{sec:coresms}). In a  second step we evaluate the effects
of the inclusion of extra-mixing such as overshooting, diffusion and
turbulent mixing (Sec.~\ref{sec:extra}). The behaviour of modes with
different $\ell$ will be briefly addressed in Sec.~\ref{sec:ell}.


\subsection{Convective core evolution: stellar mass dependence}
\label{sec:coresms}
\begin{figure}
\resizebox{\hsize}{!}{\includegraphics[angle=-90]{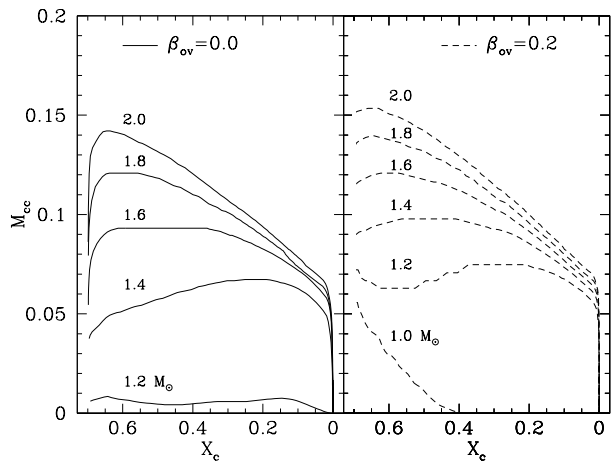}}
\caption{\small Fractional mass of the convective core as a function
of central hydrogen abundance for models computed with  (right
panel) and without (left panel) overshooting, for masses between 1.0
and 2~\msun. $\beta$ is the overshooting parameter as defined in
Sect.~\ref{sec:over0}.}\label{fig:cc}
\end{figure}

In our analysis of the signatures of the $\mu$~gradients on  g-modes, we consider three
stellar mass domains:
{\it i)} $M < \rm M_{\rm Lcc}$, with $\rm M_{\rm Lcc}$ being the minimum mass required to keep a
convective core during the main sequence;
{\it ii)}  $\rm M_{\rm Lcc}$$\leq M \leq \rm M_{\rm gc}$, for which the mass of the convective core increases
during part of the main sequence; and {\it iii)}  $M > \rm M_{\rm gc}$ for which the convective core
recedes as the star evolves. The situations described here above are presented in Fig. \ref{fig:cc}, where the size in mass of the convective core is shown as a function of the central hydrogen abundance (and therefore of the age) of the star.
The exact values of $\rm M_{\rm Lcc}$ and $M > \rm M_{\rm gc}$  depend on the chemical composition and, as we shall see
below, on extra-mixing processes.

\subsubsection{ Models with a radiative core}
\label{sec:radiative} As a first example we consider the evolution
of the period spacing on the main sequence in models without a
convective core, e.g. in a 1 \msol star.
\begin{figure}
\begin{center}
\resizebox{\hsize}{!}{\includegraphics[angle=0]{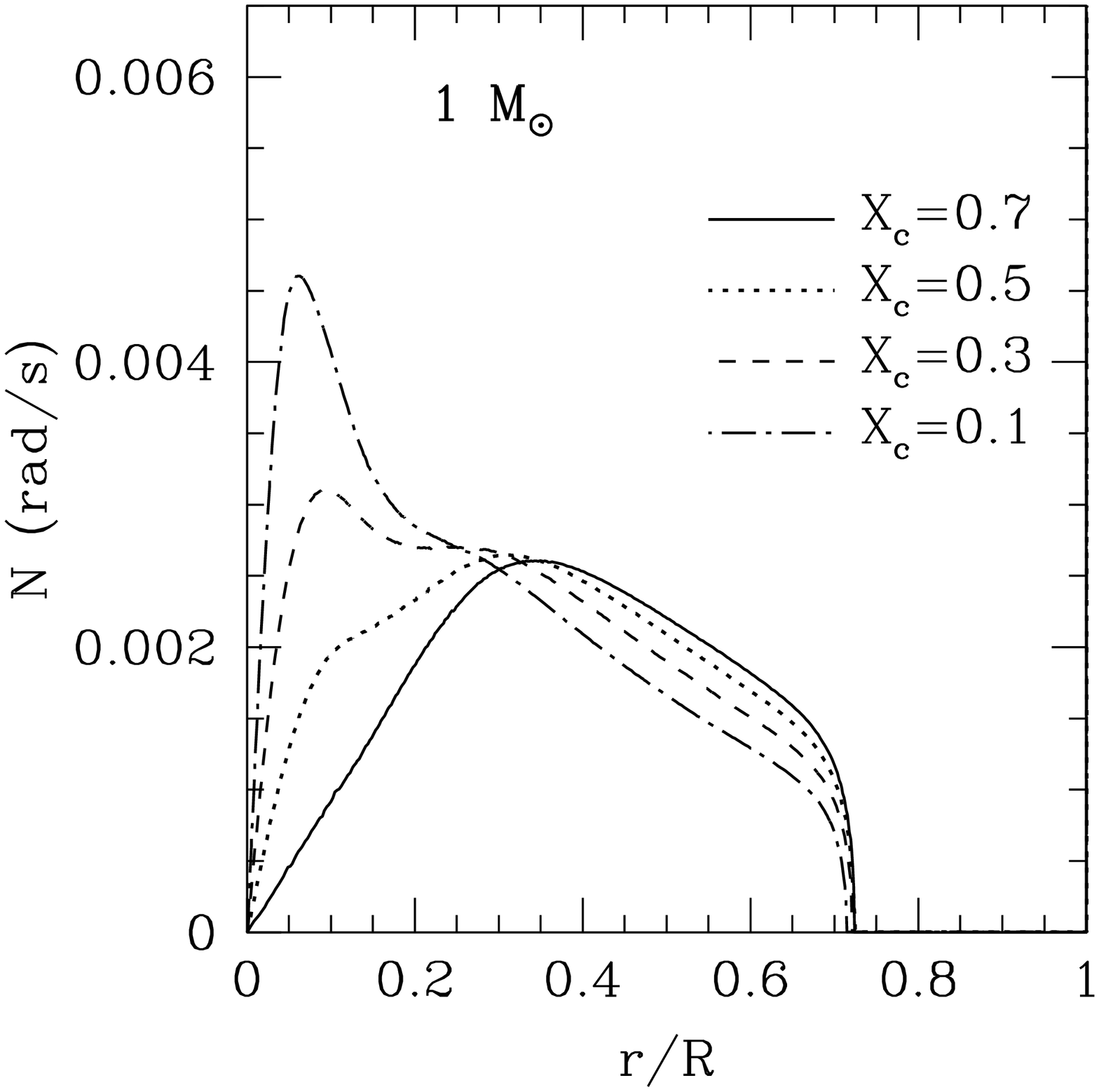}\includegraphics[angle=0]{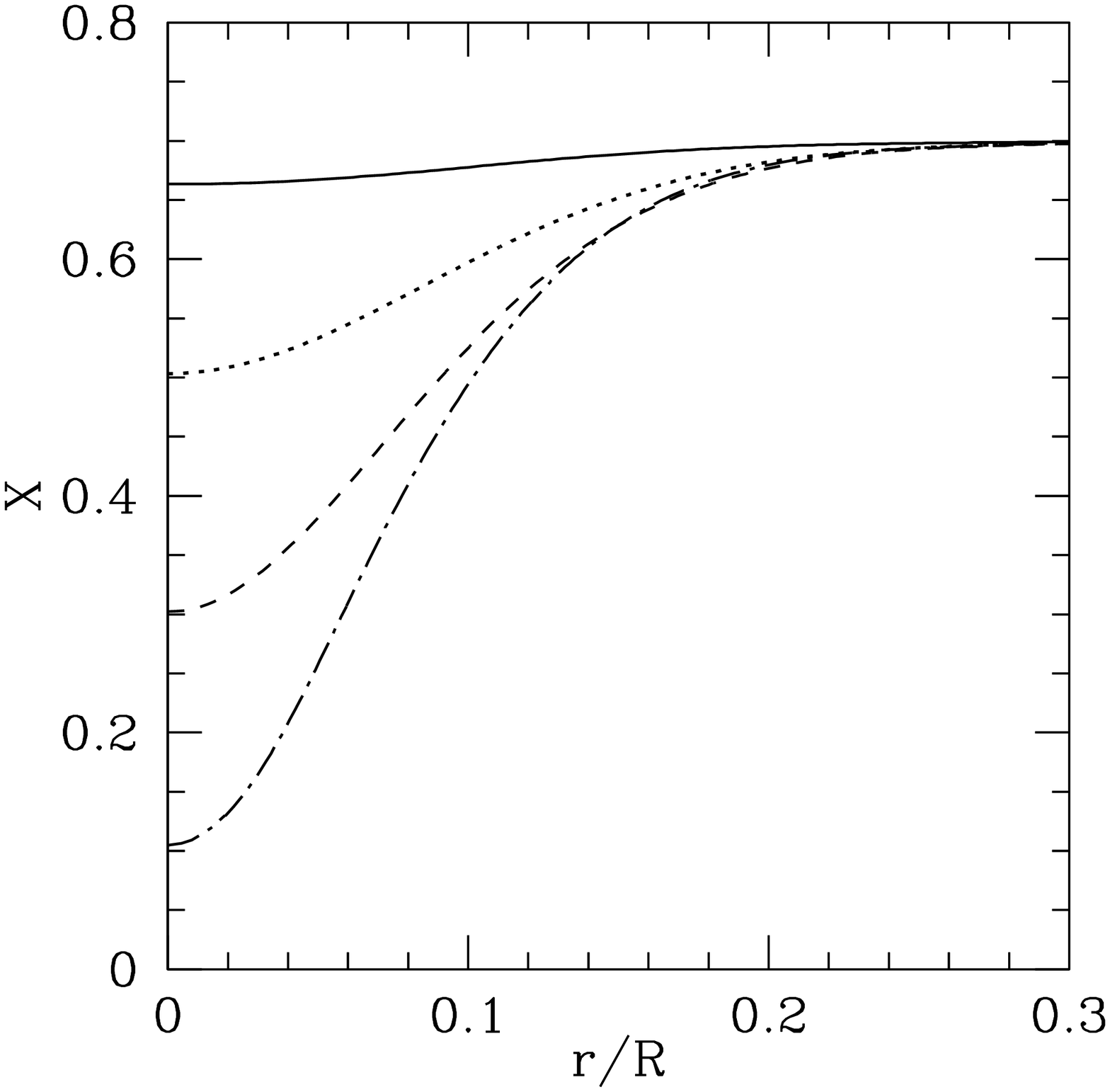}}
\resizebox{0.8\hsize}{!}{\includegraphics[angle=0]{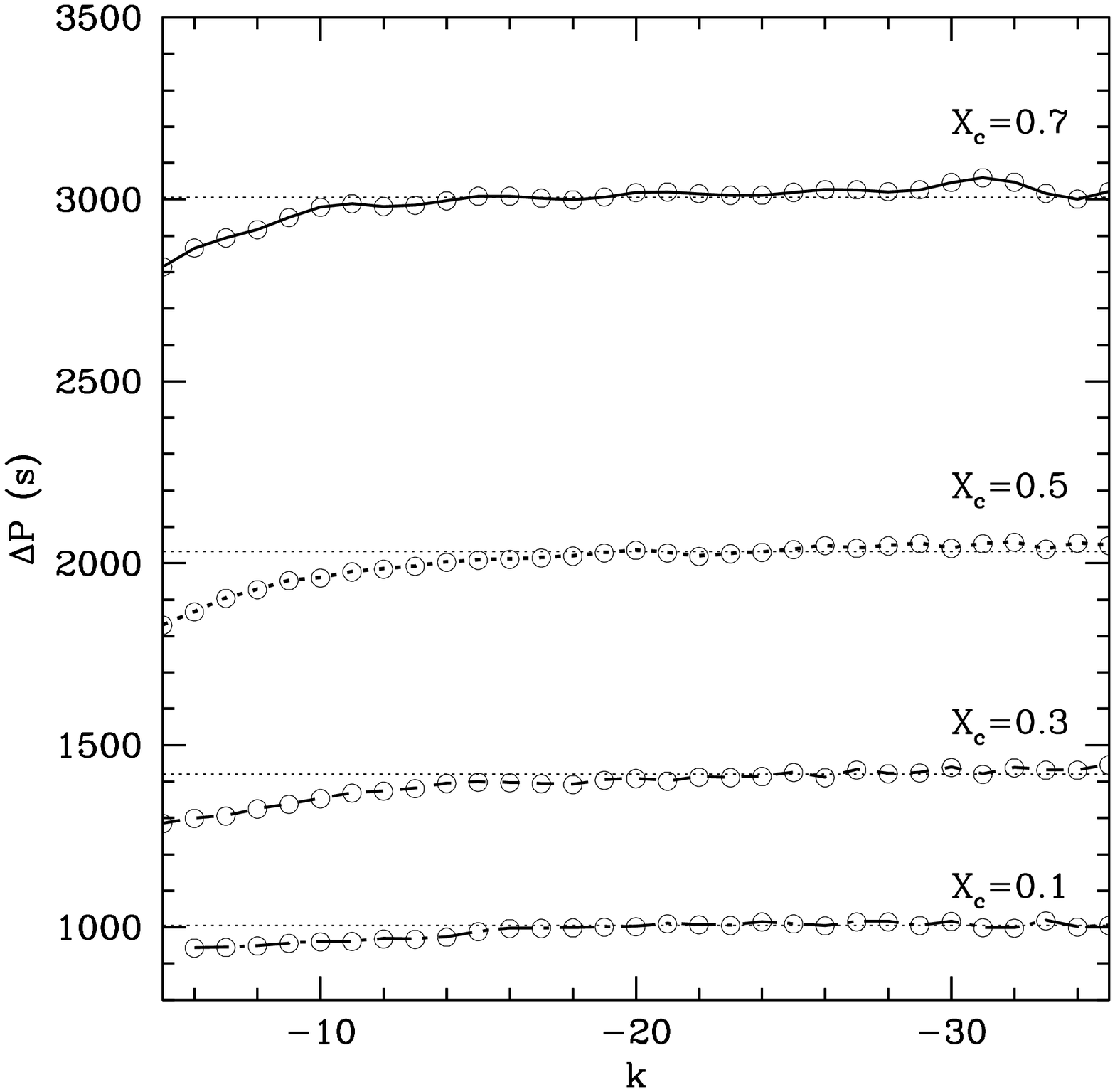}}
\caption{\small  {\it Upper left panel:} $N$ as a function of the
normalized radius in 1 \msol models with decreasing central hydrogen
abundance. {\it Upper right panel:} Hydrogen abundance profile {\it
versus} $r/R$. {\it Lower panel:} g-modes period spacing as a
function of the radial order $k$. The asymptotic value of $\Delta P$
(predicted by Eq. \ref{eq:asy}) is represented, for each model, by
dotted lines.} \label{fig:m1}
\end{center}
\end{figure}
The behavior is substantially different from higher mass models: $\Delta P$, as shown in Fig. \ref{fig:m1}, considerably decreases during the main sequence: this can easily be understood recalling the first order asymptotic expression for the mean period spacing. The increase of $N$ near the centre of the star,
 due to the mean molecular weight  gradient developing in a radiative region (see upper panel of Fig.  \ref{fig:m1}), has a larger and larger contribution to $\int{\frac{N}{x'}dx'}$, leading to a significant reduction of the mean period spacing. The increase of $N$ near the centre is however not sufficient to produce any periodic  component of appreciable amplitude in the period spacing (see lower panel of Fig. \ref{fig:m1}).


\subsubsection{Models with a growing convective core on the main-sequence.}
\label{sec:growing}
In models with masses between $\rm M_{\rm Lcc}$ and $M_{\rm gc}$,  the contribution of nuclear burning through CNO cycle becomes more and more important as the star evolves on the main sequence \citep[see e.g.][]{Gabriel77,Crowe82,Popielski05}.
 The ratio $L/m$ in the nuclear burning region becomes large enough to alter the behaviour of $\nabla_{\rm rad}$: the latter increases and so does the size of the convective core.

\begin{figure}
\begin{center}
\resizebox{0.8\hsize}{!}{\includegraphics[angle=-90]{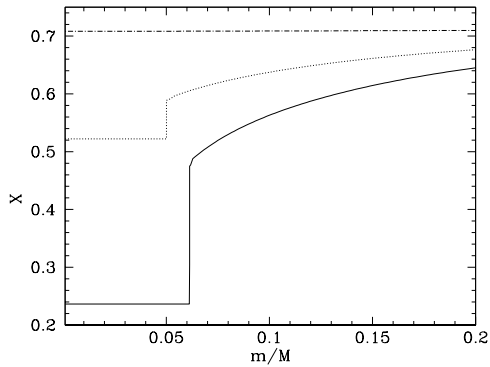}}
\caption{\small The discontinuous chemical composition profile generated by a growing convective core in 1.3 \msol (see Fig. \ref{fig:sc}) when no extra-mixing outside the core is allowed.}\label{fig:scx}
\end{center}
\end{figure}

\begin{figure}
\begin{center}
\resizebox{0.8\hsize}{!}{\includegraphics[angle=0]{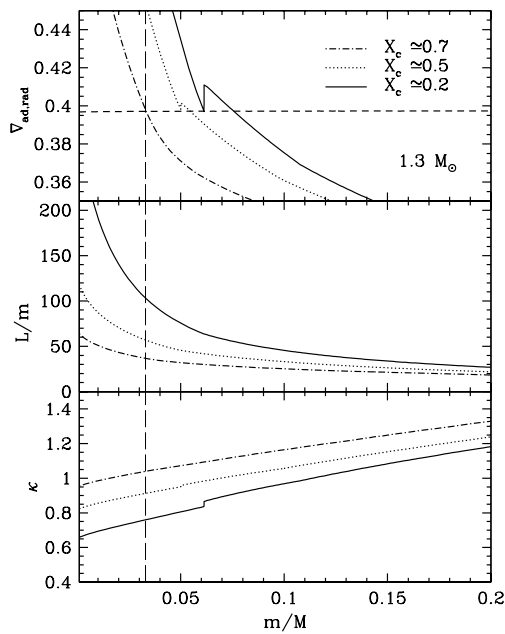}}
\caption{\small {\it Upper panel:} Radiative gradient in the inner regions of 1.3 \msol models at different stages on the main sequence. The adiabatic gradient is represented with a short-dashed line. {\it Middle panel:} Ratio $L(r)/m(r)$. During the evolution the large increase of $L/m$ at the former border of the convective core dominates the behaviour of $\nabla_{\rm rad}$. The vertical dashed line denotes the border of the convective core in the ZAMS model. {\it Lower panel:} Behaviour of $\kappa$ for the same models as in the other panels.}\label{fig:sc}
\end{center}
\end{figure}

A growing convective core generates a discontinuity in the chemical
composition at its boundary (see Fig.~\ref{fig:scx}), and may lead
to an inconsistency in the way the convective boundary is defined.
The situation is illustrated in Fig. \ref{fig:sc}: the discontinuous
hydrogen profile forces the radiative gradient to be discontinuous
and to increase outside the region that is fully mixed by
convection, and therefore, this region should be convective as well!
If this is the case, then we have a contradictory situation: if we
allow this region to have the same chemical composition as the core,
then $\nabla_{\rm rad}$ decreases and the region becomes radiative
again. The question of the semi-convection onset in models with
masses in the range 1.1-1.6 was already addressed by
\cite{Gabriel77} and \cite{Crowe82} quite some time ago.
Nevertheless, what happens in this so-called ``semi-convective''
region is still a matter of debate. Some mixing is likely to take
place, so that the composition gradients are adapted to obtain
$\nabla_{\rm rad}=\nabla_{\rm ad}$ in the semiconvective region.

Even if no specific mixing is added in the semi-convective region,
the $\mu$ gradient at the boundary of the convective core is very
sensitive to the details of the numerical algorithm used in
describing the core evolution. In fact, a strict discontinuity in
chemical composition is only obtained  if the border of the
convective region is treated with a double mesh point
(Fig.~\ref{fig:scx} for instance). This ``unphysical'' framework
leads to a problem when computing the Brunt-V\"ais\"al\"a frequency.
The numerical difficulty can however be avoided keeping a
quasi-discontinuous chemical composition, with a sharp change of $X$
in an extremely  narrow region ($\delta_0=x_\mu-x_0$) outside the
convective core ($x_0$).
 From Eqs.~(\ref{eq:pimu}) and (\ref{eq:varia}) it is evident that
the signal in the period spacing will then have an almost infinite
period.

Of course, any  treatment of the semi-convective region should
destroy the discontinuity  leading to a wider $\mu$-gradient region.
The chemical composition  discontinuity may also be removed by a
sort of ``numerical diffusion'' that appears when the grid of mesh
points (necessarily finite) in the modelling  does not follow the
convection limits. That is the case of the evolution code ({\sc
CLES},  \citealt{Scuflaire07}) used to compute most of the stellar
models presented in this paper. In these models, the region where
the discontinuity would be located is assumed to have an
intermediate chemical composition between the one in the outermost
point of the convective core and the one in the innermost point of
the radiative region. The final effect is to have a partial mixing
at the edge of the convective core, and thus to remove the
discontinuity in $\mu$.

\begin{figure}
 \begin{center}
\resizebox{0.48\hsize}{!}{\includegraphics[angle=0]{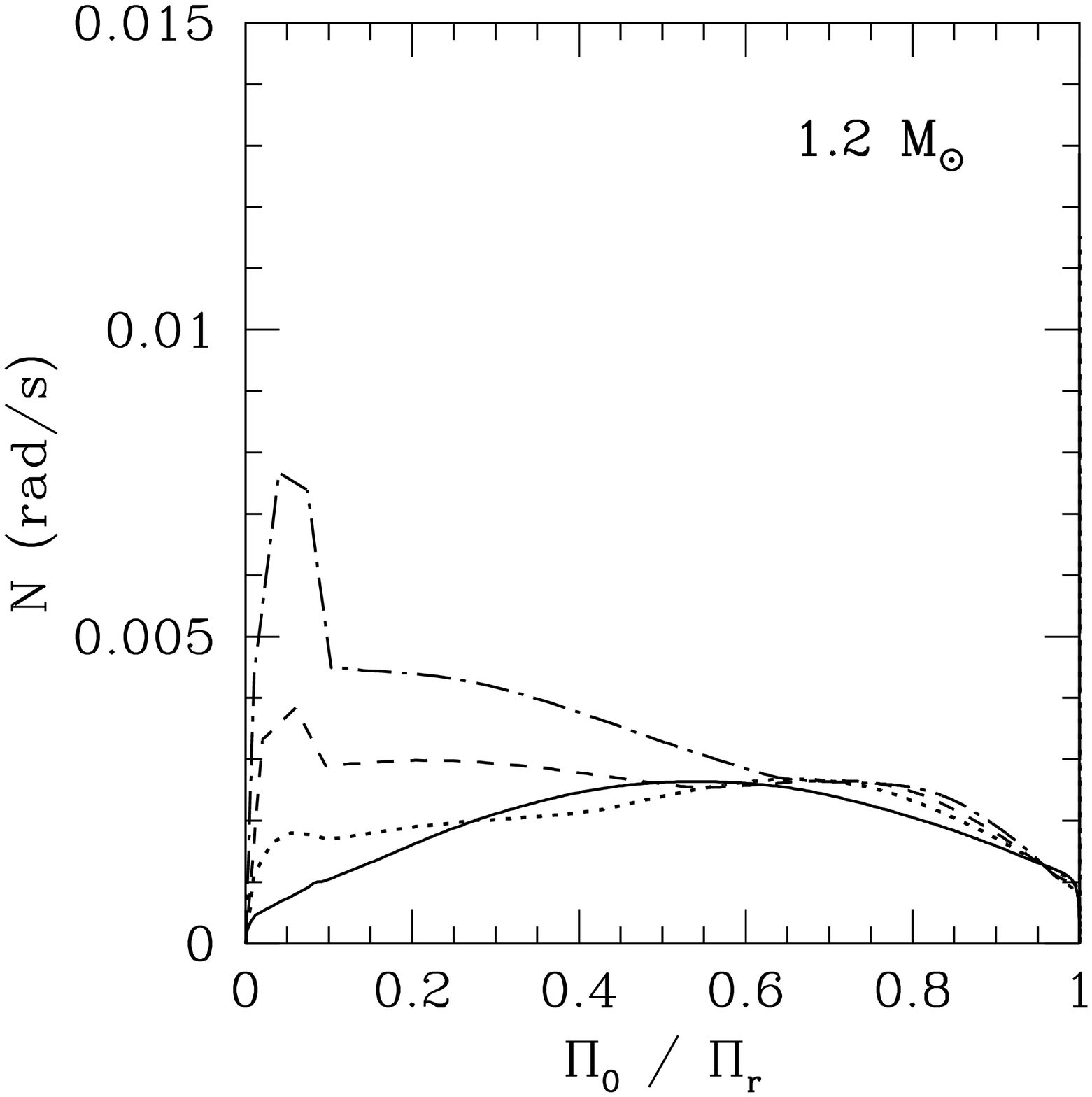}}
\resizebox{0.48\hsize}{!}{\includegraphics[angle=0]{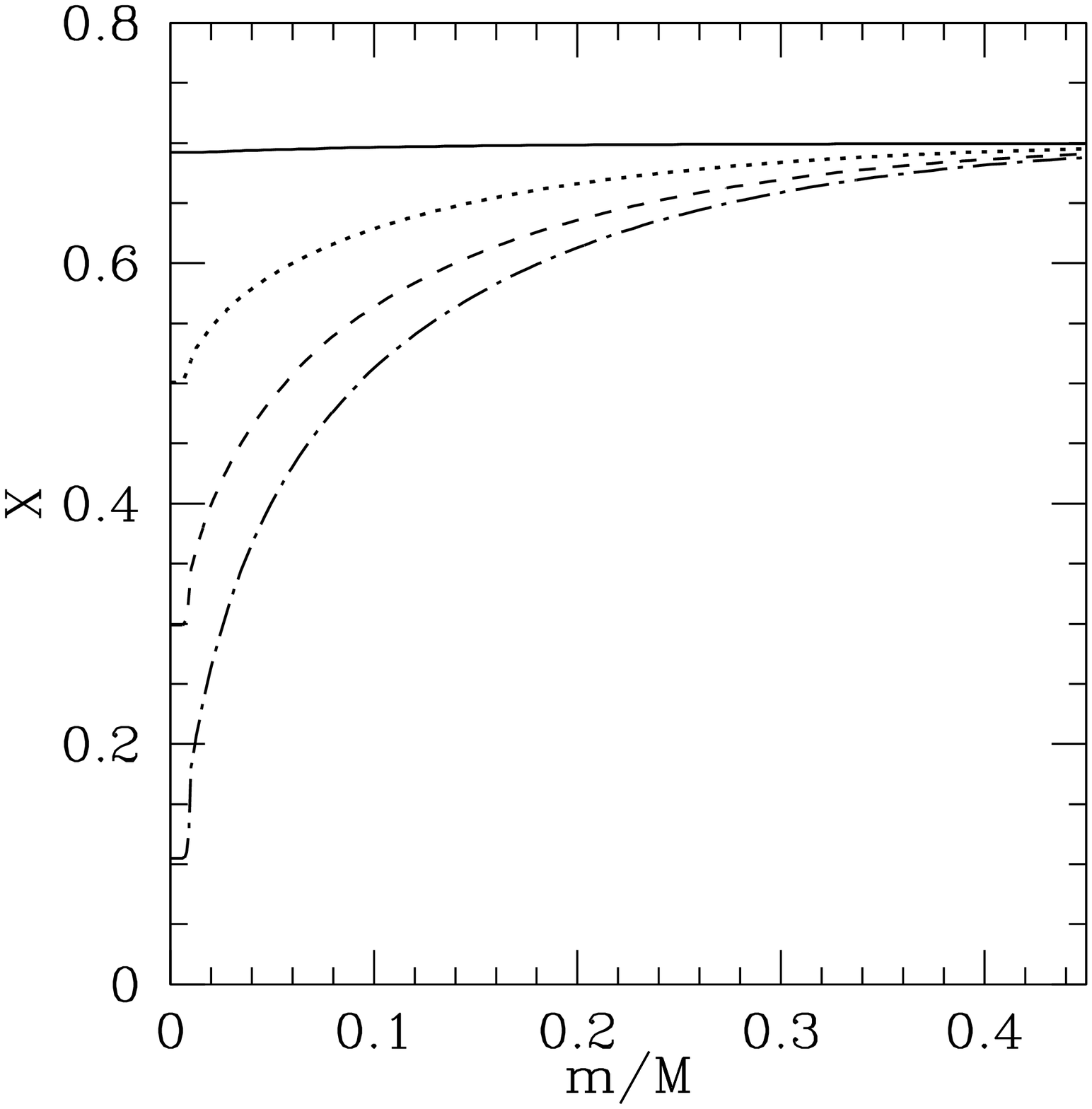}}
\resizebox{0.8\hsize}{!}{\includegraphics[angle=0]{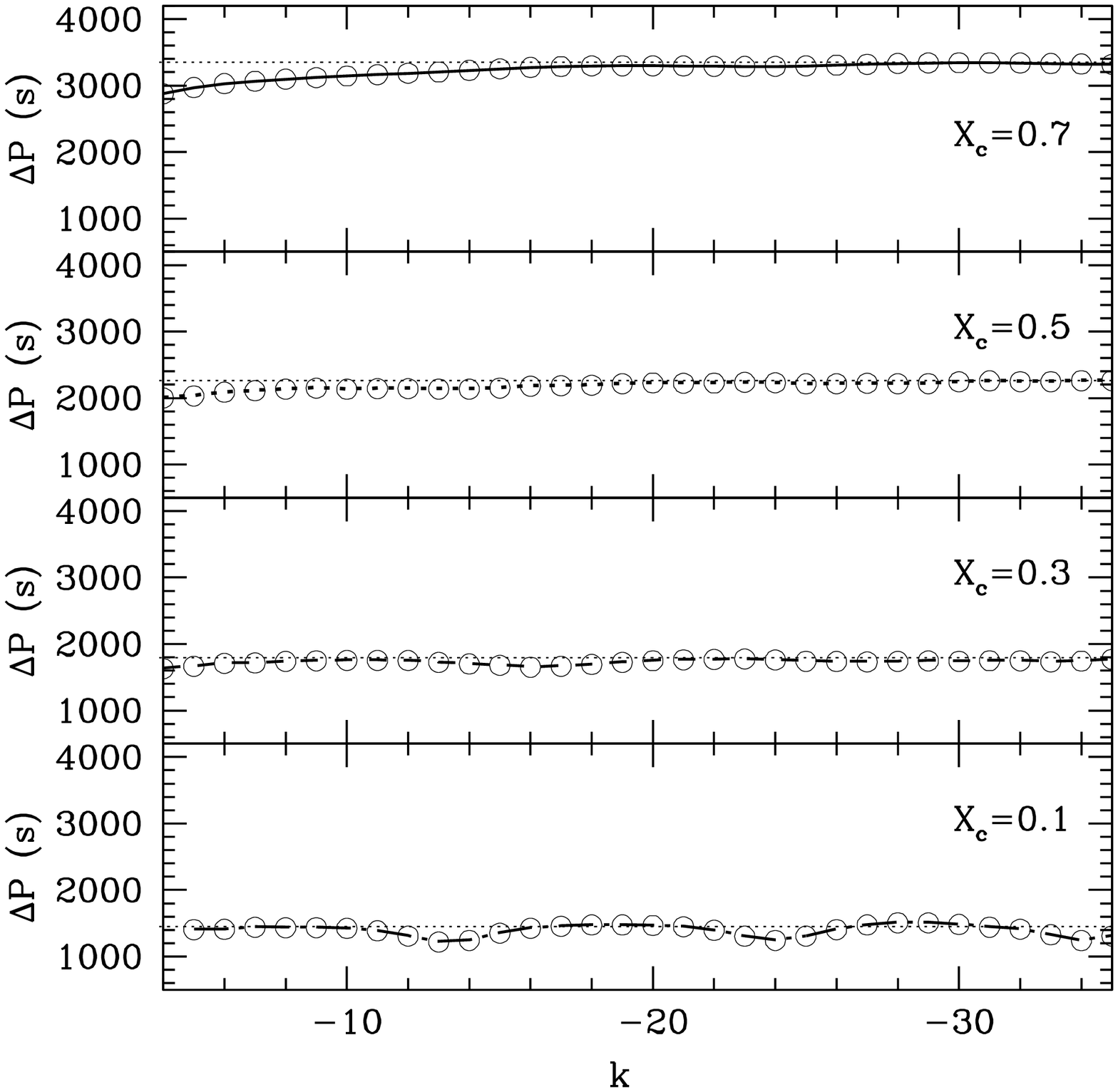}}
\caption{\small Behavior of the \BV frequency (upper left panel), of
the hydrogen abundance profile (upper right panel) and of the
$\ell=1$ g-mode period spacing in models of 1.2 \msun. We consider,
as in all the following figures, several models on the main sequence
with decreasing central hydrogen abundance ($X_c$ 0.7, 0.5, 0.3 and
0.1).} \label{fig:12}
 \end{center}
\end{figure}

\begin{figure}
 \begin{center}
\resizebox{0.48\hsize}{!}{\includegraphics[angle=0]{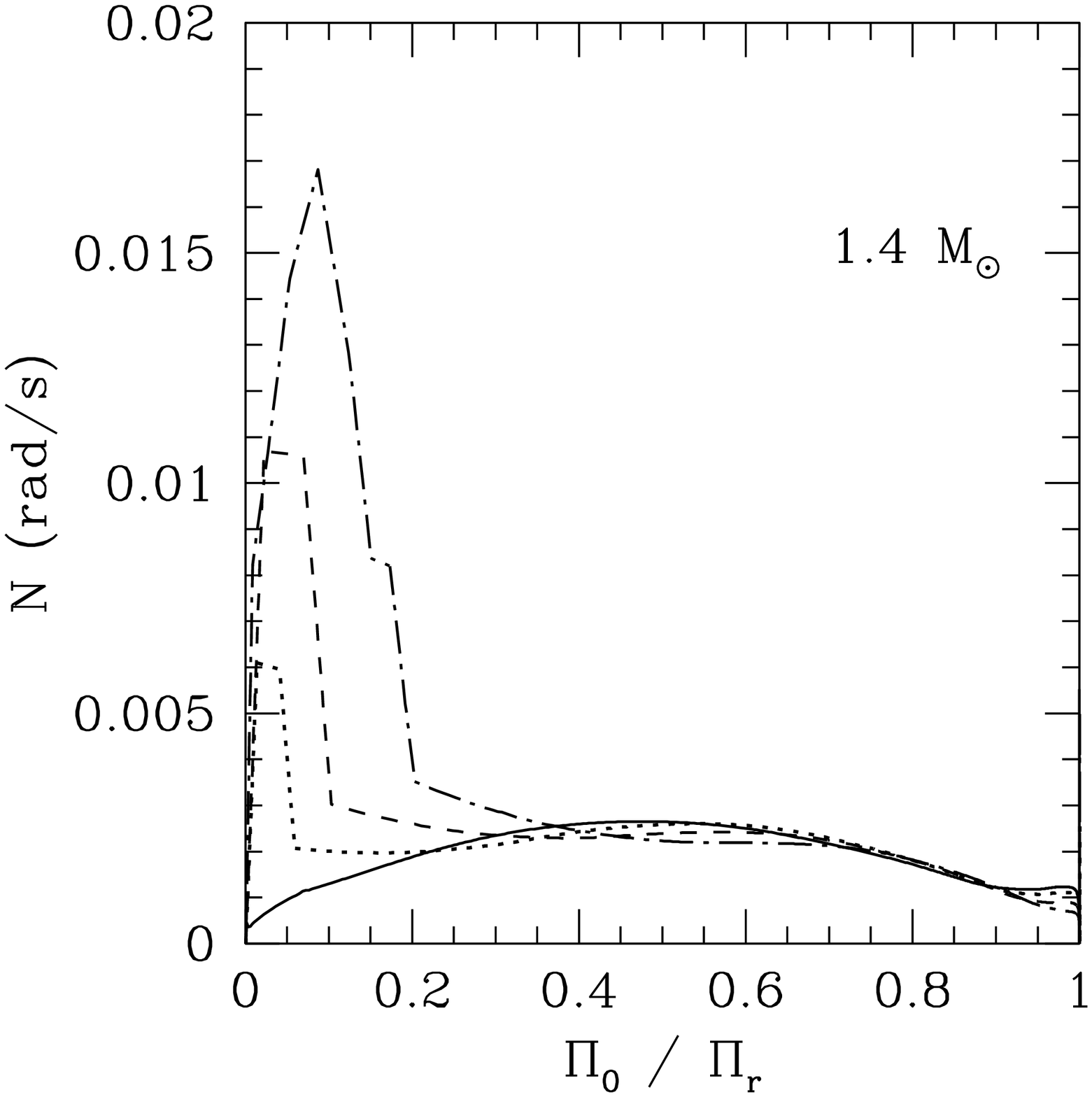}}
\resizebox{0.48\hsize}{!}{\includegraphics[angle=0]{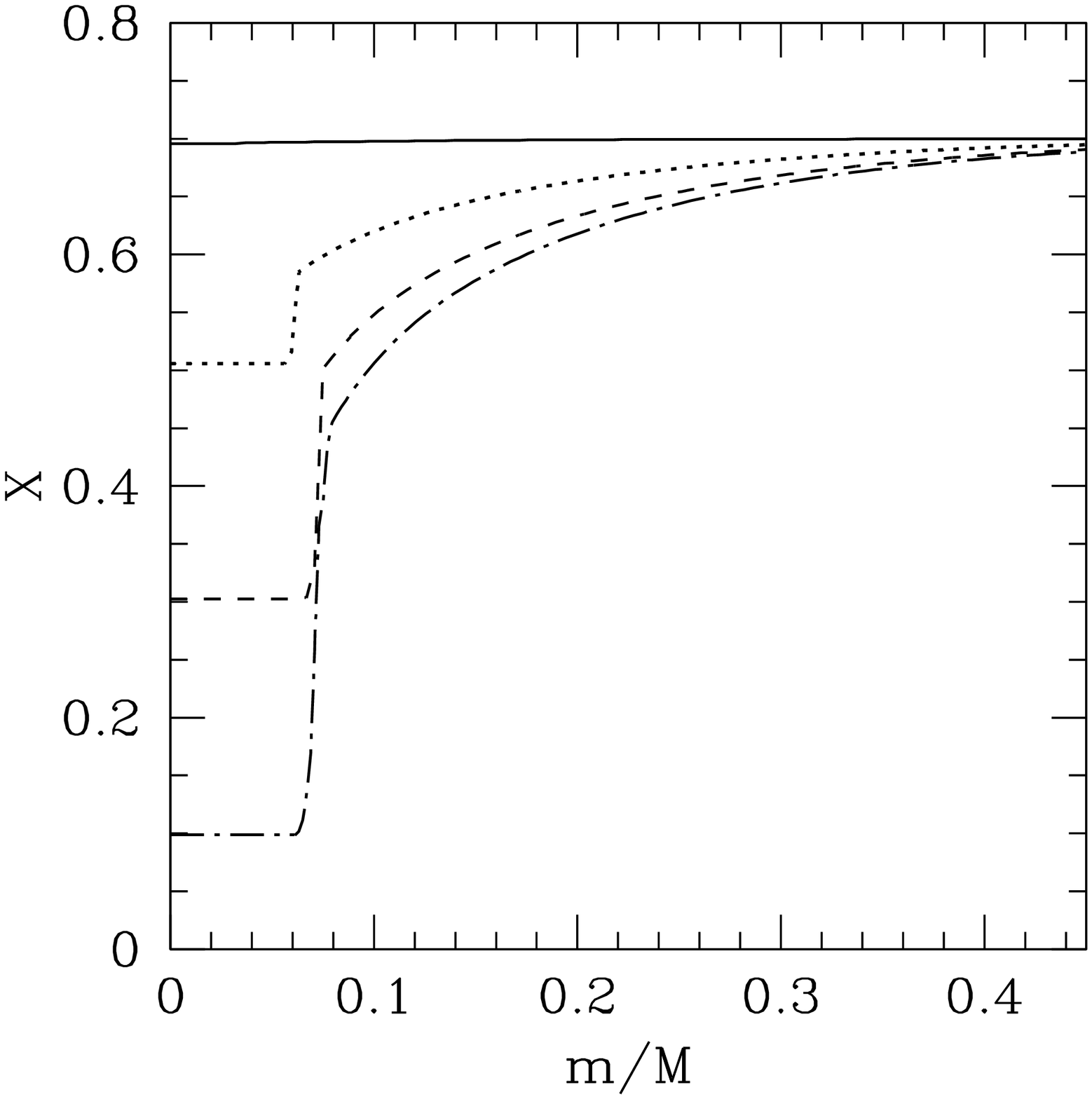}}
\resizebox{0.8\hsize}{!}{\includegraphics[angle=0]{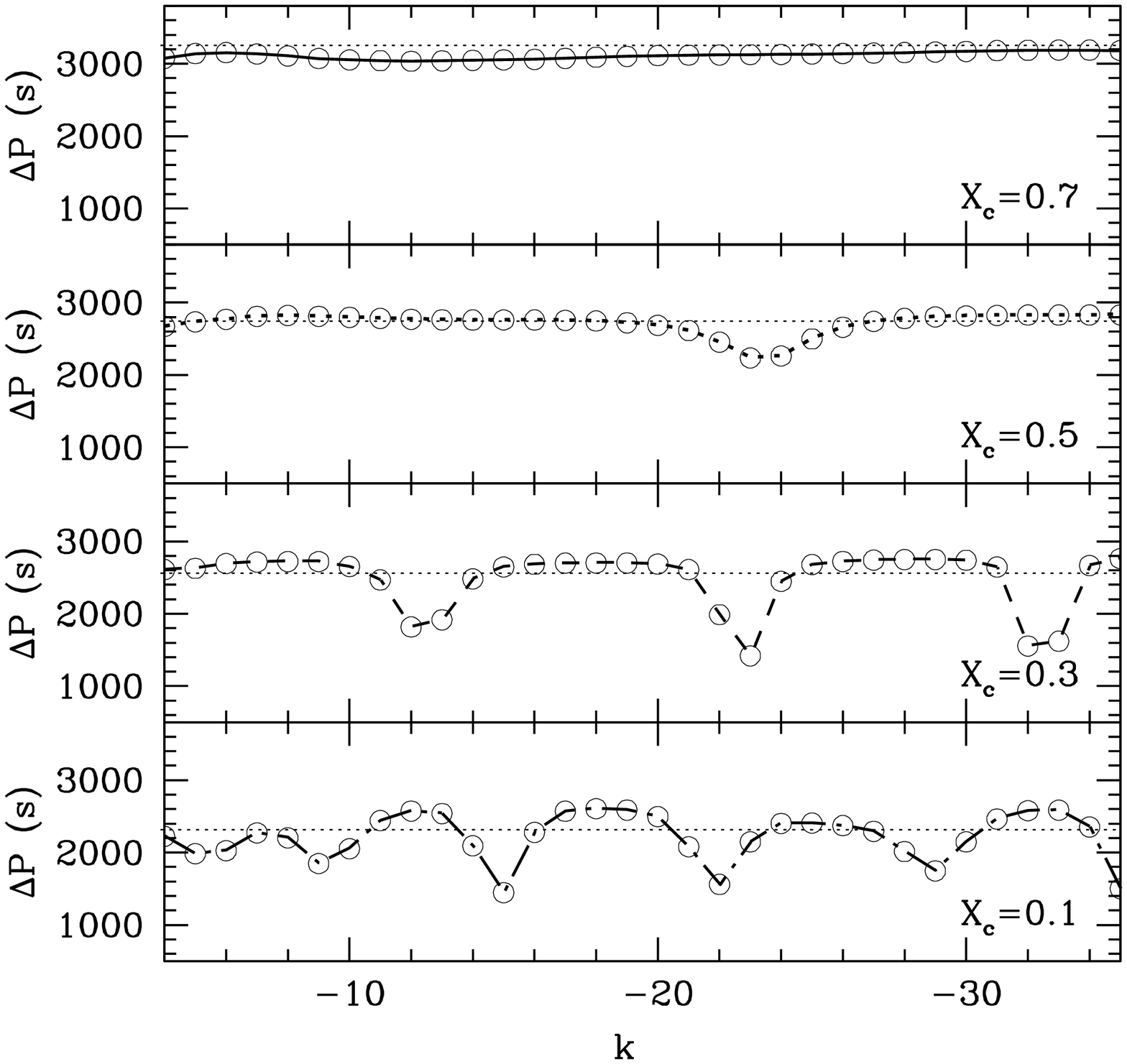}}
\caption{\small Same as Fig.\ref{fig:12} for 1.4 \msol models.}
\label{fig:14}
 \end{center}
\end{figure}

Furthermore, in   models with a mass $M\simeq \rm M_{\rm Lcc}$, e.g.
$M=1.2$ \msun, the convective core is so small ($m/M\,\sim 0.01$)
that the period spacing resembles the behaviour of the 1 \msol
model. We notice, however, the appearance of oscillatory components
in $\Delta P$ in the model with $X_c\simeq 0.1$ (see Fig.
\ref{fig:12}). The sharp variation of $N$ located at $\Pi_0/\Pi_r
\simeq 0.1$ is large enough to generate components with a
periodicity of $10$ $k$ in $\Delta P$. In more massive models,
the $\mu$ gradient becomes larger and so does the amplitude of the components in the period spacing (see e.g. Fig. \ref{fig:14}).

\subsubsection{Models with a receding convective core.}
\label{sec:receding} In models with shrinking convective cores the
situation is much simpler. If $M>M_{\rm gc}$ the dominant term in
the behaviour of the radiative gradient is the opacity and, since
$\kappa\propto(1+X)$, $\nabla_{\rm rad}$ decreases with time as X
decreases: the boundary of the convective core is displaced towards
the centre. A receding convective core leaves behind a chemical
composition gradient that is responsible for an abrupt change in the
$N$ profile (as in Fig. \ref{fig:6trapbvx}). Such a  sharp feature
which is a direct consequence of the evolution of a convective core,
leaves a clear signature in the periods of gravity modes.

\begin{figure}
 \begin{center}
\resizebox{0.48\hsize}{!}{\includegraphics[angle=0]{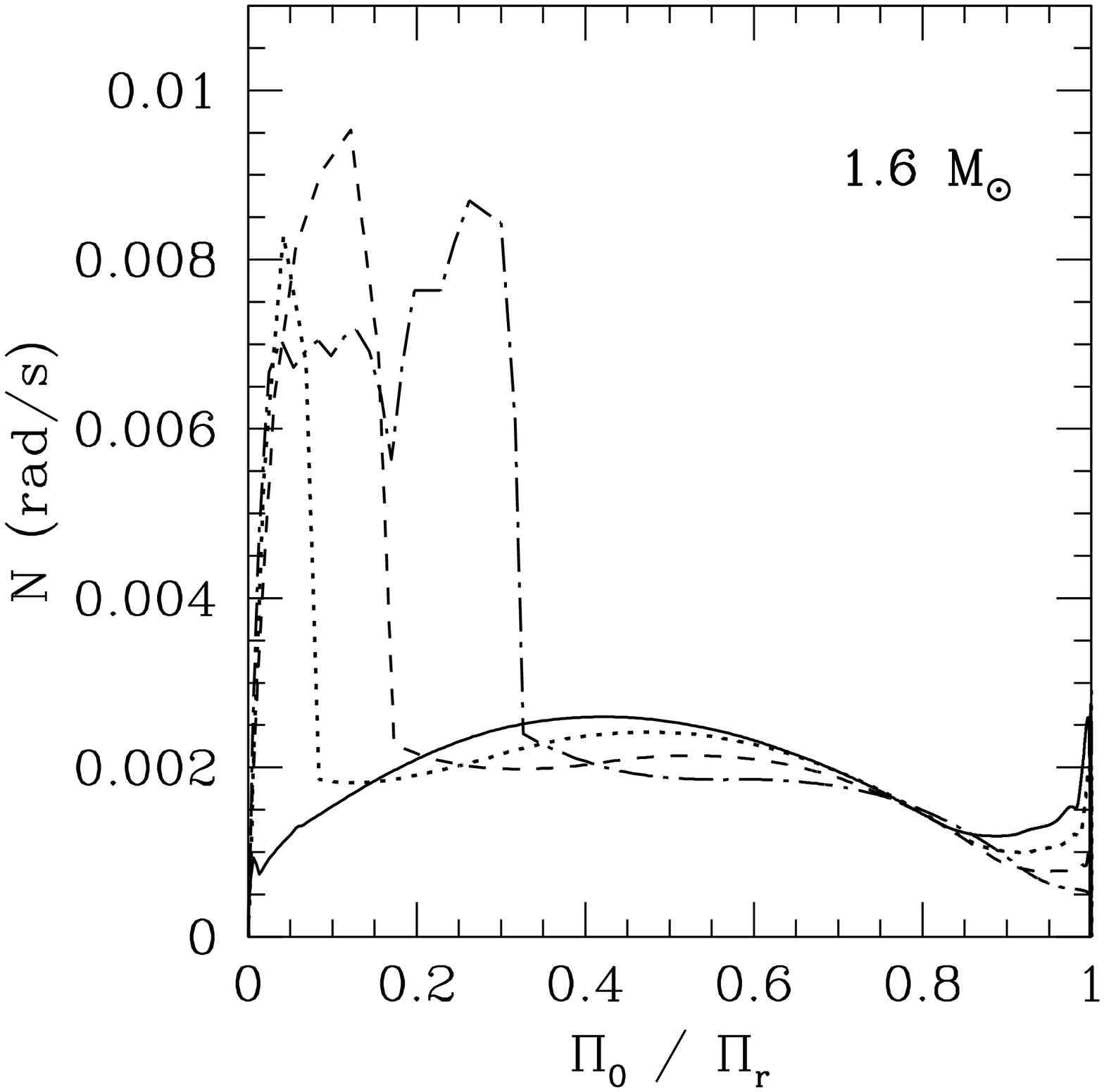}}
\resizebox{0.48\hsize}{!}{\includegraphics[angle=0]{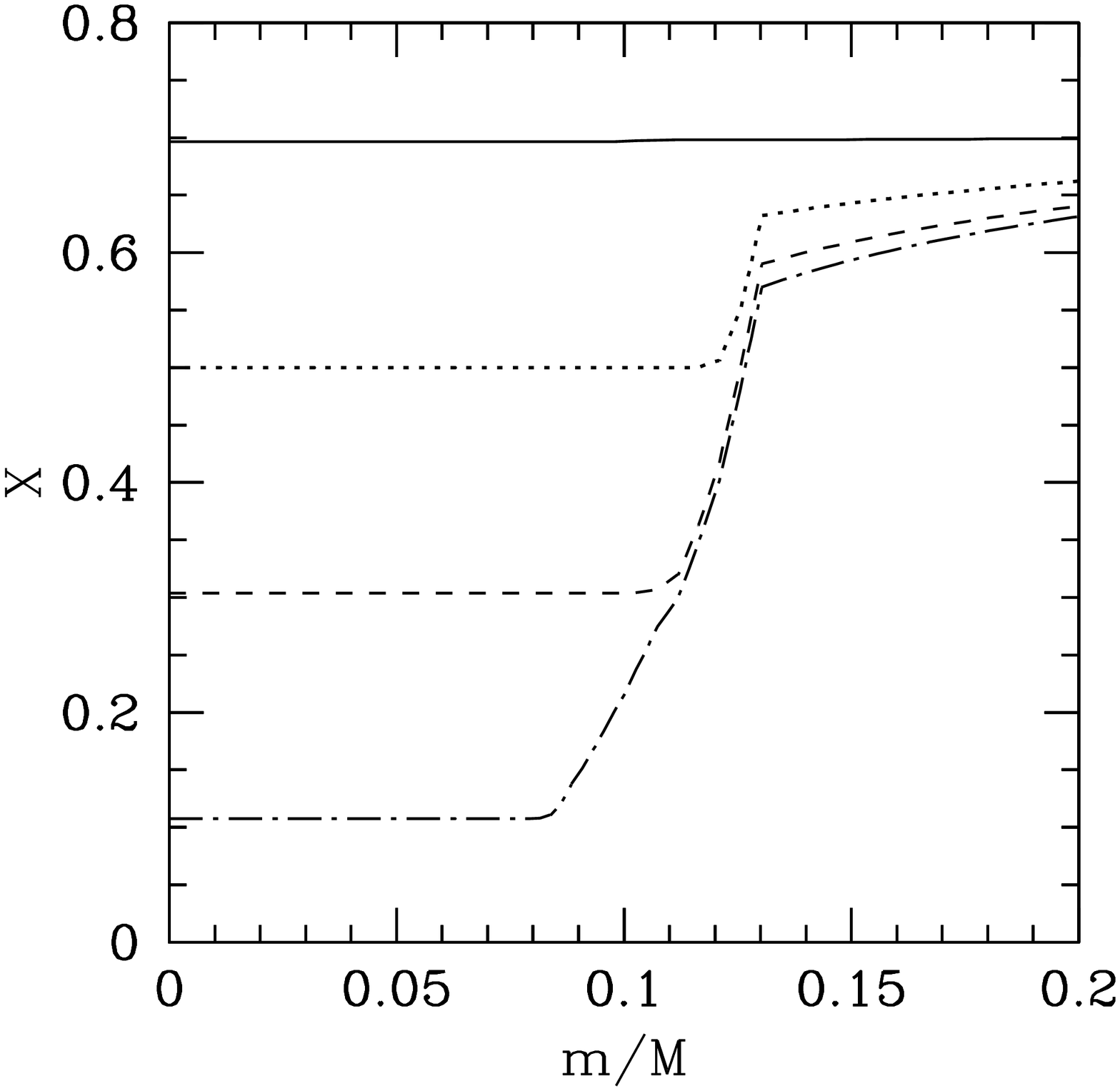}}
\resizebox{0.8\hsize}{!}{\includegraphics[angle=0]{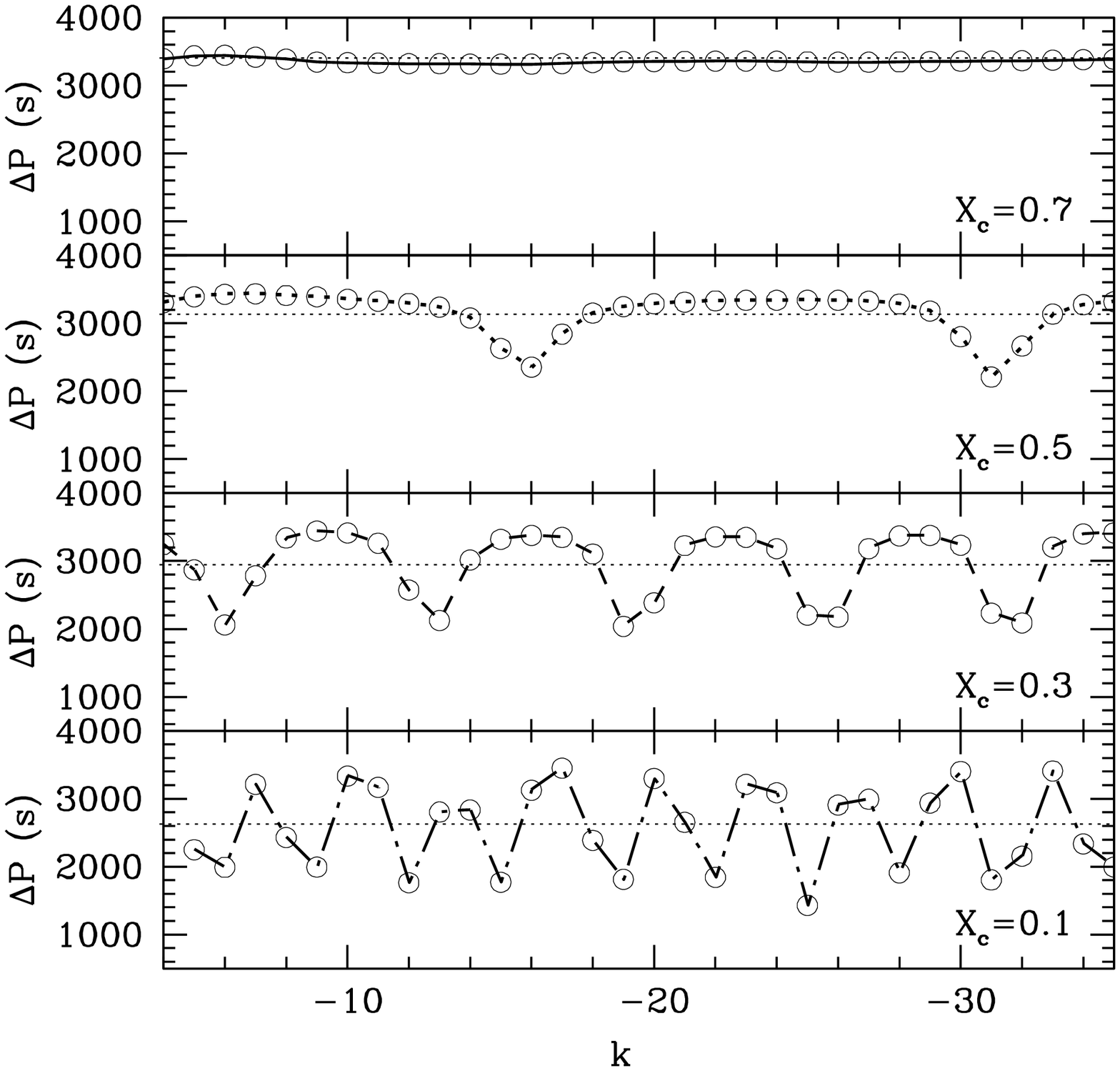}}
\caption{\small Same as Fig.\ref{fig:12} for 1.6 \msol models.}
\label{fig:16}
 \end{center}
\end{figure}

\begin{figure}
\begin{center}
\resizebox{0.48\hsize}{!}{\includegraphics[angle=0]{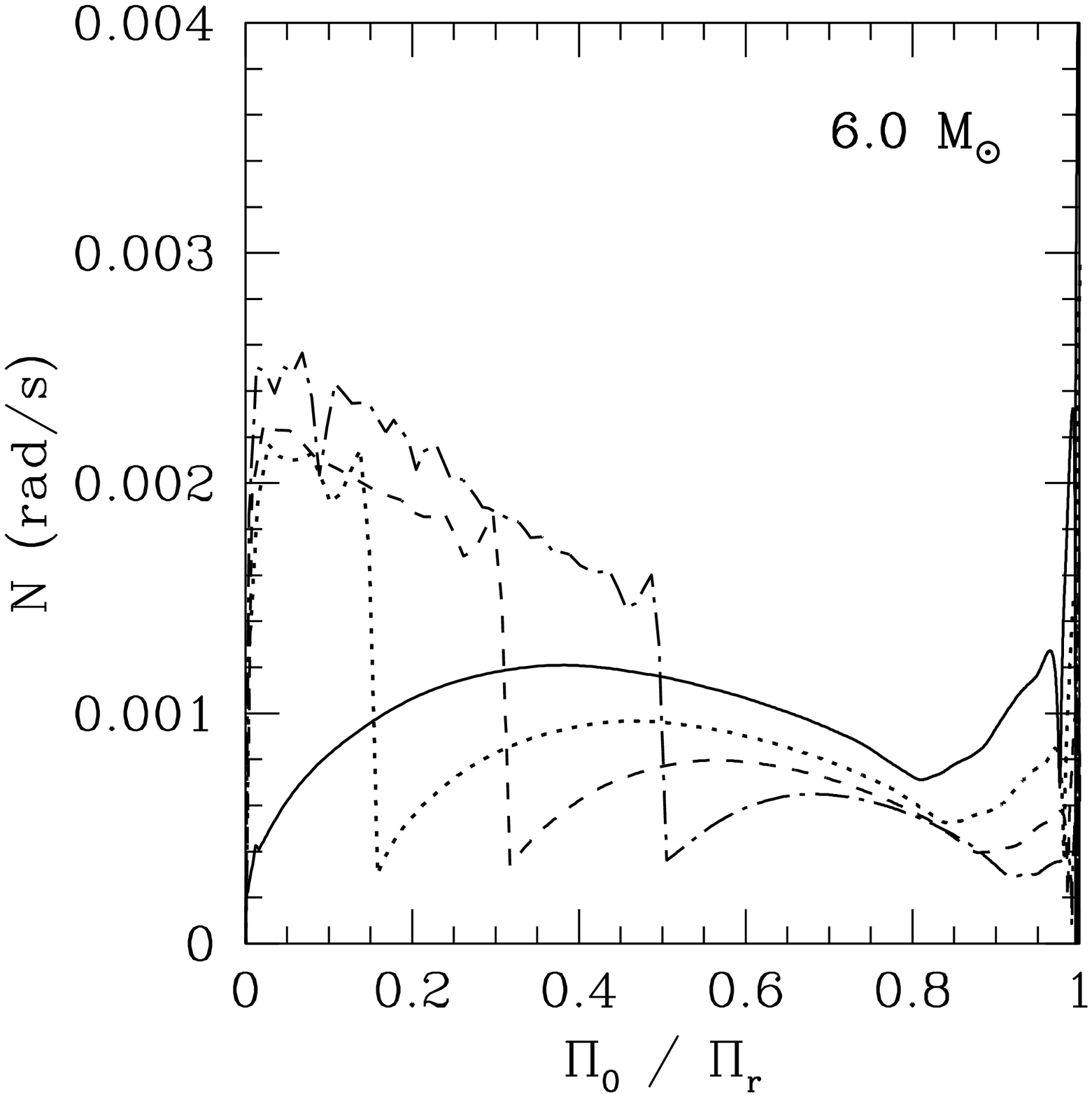}}
\resizebox{0.48\hsize}{!}{\includegraphics[angle=0]{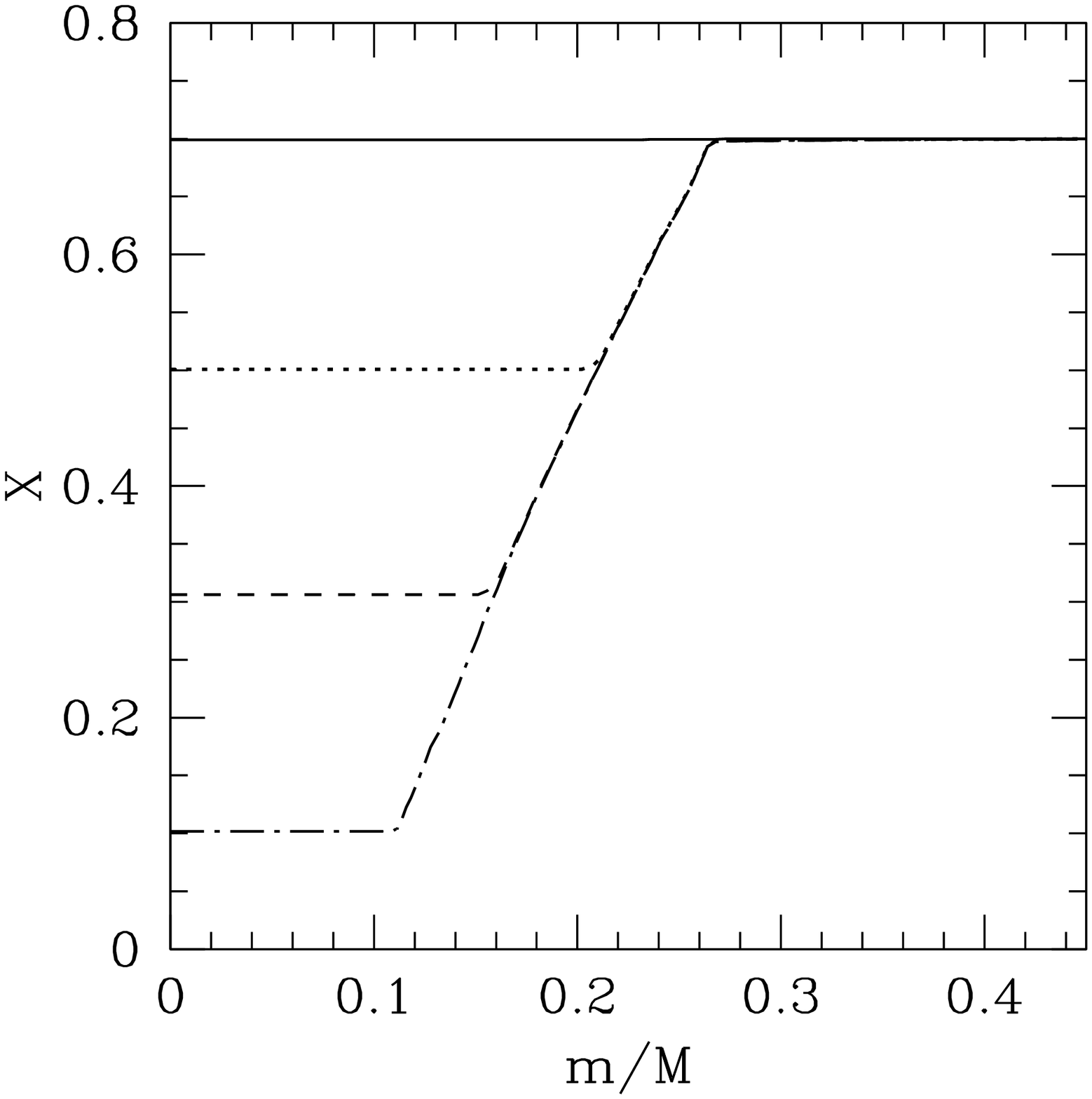}}
\resizebox{0.8\hsize}{!}{\includegraphics[angle=0]{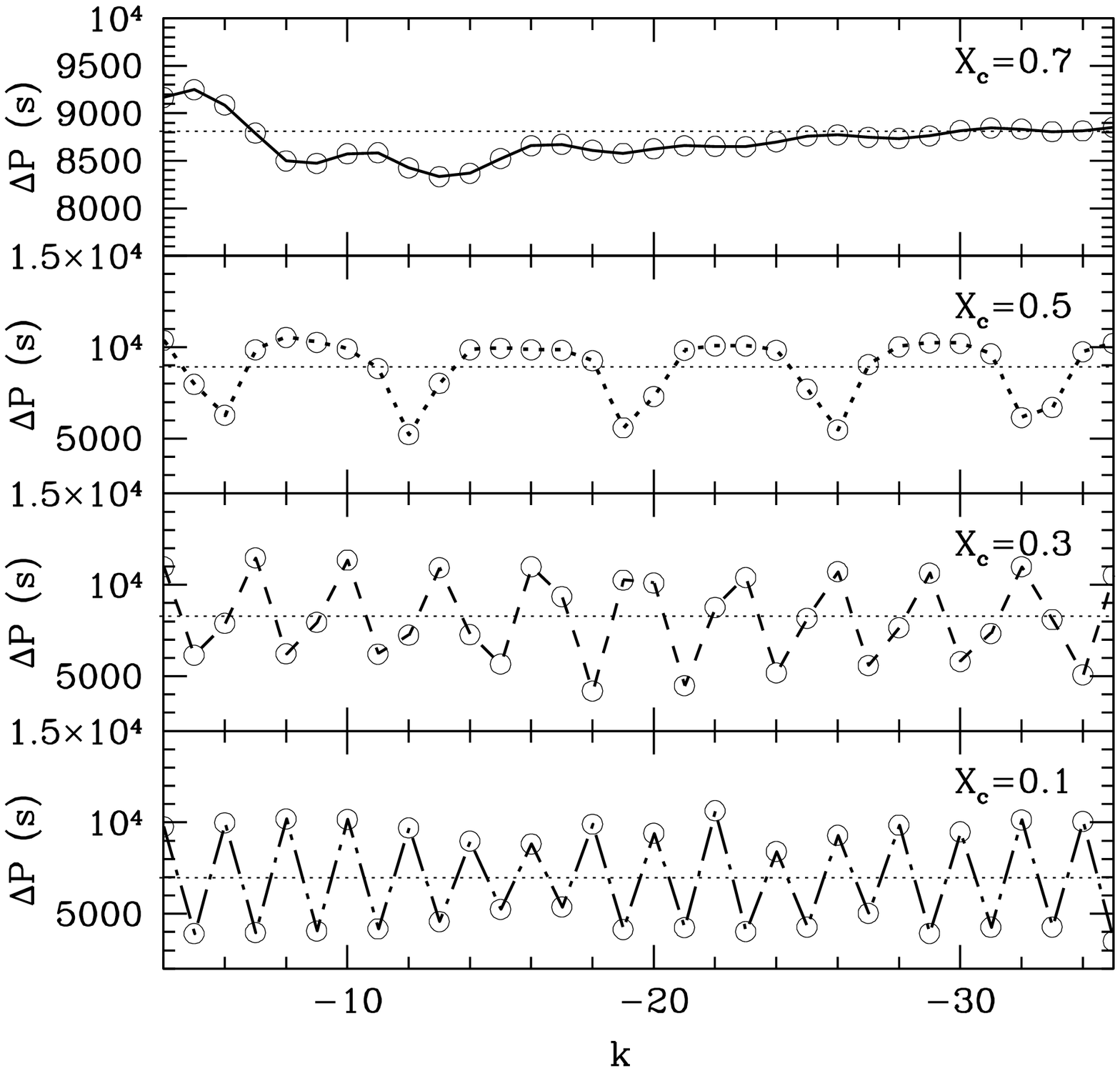}}
\caption{\small Same as Fig.\ref{fig:12} for 6.0 \msol models. In the homogeneous ZAMS model ($X_{\rm c}\simeq 0.7$, full lines) a sharp variation in $N$ located at $\Pi_0/\Pi_{\rm r} \simeq 0.8$ generates a periodic signal of small amplitude in the period spacing. Note that for this model, to make such a component more visible in $\Delta P$, the y-axis scale has been modified.}
\label{fig:6}
 \end{center}
\end{figure}

That behaviour is shown in Figures \ref{fig:16} and \ref{fig:6} for
models of 1.6 and 6~\msun (for larger masses, the behavior is almost
identical to that of 6~\msun). The periodicity of the components in
$\Delta P$ can be easily related to the profile of the \BV frequency
by means of expression (\ref{eq:variak}). For instance, the $\Delta
k = 7$ in Fig.~\ref{fig:6} (lower panel, $X_{\rm c}=0.50$),
corresponds to the sharp signal in $N$ at $\Pi_0/\Pi_{\rm r}\sim
0.14$. As the star evolves, the sharp feature in $N$ is shifted to
higher $\Pi_0/\Pi_{\rm r}$, and when  $\Pi_0/\Pi_\mu \simeq 0.5$ a
kind of beating occurs in the period spacing due to the fact that
the sampling frequency is about half the frequency of the periodic
component.

The amplitude of the variation of $\Delta P$ as a function of the
mode order is well reproduced by Eq. (\ref{eq:relaz}) (compare e.g.
Figs. \ref{fig:6} and \ref{fig:solutions}), but not by Eq.
(\ref{eq:varia}) that predicts a sinusoidal behaviour. However,
having a simple analytical relation between the amplitude of the
components and the sharpness of $\delta N/N$ is not straightforward
from Eq. (\ref{eq:relaz}).

It can also be noticed that oscillatory components of small
amplitude occur already in zero age main-sequence stars with $M \ga
6\; \msol$  (see e.g. Fig. \ref{fig:6}, solid lines). Although a
chemical composition gradient is not yet present in these models,
the bump in the \BV frequency due to an increase of the
opacity\footnote{Mainly due to C, O, Ne and Fe transitions
\citep[see e.g.][]{Rogers92,Seaton04}.} at a temperature of $\sim
3\times 10^6$~K ($\Pi_0/\Pi_{\rm r}\simeq 0.8$) is able to produce
such a deviation from constant $\Delta P$.
 It is not surprising that the effects of a sharp feature located near the surface can mimic the effect of a perturbation in the core: as shown by \cite{Montgomery03} the signature in high order g-modes of a perturbation in $N$ located at a normalized buoyancy radius $r_{\rm BV}=\Pi_0/\Pi_r$ is aliased to a signal whose source is located at $1-r_{\rm BV}$. The signal shown in Fig \ref{fig:6} could indicate a source at 0.2 $\Pi_0/\Pi_r$ which is in fact approximatively an alias $(1-0.2)$ of the source at 0.8 $\Pi_0/\Pi_r$.
The amplitude of this signal increases with the stellar mass as the
contribution of this opacity bump becomes dominant in the behaviour
of  $\nabla_{\rm rad}$ (and therefore of $N$). In fact for large
enough stellar masses, a convective shell can appear at a
temperature $\sim 3\times 10^6$~K. The amplitude of such components
is however less than 1000~s and therefore much smaller than the
amplitude due to the chemical composition gradient at the edge of
the convective core.

The g-mode period spacing is clearly different depending on the mass
and age of the models. While in models without (or with a very
small) convective core the mean value of the period spacing
decreases with age (see Sec. \ref{sec:radiative}), for more massive
models the period spacing does not significantly change with age. In
stars with larger convective cores the chemical composition gradient
is located at a larger fractional radius, giving a smaller
contribution to $\int{\frac{N}{x'}dx'}$, and thus to $\Delta P$, as
predicted from the asymptotic expression. In these models the age
effect is made evident, however, through the appearance of the
periodic signal in the period spacing, whose periodicity is directly
linked to the chemical gradient left by the evolution of the
convective core.

\subsection{Effects of extra-mixing}
\label{sec:extra} The comparison between theoretical models and
observations clearly shows that the standard stellar modelling
underestimates the size of the central mixed region \citep[see
e.g.][]{Andersen90, Ribas00}. This fact is generally accepted, but
there is no consensus about the physical processes responsible for
the required extra-mixing that is missing in the standard evolution
models: overshooting \citep[e.g. ][]{Schaller92}, microscopic
diffusion \citep{Michaud04}, rotationally induced mixing \citep[see
e.g. ][ and references therein]{Maeder00, Mathis04}, or mixing
generated by propagation of internal waves \citep[e.g. ][]{Young05}.
The shape of the composition transition zone is a matter of great
importance as far as asteroseismology is concerned. In particular it
significantly affects the term $\nabla_\mu$ appearing in the \BV
frequency and plays a critical role in the phenomenon of mode
trapping.

It is therefore evident that the size and  evolution of the
convective core, as well as the  $\mu$ gradients that it generates,
can be strongly affected by the occurrence of mixing processes. In
the following paragraphs  we study how these effects are reflected
on the high order g-modes. We have computed  models with
overshooting, microscopic diffusion, turbulent mixing, and we have
compared their adiabatic g-mode periods with those derived for
models computed without mixing and with the same central hydrogen
abundance.

\subsubsection{Overshooting}
\label{sec:over0}
\begin{figure*}
 \begin{center}
\resizebox{0.33\hsize}{!}{\includegraphics[angle=0]{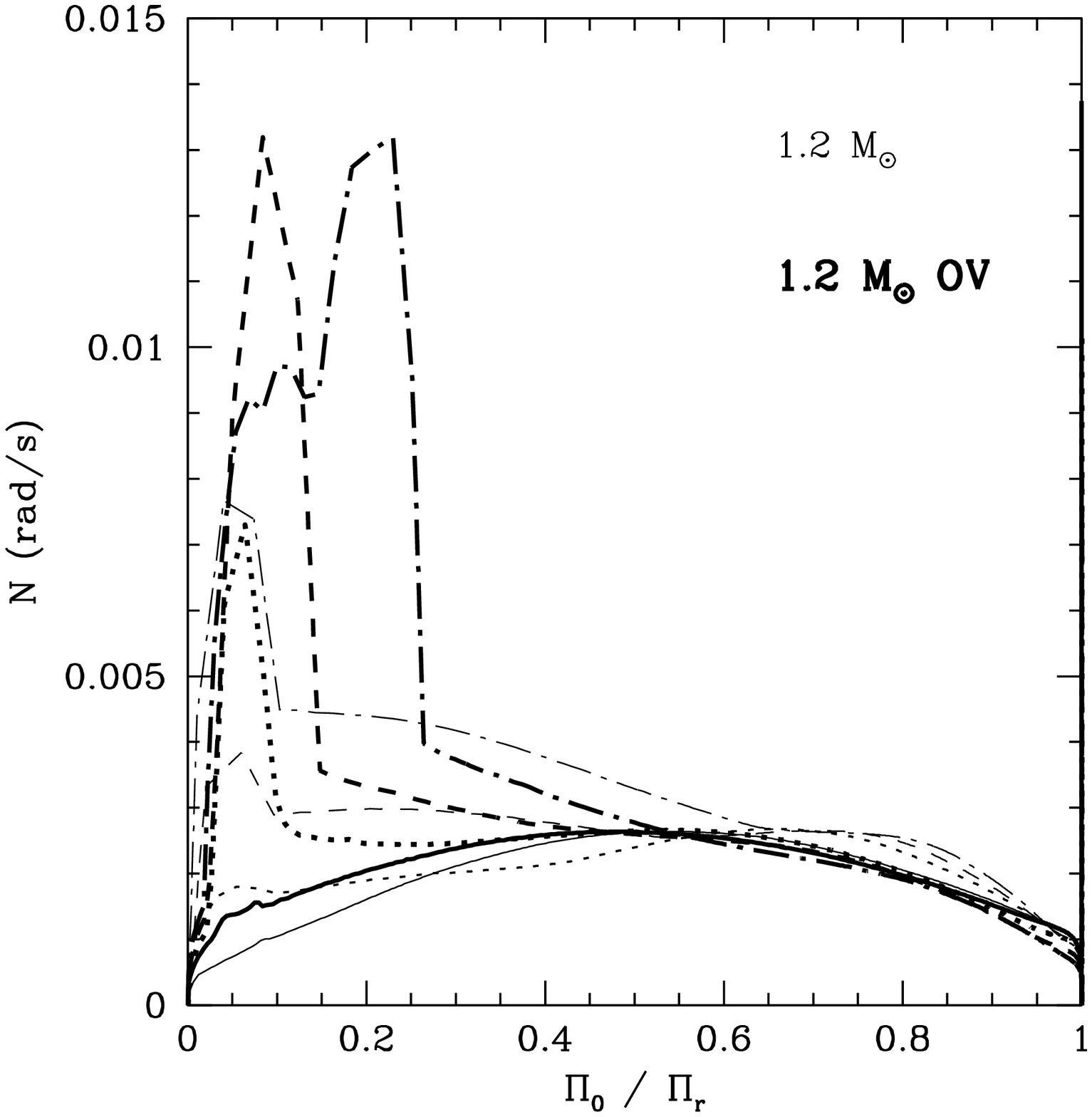}}
\resizebox{0.33\hsize}{!}{\includegraphics[angle=0]{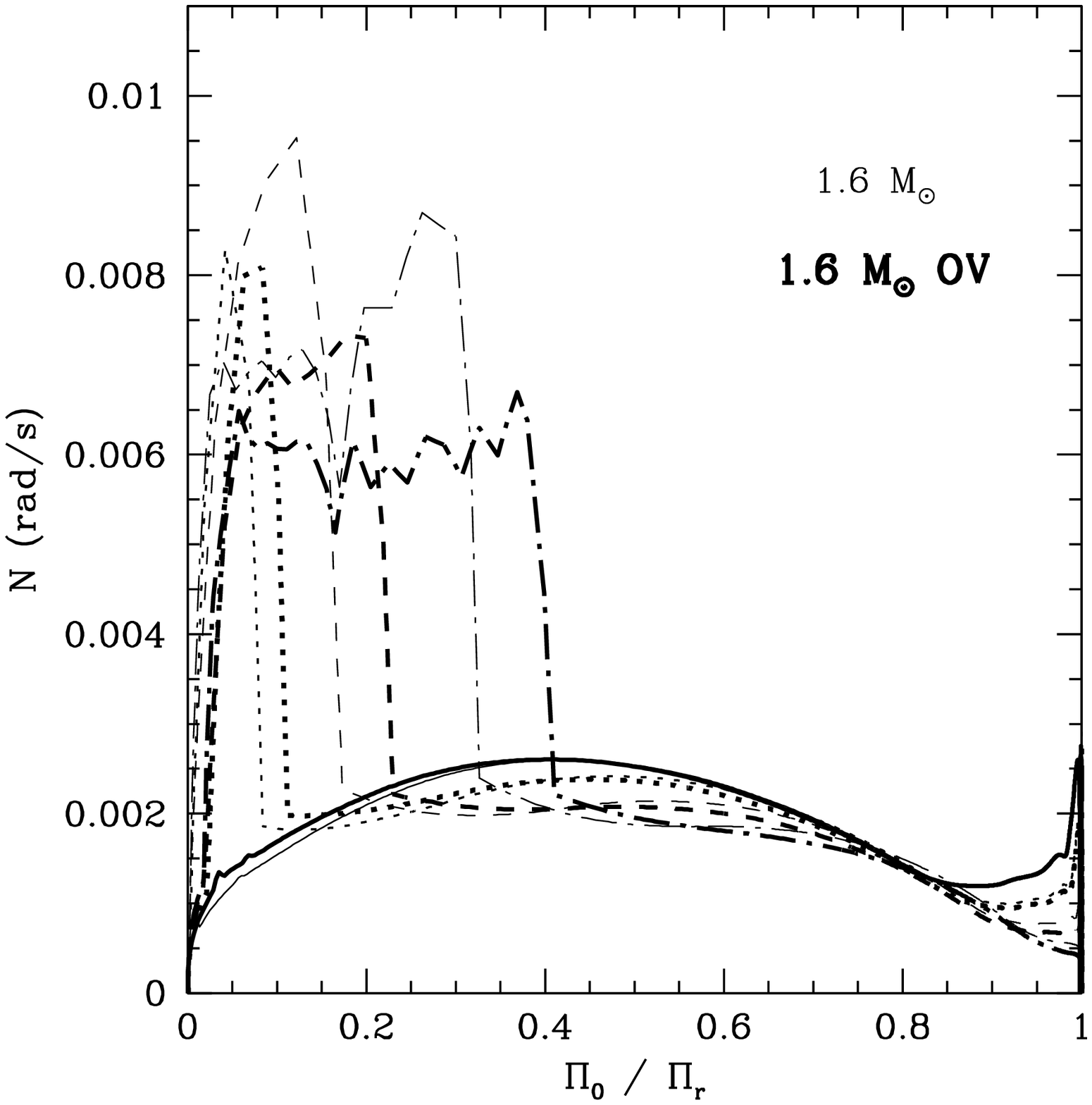}}
\resizebox{0.33\hsize}{!}{\includegraphics[angle=0]{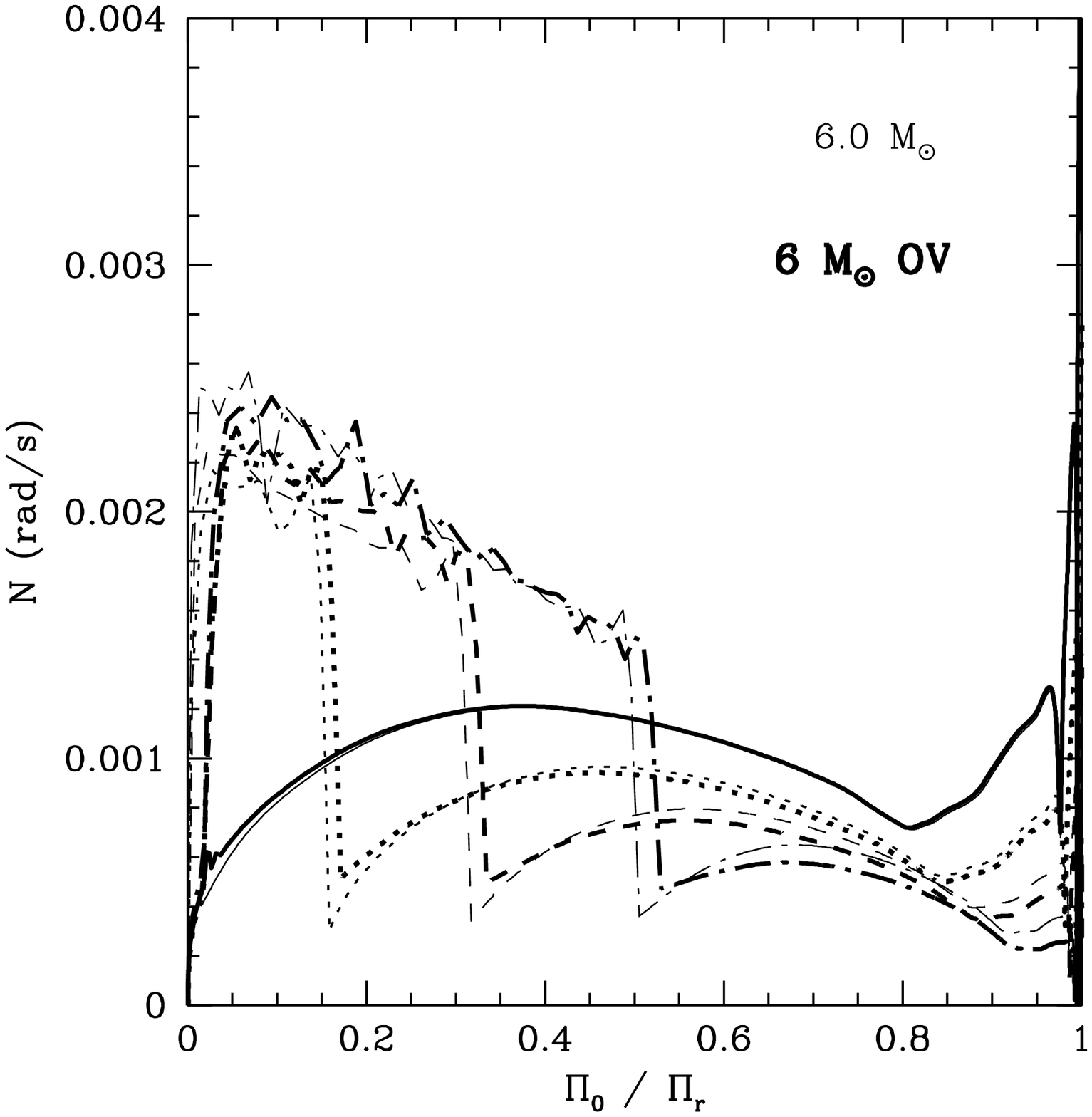}}\\

\resizebox{0.33\hsize}{!}{\includegraphics[angle=0]{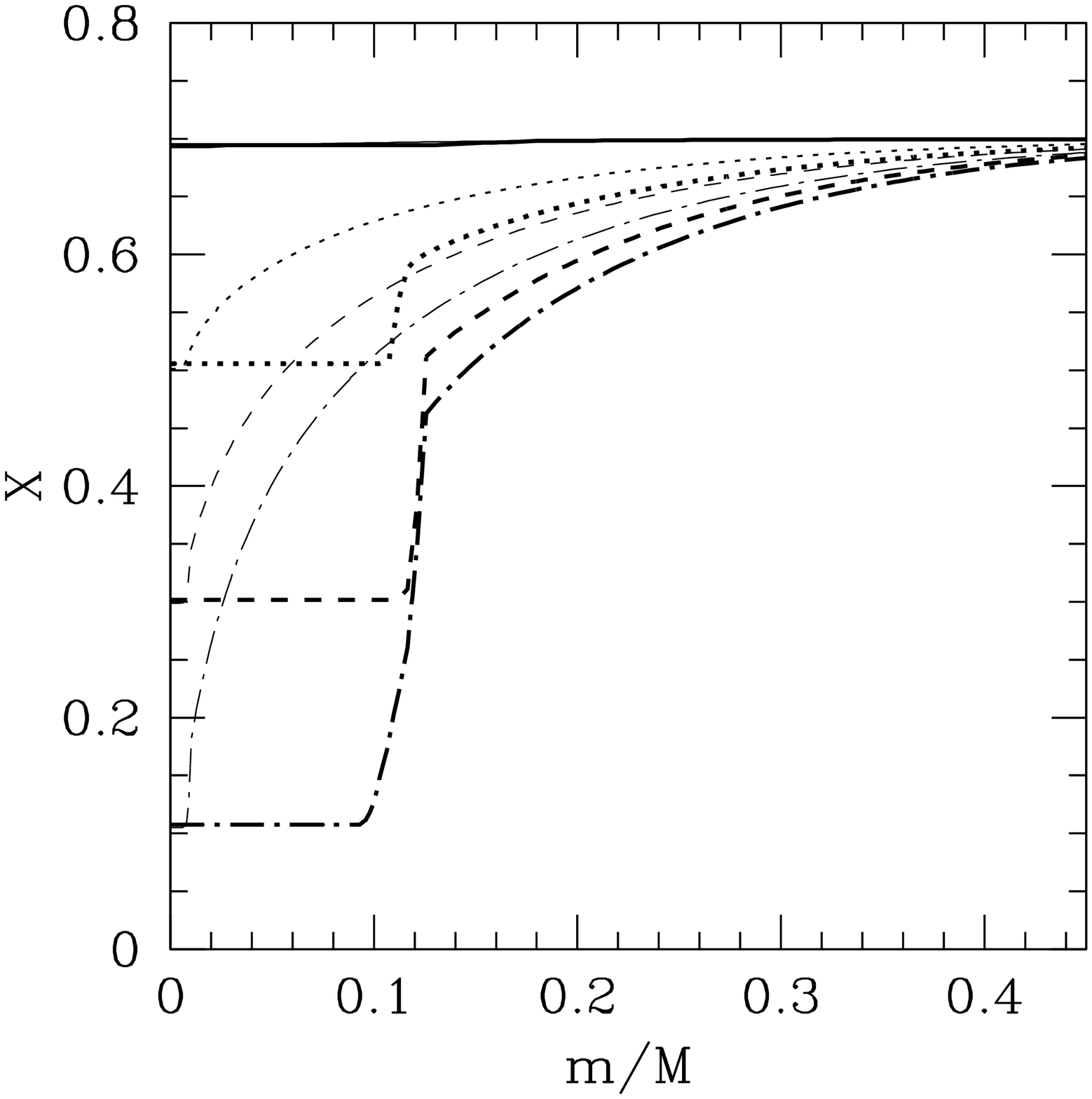}}
\resizebox{0.33\hsize}{!}{\includegraphics[angle=0]{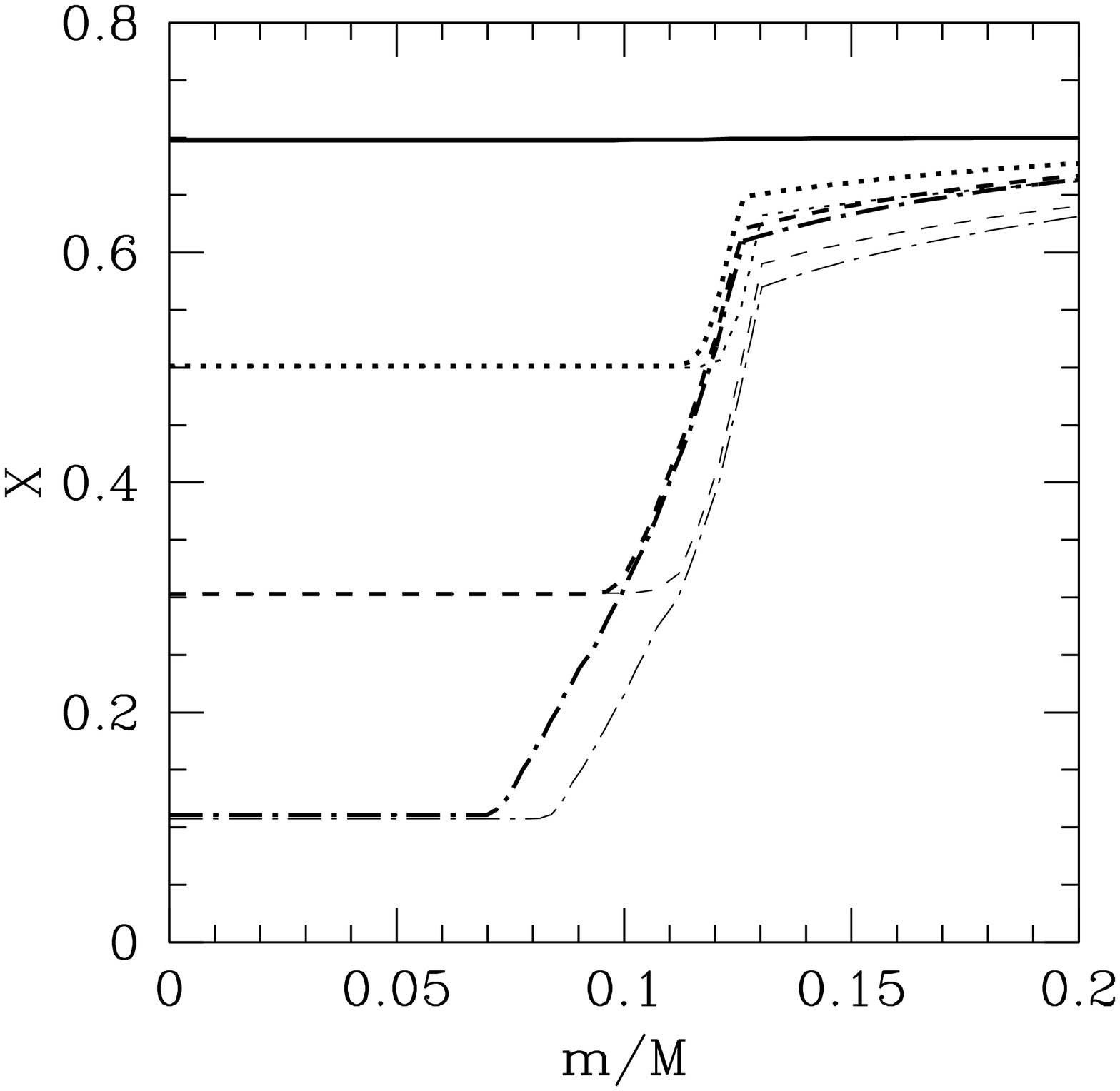}}
\resizebox{0.33\hsize}{!}{\includegraphics[angle=0]{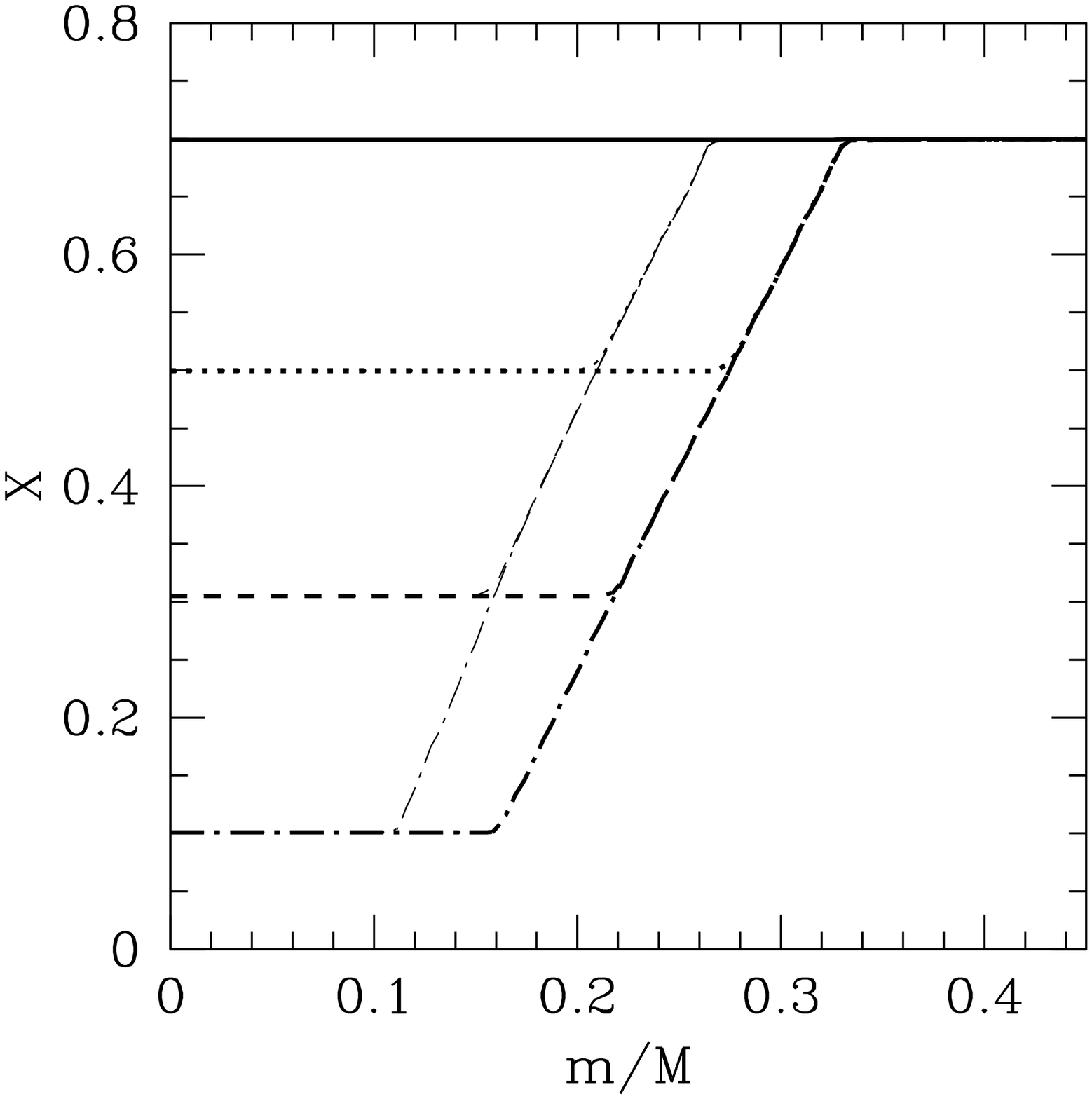}}\\

\resizebox{0.33\hsize}{!}{\includegraphics[angle=0]{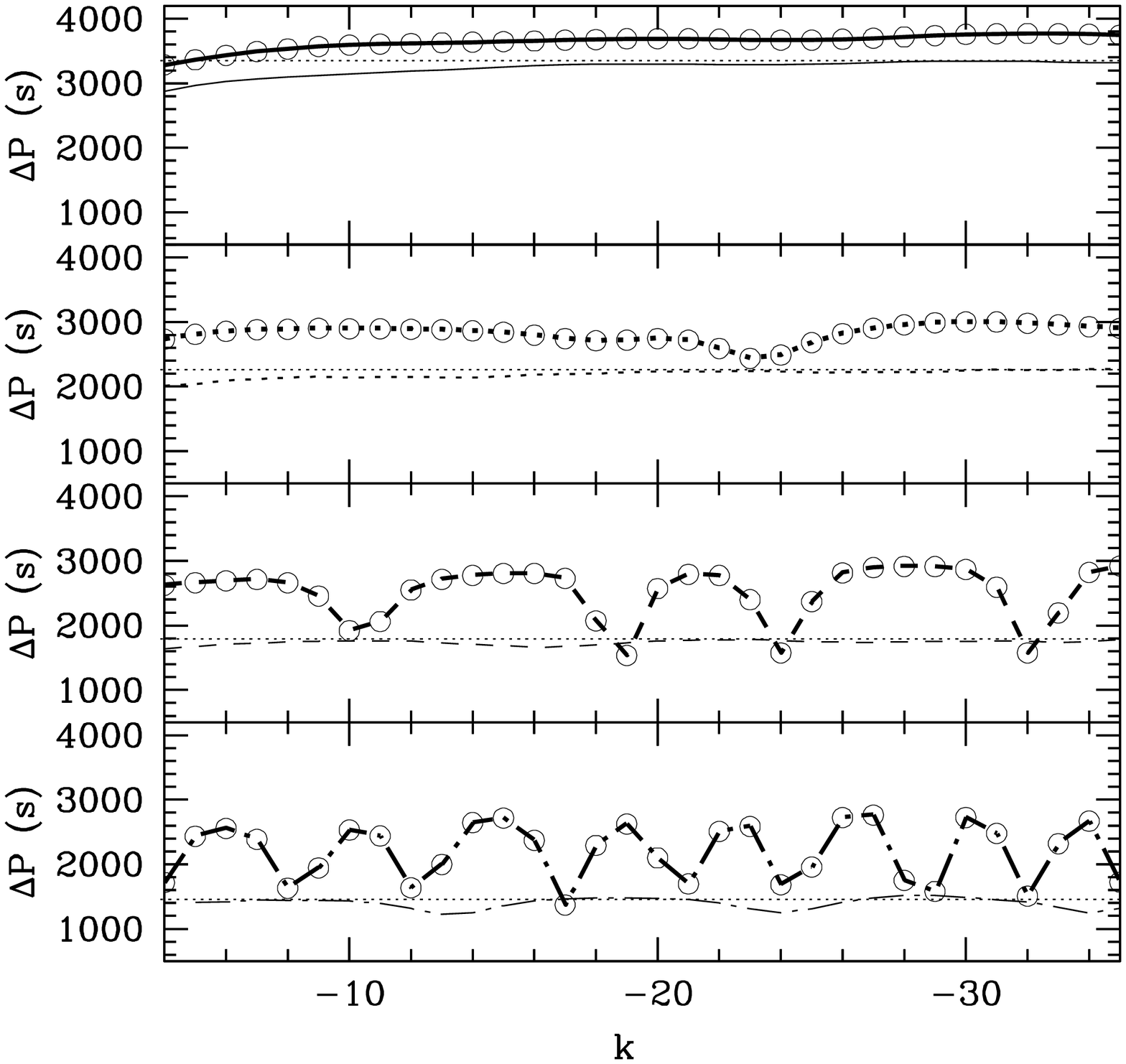}}
\resizebox{0.33\hsize}{!}{\includegraphics[angle=0]{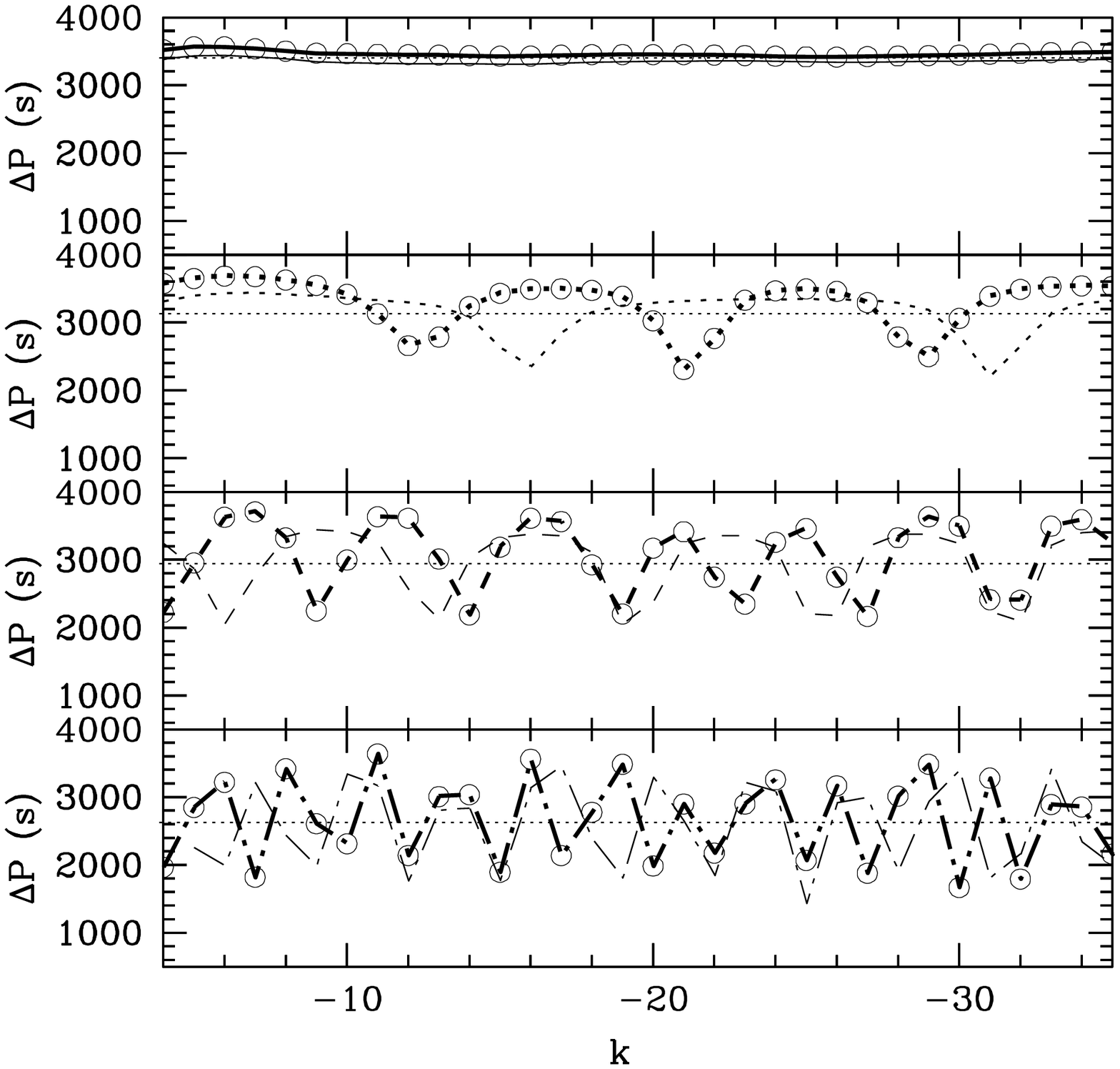}}
\resizebox{0.33\hsize}{!}{\includegraphics[angle=0]{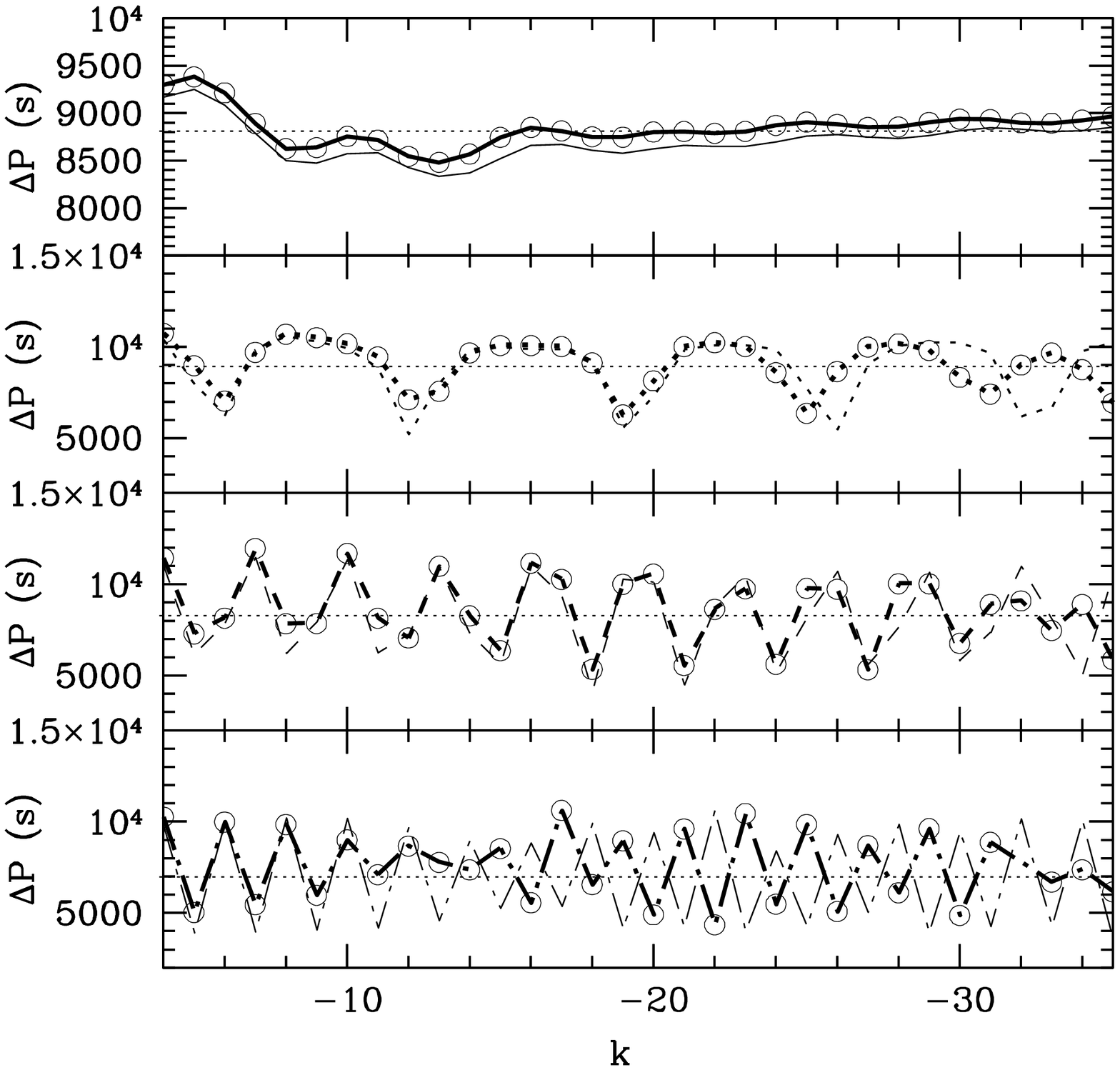}}\\
\caption{\small  Behavior of the \BV frequency (upper row), of the hydrogen abundance profile (central row) and of the $\ell=1$ g-mode period spacing (lower row) in models of 1.2,1.6 and 6 $\rm M_\odot$ computed with (thick lines) or without (thin lines) overshooting.}
\label{fig:ovall}
 \end{center}
\end{figure*}

Penetration of motions beyond the boundary of convective zones defined by the Schwarzschild stability criterion has been the subject of many studies in an astrophysical context \citep[see e.g.][]{Zahn91}.
Unfortunately,
features such as extension, temperature gradient and efficiency of the mixing
in the  overshooting region  cannot be derived from the local model of convection currently used in stellar evolution computations.
As a consequence, this region is usually described in a parametric way.
 In the models considered here the thickness of the overshooting layer $ov$ is parameterized in terms of the local pressure
scale height $H_{\rm p}$:  $ov=\beta\times(\min(r_{\rm cv}, H_{\rm
p}(r_{\rm cv})))$ (where $r_{\rm cv}$ is the radius of the
convective core and $\beta$ is a free parameter). We assume
instantaneous mixing both in convective and in overshooting regions.
The  temperature gradient in the overshooting region is left
unchanged (i.e. $\nabla=\nabla_{\rm rad}$). Therefore overshooting
simply extends the region assumed to be fully mixed by convection.
The
 larger hydrogen reservoir, due to an increase of the mixed region, translates into a longer core-hydrogen burning phase.

The adopted amount of overshooting also determines the lowest stellar mass where a convective core appears. 
For sufficiently large values of $\beta$, the convective core that develops in the pre-main sequence phase persists during the main sequence in models with $\rm M <$$M_{\rm Lcc}$ (as in Fig. \ref{fig:cc}). In these models the convective core is maintained thanks to the continuous supply of ${\rm ^3He}$ that sustains the highly temperature-dependent nuclear reaction ${\rm ^3He + {^3He} \rightarrow {^4He}+{^1H} +{^1H}}$, keeping the pp chain in an out-of-equilibrium regime. The inclusion of overshooting changes the value of the mass corresponding to the transition between models with a convective core that grows/shrinks during the main sequence (Fig. \ref{fig:cc}, right panel).
And finally, the effect of overshooting on the $\mu$ gradients depends on whether the nuclear reactions occur only inside the convective core or also outside. 

In Figure \ref{fig:ovall} we present the chemical composition
profile, the behaviour of $N$ and of the period spacing, in models
computed with overshooting ($\beta=0.2$). These models have a larger
fully mixed region than those computed without overshooting. The
chemical composition gradient is then displaced to a higher mass
fraction. If we compare with models of similar central hydrogen
abundance, however, this does not necessarily imply that the sharp
feature in $N$ is located at a different normalized buoyancy radius
($\Pi_0/\Pi_\mu$).

\begin{itemize}
\item In 6~\msol\ models (right column in Fig. \ref{fig:ovall}), for instance, neither the sharpness of the abrupt variation in $N$, nor its location in terms of $\Pi_\mu$, change when comparing models computed with and without overshooting.
\item The situation changes in lower mass models, e.g. in 1.6~\msol\ models (central column in Fig. \ref{fig:ovall}). Here the periodicity of the components in $\Delta P$ differs if we include overshooting or not. A change in the location, but also in the value of the $\mu$ gradient (as nuclear reactions take place outside the core as well), is responsible for a different behaviour of the oscillatory components in $\Delta P$.
\item In models with a mass $M\simeq M_{\rm Lcc}$, e.g. $M=1.2$ \msun , the inclusion of overshooting dramatically increases the size of the fully mixed region. The oscillatory components in $\Delta P$ have different periods and much larger amplitudes than in the case of models without overshooting (see left column of Fig. \ref{fig:ovall}).
\end{itemize}

As we mentioned above, not only the extension of the overshooting region is uncertain, but also its temperature stratification. However, the effect on $\Delta P$ of considering convective penetration ($\nabla=\nabla_{\rm ad}$ in the ``extended'' convective core) instead of simple overshooting is found to be small \citep[see][]{Straka05, Godart07}.

\subsubsection{Effects of microscopic diffusion}
\label{sec:diff}
Other physical processes, different from overshooting, can lead to an increase of the central mixed region or modify the chemical composition profile near the core.

\citet{Michaud04} and \citet{Richard05} have shown that microscopic
diffusion can induce an increase of the convective core mass for a
narrow range of masses, from 1.1 to 1.5 \msun, and that the effect
decreases rapidly with increasing  stellar mass. In this mass range,
as previously described, the mass of the convective core  increases
during the MS evolution instead of decreasing, as it occurs for
larger masses. As a consequence, a sharp gradient of chemical
composition appears at the border of the convective core, making the
diffusion process much more efficient in that region.


\begin{figure}
 \begin{center}
\resizebox{0.8\hsize}{!}{\includegraphics[angle=0]{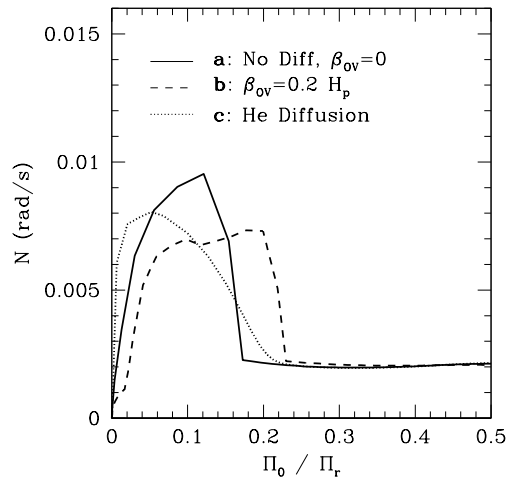}}
\caption{\small Behavior of the \BV frequency in models of 1.6 \msol
with $X_c\simeq 0.3$. The different lines correspond to models
calculated with no extra-mixing (continuous lines), overshooting
(dashed) and helium diffusion (dotted). The different location and
sharpness of the chemical composition profile determine the
behaviour of $N$.} \label{fig:16d}
 \end{center}
\end{figure}

\begin{figure}
 \begin{center}
\resizebox{0.8\hsize}{!}{\includegraphics[angle=0]{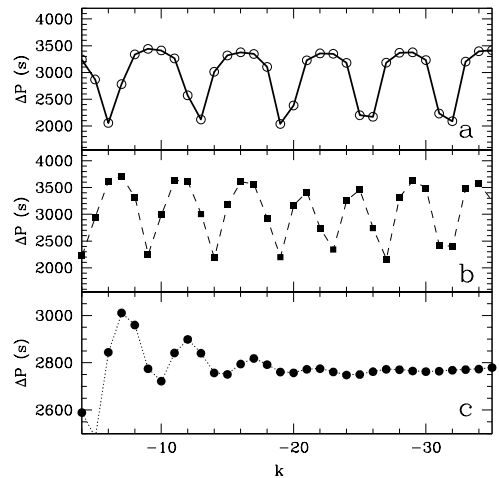}}
\caption{\small Period spacing of $\ell=1$ g modes as a function of the radial order $k$ for the 1.6 \msol models presented in Fig. \ref{fig:16d}.}
\label{fig:16dg}
 \end{center}
\end{figure}

Figure \ref{fig:16d} shows the effect of including
gravitational settling, thermal and chemical diffusion of hydrogen and helium in a 1.6~\msun\ model.
If only He and H diffusion is considered in the modelling we find that a smoother chemical composition profile
 is built up at the edge of the convective core (see Fig. \ref{fig:16d}) preventing the presence of a
 discontinuity in $\mu$ caused by a growing convective core. Comparing models
calculated with or without diffusion, the location of the convective
core boundary is not significantly changed. Nevertheless,
a less sharp variation in the \BV frequency is responsible for a considerable reduction of the amplitude of the components
in the period spacing (see Fig. \ref{fig:16dg}). It was in fact predicted in Section \ref{sec:varia} that a discontinuity not in $N$ itself, but in its first derivative, would  generate a periodic component whose amplitude decreases with the period: this simplified approach is therefore sufficient to account for the behaviour  of $\Delta P$ in models with a smoother chemical composition gradient (see Eq. \ref{eq:senal}).

When diffusion of Z is also included,
 the overall effect of diffusion is no longer able to erode the sharp chemical composition gradient and to prevent the formation of a semi-convective zone, on the contrary, diffusion of Z outside the core makes the
occurrence of semi-convection easier \citep[see also ][]{Richard01, Michaud04, Montalban07}.

If chemical diffusion is accountable for a smoother chemical
composition profile in intermediate- and low-mass stars, we find
that such an effect disappears as higher masses are considered and
evolutionary time-scales decrease. In fact we find that in models
with $M\ga 4\;$ \msol diffusion has no effect on the \BV frequency
profile near the hydrogen burning core, nor on the behaviour of the
period spacing.

As higher masses are considered, the effect of microscopic diffusion on the stellar structure near the core becomes negligible but, other mixing processes can partly erode the chemical composition gradients.

\subsubsection{Rotationally induced mixing}
\label{sec:rotation}
Rotationally induced mixing can influence the internal distribution of $\mu$ near the energy generating core. Different approaches have been proposed to treat the effects of rotation on the transport of angular momentum and chemicals \citep[see e.g.][ and references therein]{Maeder00, Heger00, Pinsonneault89}. Such an additional mixing has an effect on the evolutionary tracks which is quite  similar to that of overshooting, but  it leads also to a smoother chemical composition profile at the edge of the convective core.

Since our stellar evolution code does not include a consistent treatment of rotational effects,
we simply include the chemical turbulent mixing by adding a turbulent diffusion coefficient
($\rm D_{\rm T}$) in the diffusion equation. In our parametric approach $\rm D_{\rm T}$ is assumed to
 be constant inside the star and independent of age.

\begin{figure}
\begin{center}
\resizebox{0.85\hsize}{!}{\includegraphics[angle=0]{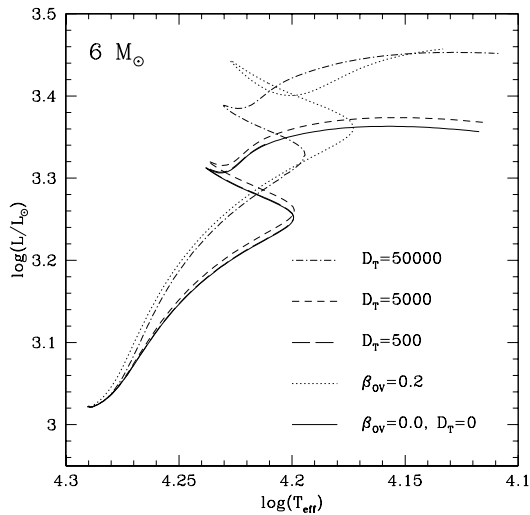}}
\caption{\small HR diagram showing evolutionary tracks of 6 \msol
models calculated with different extra-mixing processes. The
evolutionary track computed with $\rm D_{\rm T}=500$ is superposed
to that without extra-mixing (continuous line).} \label{fig:6hr}
\end{center}
\end{figure}

 The simplified parametric treatment of rotationally induced mixing used in this work has the aim of
showing that if an extra-mixing process, different from overshooting, is acting near the core it will
produce a different chemical composition profile in the central regions of the star and leave a
different signature in the periods of gravity modes.
The effects of such a mixing on the evolutionary tracks (see Fig. \ref{fig:6hr}) and on the internal structure clearly depend on the value of $\rm D_{\rm T}$.

\begin{figure}
 \begin{center}
\resizebox{0.48\hsize}{!}{\includegraphics[angle=0]{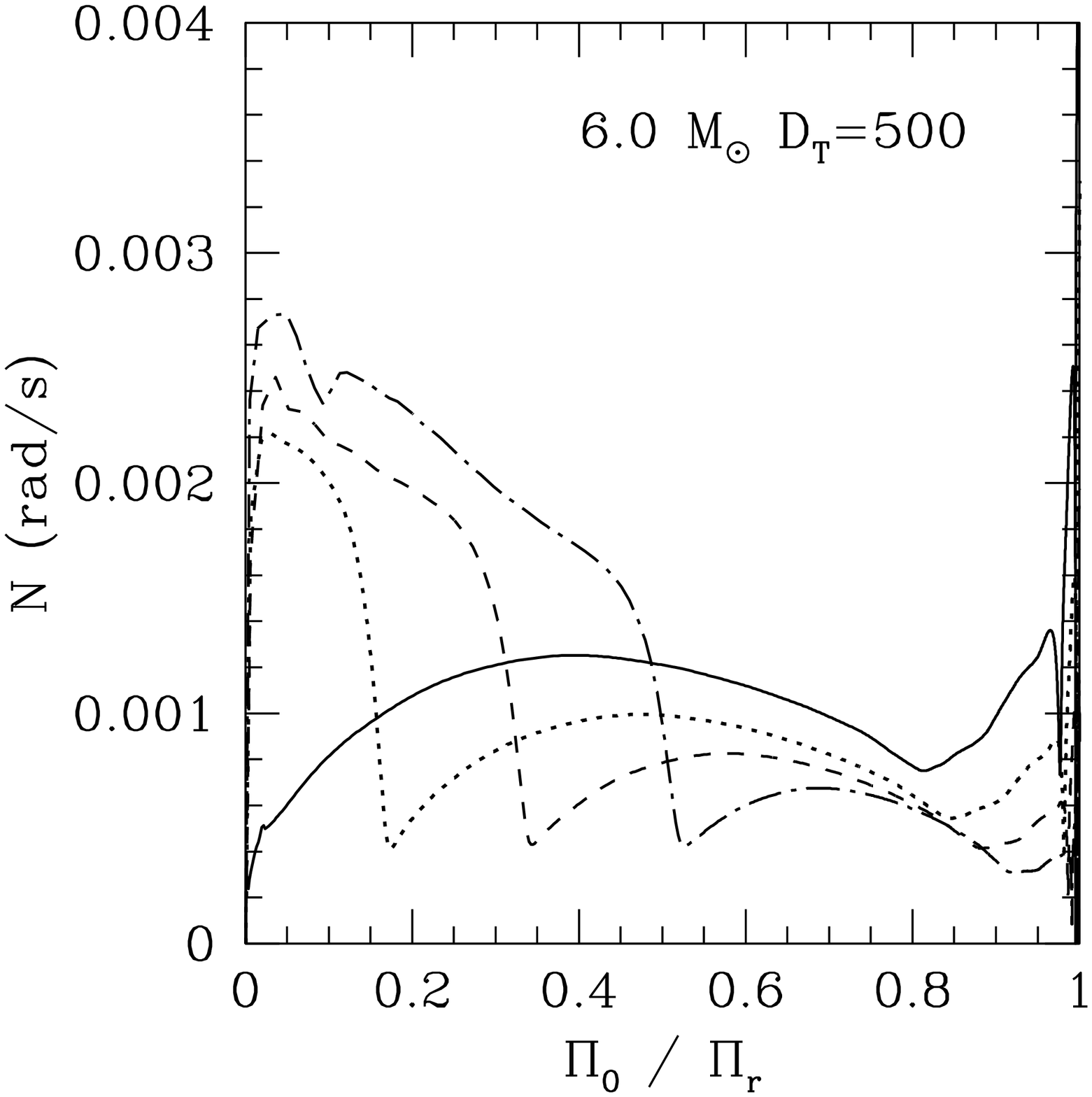}}
\resizebox{0.48\hsize}{!}{\includegraphics[angle=0]{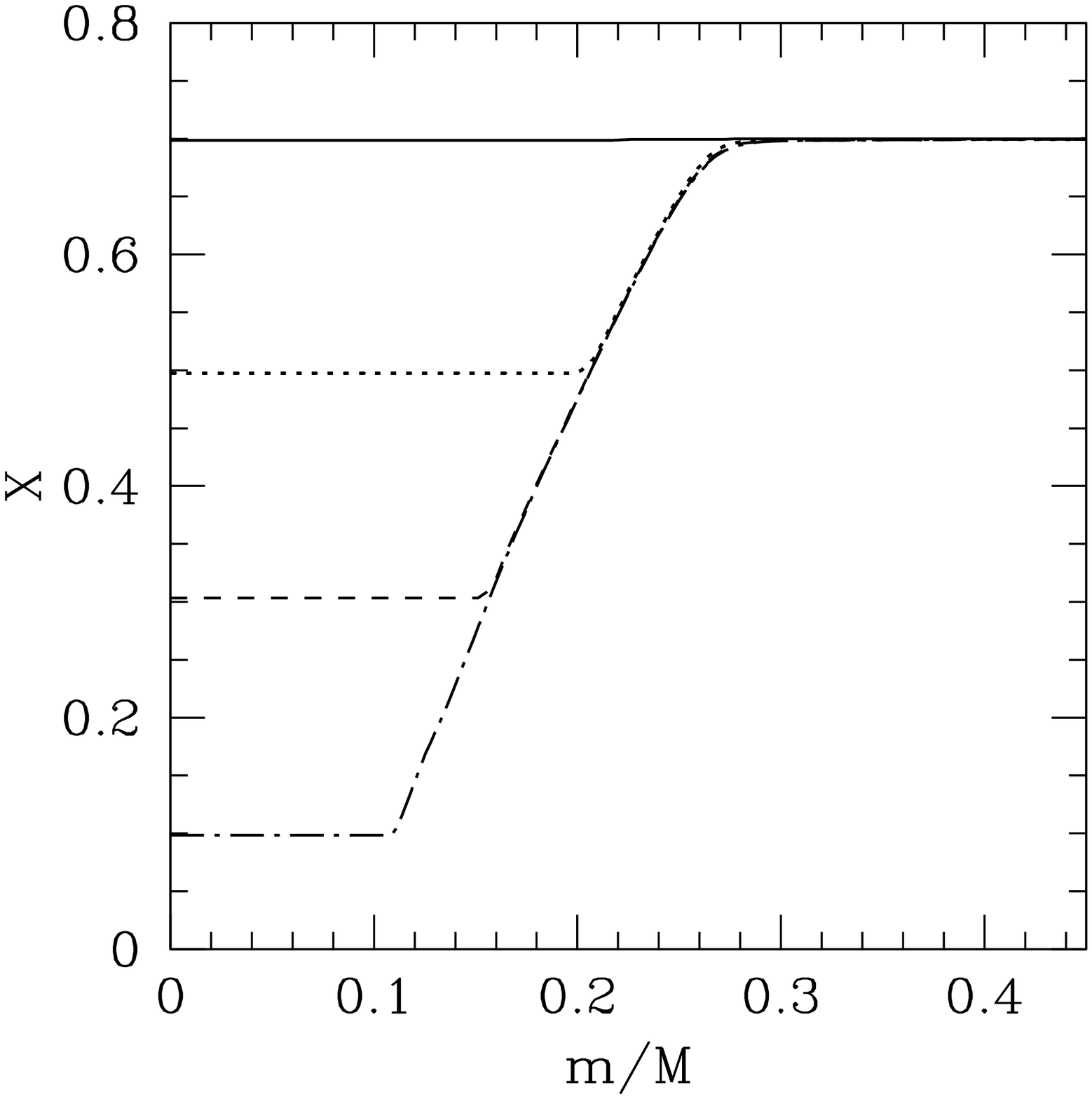}}
\resizebox{0.8\hsize}{!}{\includegraphics[angle=0]{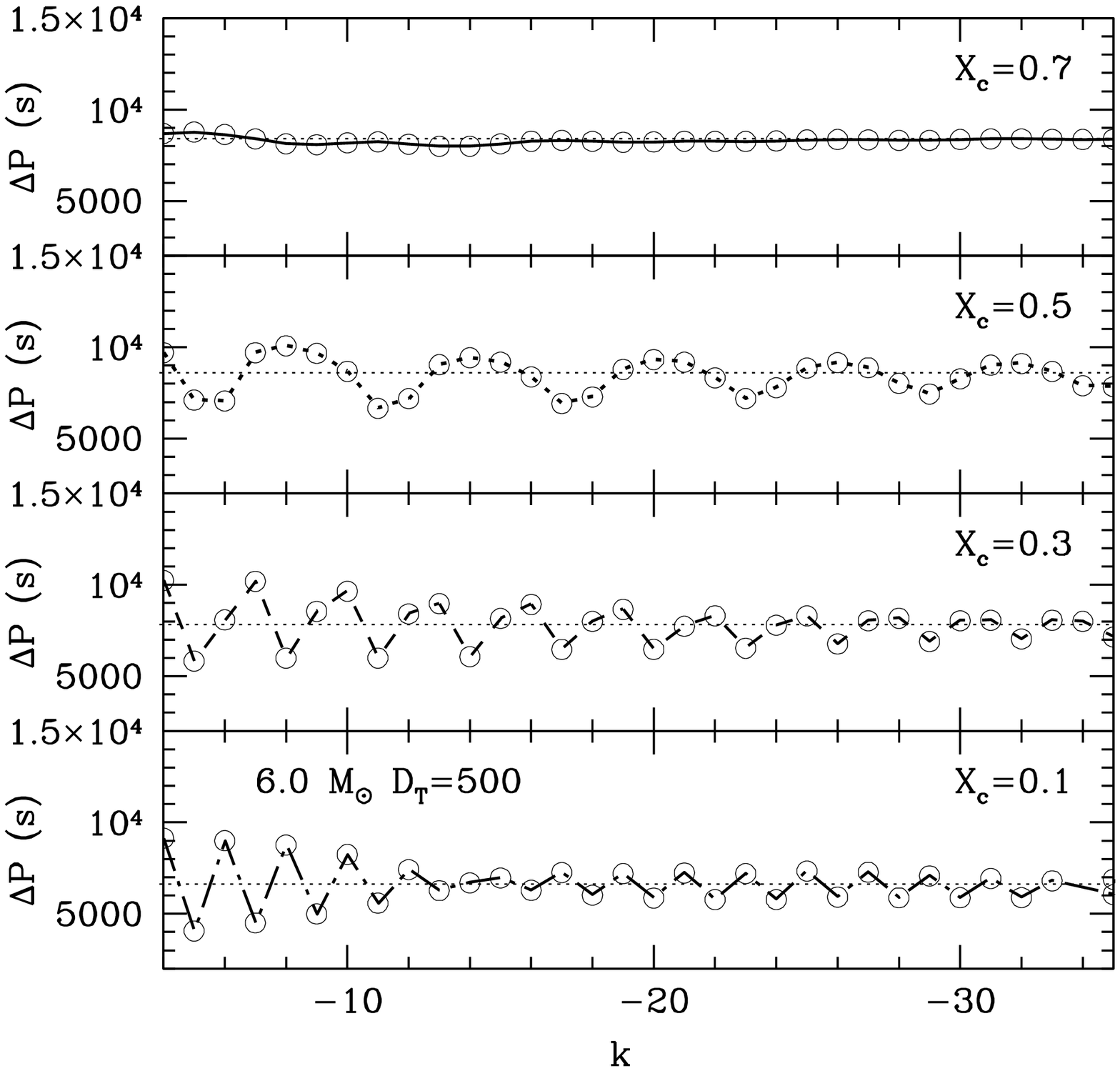}}
\caption{\small Behavior of the \BV frequency (upper left panel), of the hydrogen abundance profile (upper right panel) and of the $\ell=1$ g-mode period spacing in models of 6.0 \msol computed with a turbulent diffusion coefficient $\rm D_{\rm T}=500$.}
\label{fig:6r4}
 \end{center}
\end{figure}

\begin{figure}
\begin{center}
\resizebox{0.48\hsize}{!}{\includegraphics[angle=0]{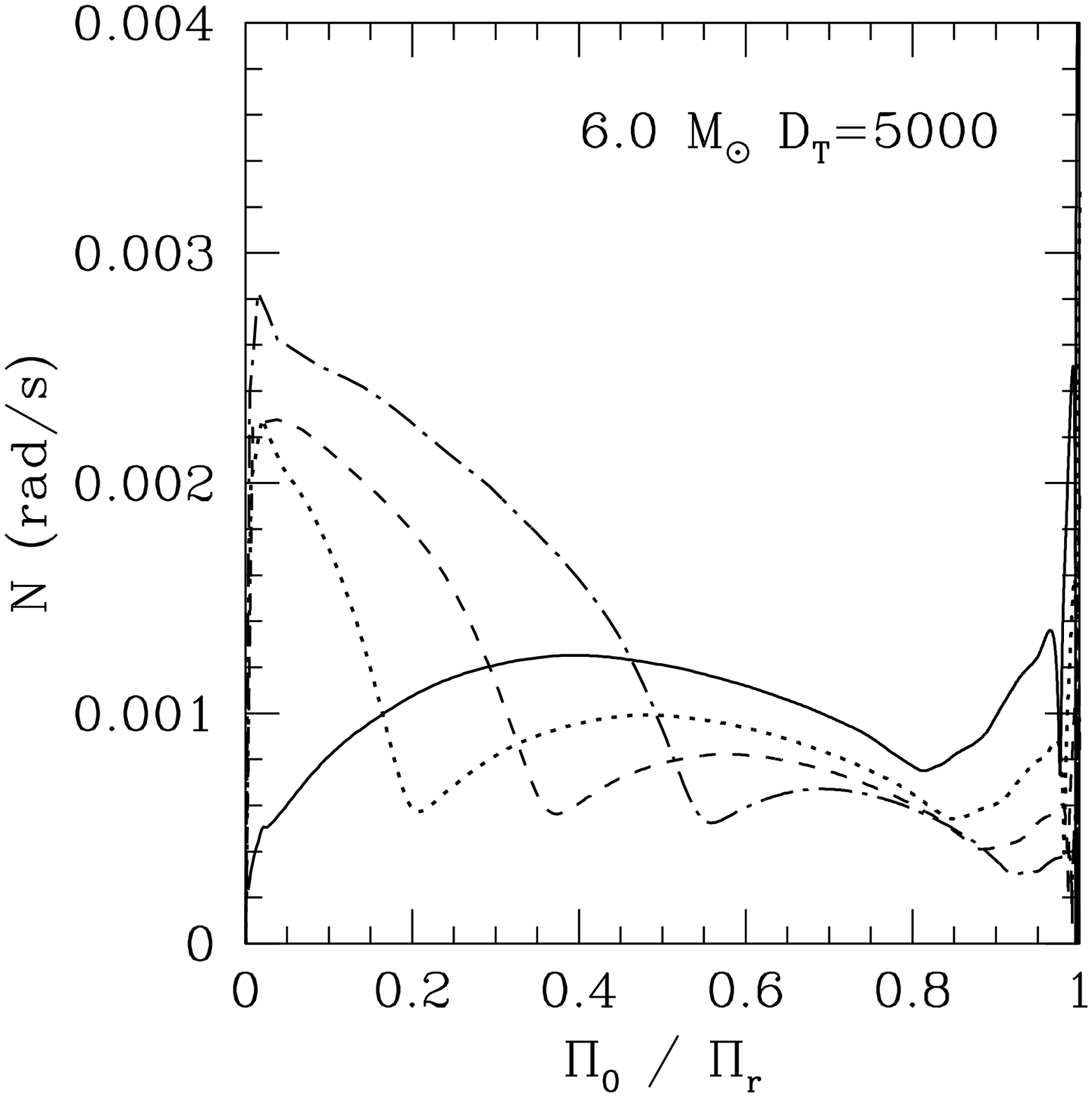}}
\resizebox{0.48\hsize}{!}{\includegraphics[angle=0]{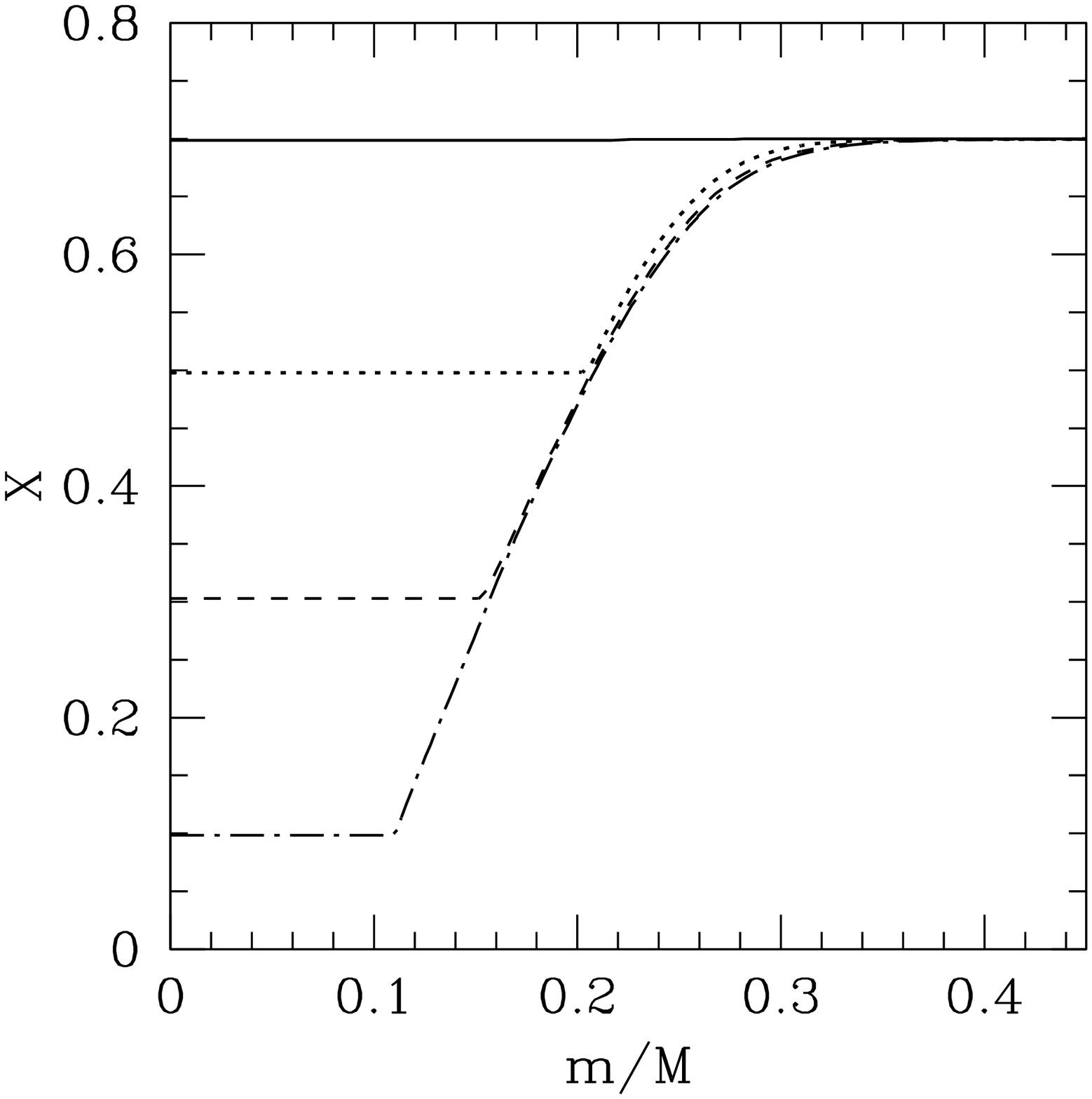}}
\resizebox{0.8\hsize}{!}{\includegraphics[angle=0]{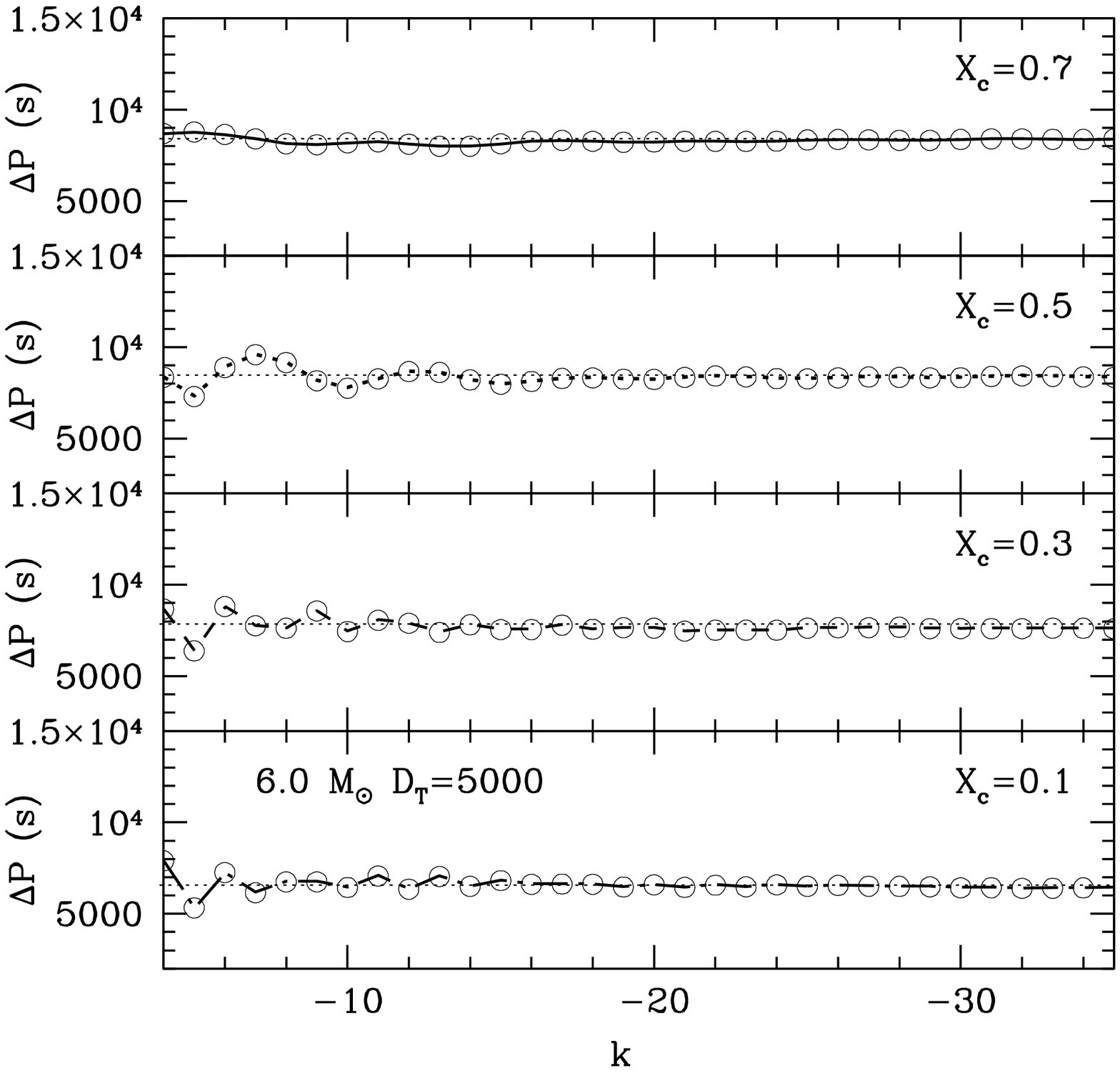}}
\caption{\small As in Fig. \ref{fig:6r2} but for models with $\rm D_{\rm T}=5000$}
\label{fig:6r2}
 \end{center}
\end{figure}

As an example we consider  models of a 6 \msol star computed with three different values of the turbulent-diffusion coefficient $\rm D_{\rm T}=5\times10^2,\, 5\times10^3$ and $5\times10^4$~cm$^2$ s$^{-1}$.
The lowest $\rm D_{\rm T}$ value provides an evolutionary track that overlaps  the one computed with no mixing, however, its effect on the chemical composition gradient (see Fig. \ref{fig:6r4}) is sufficient  to affect $\Delta P$ for high order modes.
In order to significantly change the periods of low-order modes a much more effective mixing is needed. The value $\rm D_{\rm T}=5\times10^3$ leads to a slightly more luminous evolutionary track,
but  such a  mixing has a substantial effect on the period spacing: the amplitude of the periodic components in $\Delta P$ becomes a decreasing function of the radial order $k$ (see Figures \ref{fig:6r4} and \ref{fig:6r2}). As in the case of helium diffusion (see Sec. \ref{sec:diff}) this behaviour can be easily explained by the analytical approximation presented in Sec.~\ref{sec:varia} (Eq.~\ref{eq:senal}), provided the sharp feature in $N$ is modelled not as a step function but, for instance, as a ramp (Eq.~\ref{eq:ramp}).

\begin{figure}
\begin{center}
\resizebox{0.48\hsize}{!}{\includegraphics[angle=0]{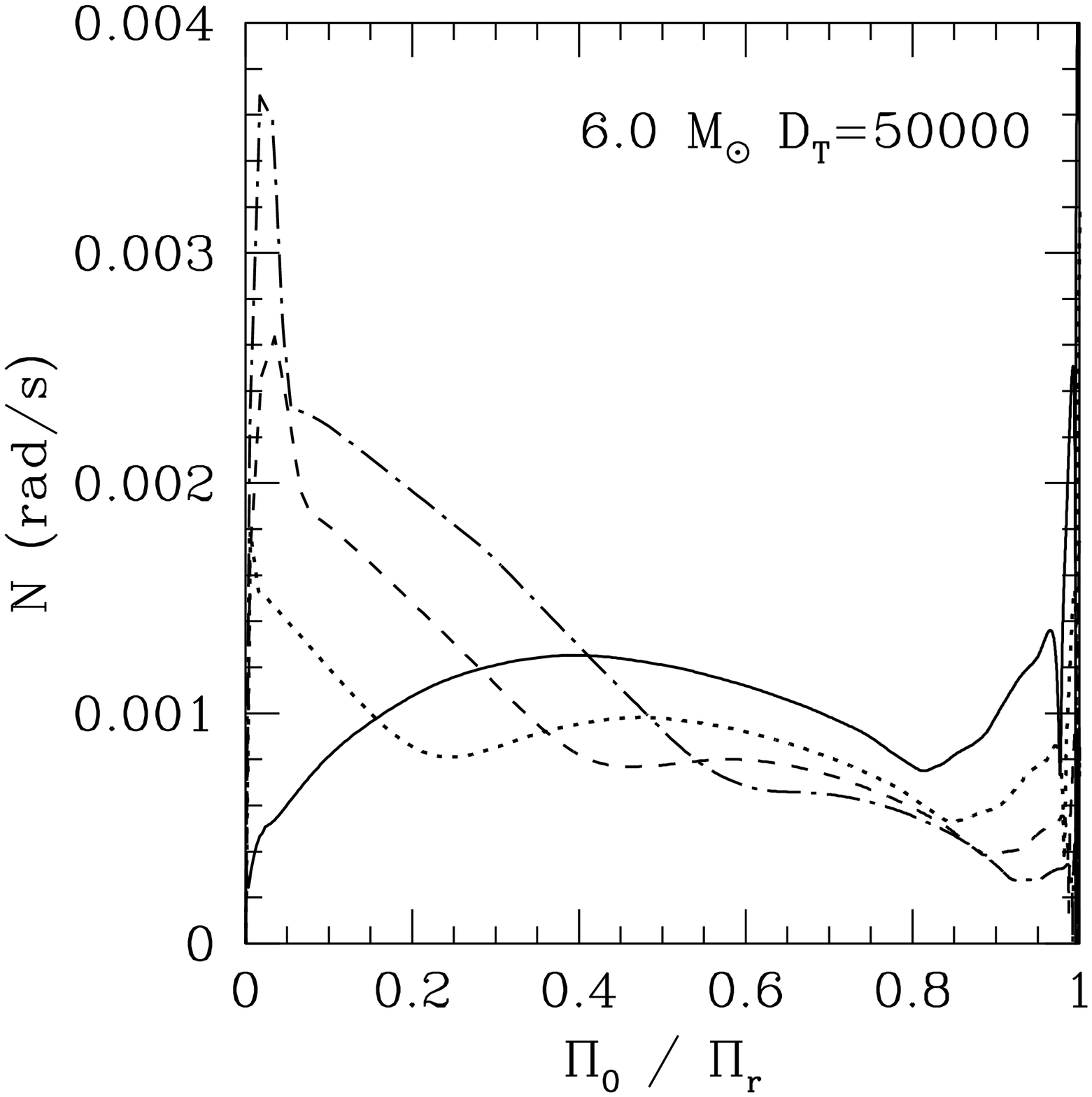}}
\resizebox{0.48\hsize}{!}{\includegraphics[angle=0]{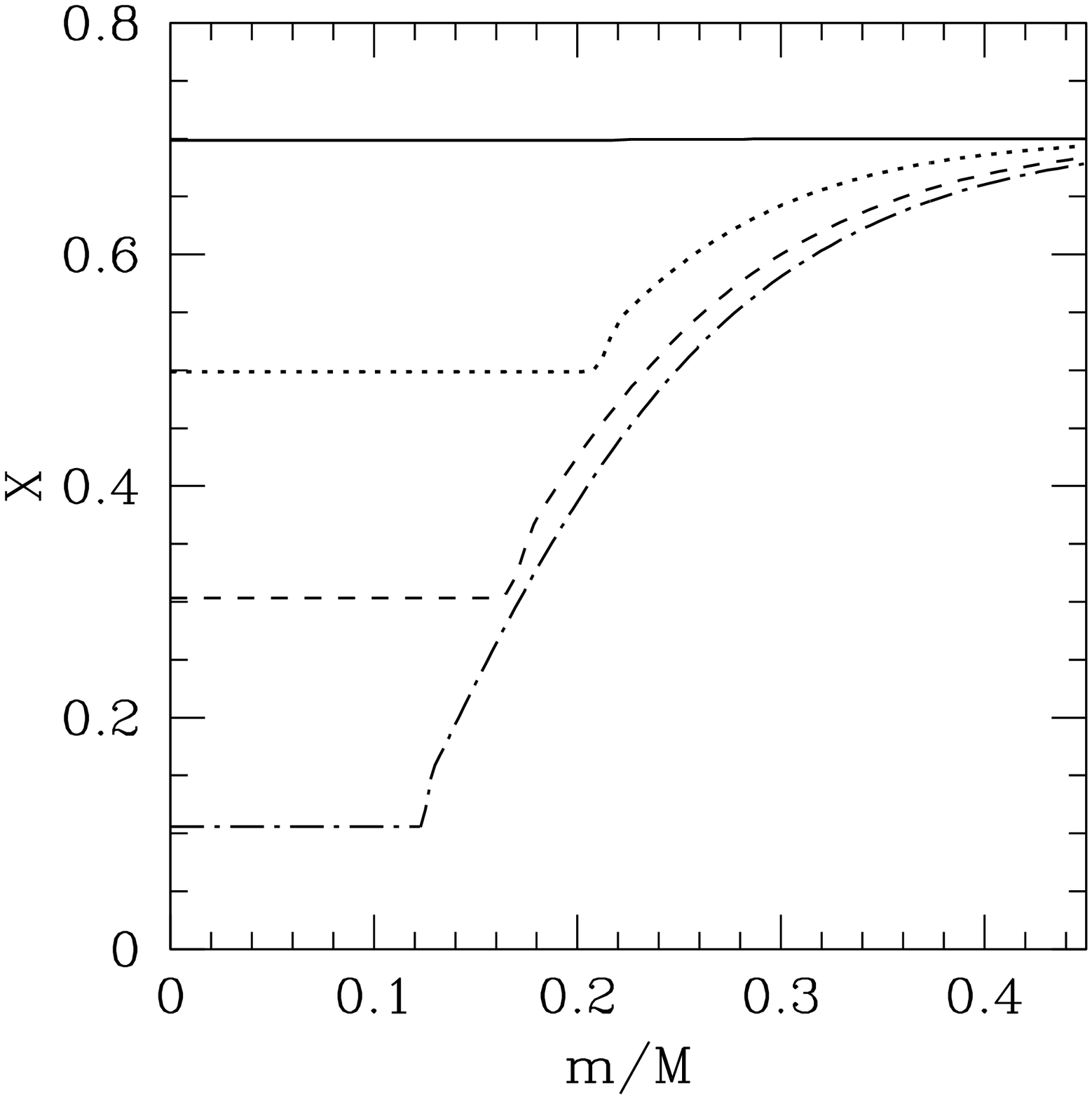}}
\resizebox{0.8\hsize}{!}{\includegraphics[angle=0]{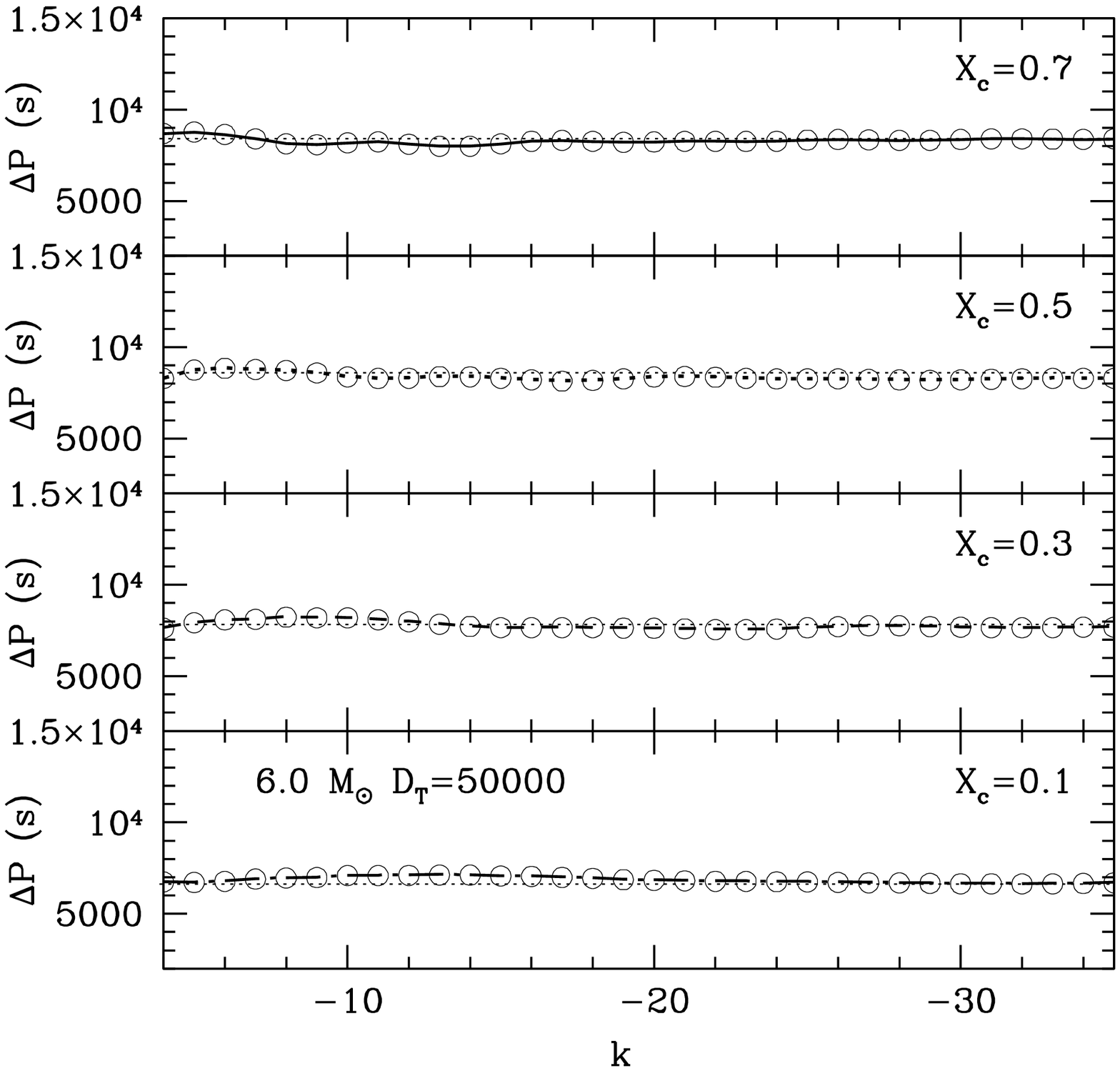}}
\caption{\small \small As in Fig. \ref{fig:6r2} but for models with $\rm D_{\rm T}=50000$. The effect of such a mixing on the evolutionary track is shown in Fig. \ref{fig:6hr}}
\label{fig:6r5}
 \end{center}
\end{figure}

If a significantly more effective mixing is considered (e.g. $\rm D_{\rm T}=50000$, see Fig. \ref{fig:6hr} and \ref{fig:6r5}), the corresponding evolutionary track is close to that obtained by including a classical overshooting but  the periodic components in $\Delta P$ are no longer present.

 The effects of such a turbulent mixing on low-order gravity modes and avoided crossings will be addressed in detail in a future work.


\begin{figure}
\begin{center}
\resizebox{0.85\hsize}{!}{\includegraphics[angle=0]{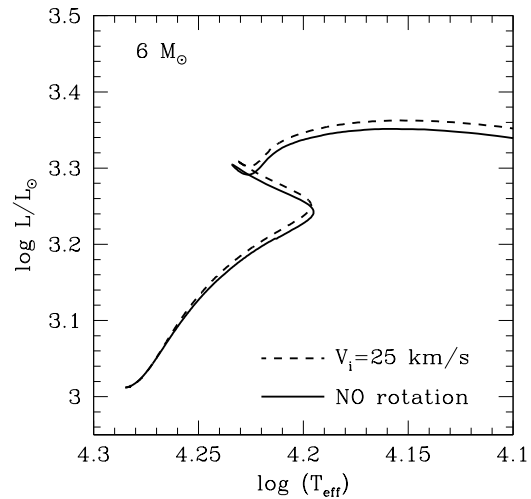}}
\caption{\small HR diagram showing evolutionary tracks of 6 \msol models calculated with the Geneva evolutionary code. The full line evolutionary track is computed without rotation, whereas in the models evolving on the dotted track an initial surface velocity of 25 $\rm km\;s^{-1}$ is assumed.}
\label{fig:6PatHR}
\end{center}
\end{figure}

\begin{figure}
\begin{center}
\resizebox{0.85\hsize}{!}{\includegraphics[angle=0]{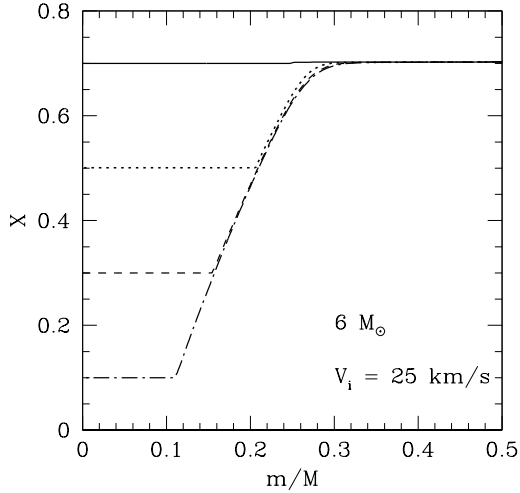}}
\caption{\small  Hydrogen abundance in the core of 6 \msol models with an initial surface velocity of 25 $\rm km\;s^{-1}$.}
\label{fig:6PatX}
\end{center}
\end{figure}

\begin{figure}
\begin{center}
\resizebox{0.95\hsize}{!}{\includegraphics[angle=0]{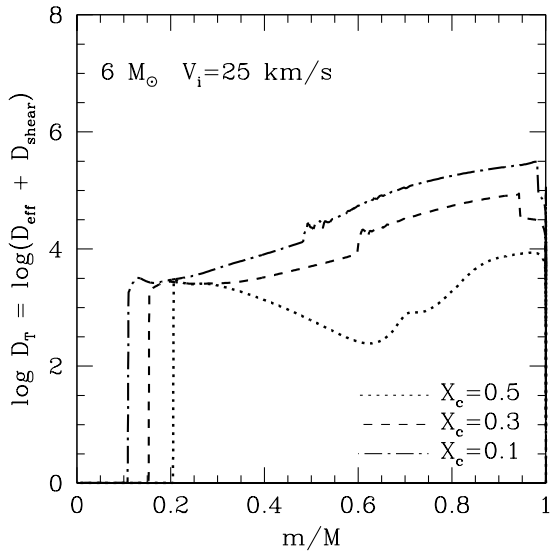}}
\caption{\small Total diffusion coefficient $\rm D_{\rm T}$ as a function of the normalized mass in 6 \msol models evolving on the main sequence. The initial surface velocity assumed is 25 $\rm km\;s^{-1}$. $\rm D_{\rm T}$ includes both the effects of shear ($\rm D_{\rm shear}$) and meridional circulation ($\rm D_{\rm eff}$).}
\label{fig:6PatD}
\end{center}
\end{figure}

In order to check that our parametric approach is at least in qualitative agreement with the outcome of models where rotationally induced mixing is treated consistently, we present for comparison a sequence of models computed with the Geneva evolutionary code \citep{Meynet00,Eggenberger07}. We considered a 6 \msol model with an initial surface rotational velocity of 25 $\rm km\;s^{-1}$, the typical $vsin i$ for SPBs stars \citep{Briquet07}. As the model evolves on the MS, the effects on the HR diagram (Fig. \ref{fig:6PatHR}) and on the chemical composition gradient in the core (Fig \ref{fig:6PatX}) are very similar to those of the model computed with a uniform turbulent diffusion coefficient $\rm D_{\rm T}=5\times10^3$. This is not surprising since, in the central regions, the total turbulent diffusion coefficient shown in Fig. \ref{fig:6PatD} does not change considerably in time and its magnitude is of the order of a few thousands cm$^2$ s$^{-1}$.

\subsection{Modes of different degree and low radial order}
\label{sec:ell}
\begin{figure}
\resizebox{\hsize}{!}{\includegraphics[angle=0]{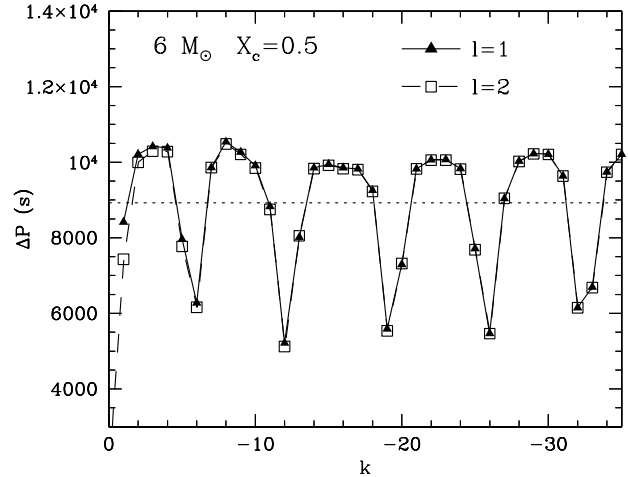}}
\caption{\small Period spacing for a 6 \msol models with $Xc=0.5$. The period spacing computed with $\ell=2$ modes is multiplied by $\sqrt{3}$, as suggested by Eq. \ref{eq:asy}.}
\label{fig:l2}
\end{figure}

In the previous sections we analyzed the properties of $\Delta P$ for $\ell=1$, high order g-modes. Does the period spacing computed with modes of different degree $\ell$, or with low radial order $k$, have a similar behaviour? In Fig. \ref{fig:l2} we consider the period spacing computed with $\ell=2$ periods $P_{k,\ell=2}$ scaled by the $\ell$-dependent factor suggested by Eq. \ref{eq:asy}, i.e. $P'_{k,\ell=2}=\sqrt{3}\;P_{k,\ell=2}$. Except for the oscillation modes with the lowest radial orders, we notice in Fig. \ref{fig:l2} that the period spacing of modes of degree $\ell=1$ and 2 have the same behaviour, provided that the dependence on $\ell$ given by Eq. \ref{eq:asy} is removed.

Though the asymptotic approximation is valid for high order modes, we see in Fig. \ref{fig:l2} (as well as in the figures presented in the previous sections) that the description of the period spacing as the superposition of the constant term derived from Eq. \ref{eq:asy} and periodic components related to $\nabla_\mu$, is able to accurately describe the periods of gravity modes even for low orders $k$. This suggest that, at least qualitatively, the description of g-modes presented in this work could represent a useful tool to interpret the behaviour of low order g-modes observed in other classes of pulsators, such as $\beta$ Cephei and $\delta$ Scuti stars. This subject will be addressed in a second paper.

\section{Challenges for asteroseismology}
\label{sec:realworld}
\begin{figure}
\resizebox{\hsize}{!}{\includegraphics[angle=-90]{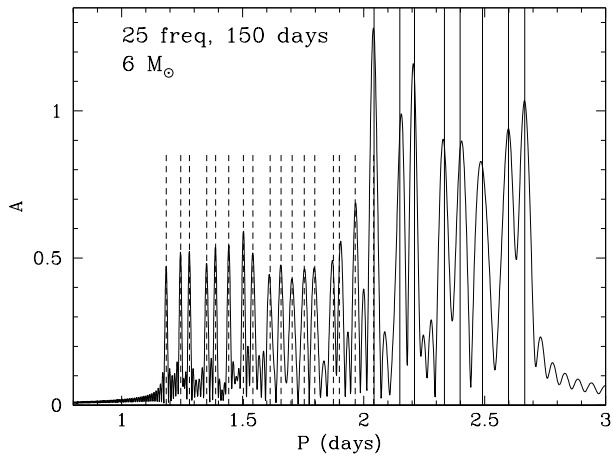}}
\caption{\small Power spectrum of simulated time series for a 6 \msol model. The vertical solid and dashed lines represent the input oscillation frequencies of, respectively, $\ell=1$ and $\ell=2$ modes.}
\label{fig:time}
\end{figure}

Asteroseismology of high-order g-mode main-sequence pulsators is not an easy task. 
The long oscillation periods and the dense frequency spectrum in
these stars require long and continuous observations in order to
resolve single oscillation frequencies. As clearly stated already in
\citet{Dziembowski93}, observations of at least 70 days are needed
to resolve the period spacing in a typical SPB star. In addition to
large observational efforts that have been made from the ground (see
e.g. \citealt{DeCat02}, \citealt{DeCat07} for SPB stars and
\citealt{Mathias04} and \citealt{Henry07} for $\gamma$ Dor), long,
uninterrupted space-based photometric time series will soon be
provided by CoRoT. In order to estimate if the CoRoT 150 day-long
observations have a frequency resolution sufficient to resolve the
period spacing of a typical SPB star, we simulated a time series of
$\ell=1$ and $\ell=2$ high-order g modes that are expected to be
excited in a 6 \msol star with $Xc=0.2$ computed without
overshooting. The excitation of oscillation modes has been computed
using the non-adiabatic code MAD \citep{Dupret03}. Random initial
phases have been considered in the sine curves describing the 25
frequencies included in the simulations, and amplitudes of 1 and 0.5
(on an arbitrary scale) have been assumed for $\ell=1$ and $\ell=2$
modes respectively. The generated time series has been analyzed
using the package Period04 \citep{Lenz05}. As it is shown in the
resulting power spectrum (see Fig. \ref{fig:time}), for such long
observational runs (150 days) the input frequencies can be
accurately recovered even for the largest periods of oscillation.
Thanks to the high frequency resolution, the departures from a
constant period spacing are also evident in the power spectrum.

\begin{figure}
\resizebox{\hsize}{!}{\includegraphics[angle=-90]{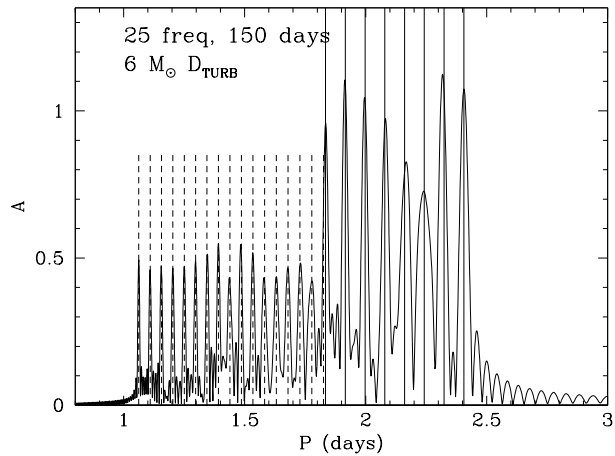}}
\caption{\small Same as Fig. \ref{fig:time} for a 6 \msol model computed with turbulent diffusion in the core.}
\label{fig:timeD}
\end{figure}

We then simulate a time series for an additional 6 \msol model that, despite having the same surface properties ($\log(\rm L/L_\odot)=3.28$, $\log{(\teff)}=4.22$) as the previous one, is computed with turbulent diffusion ($\rm D_{\rm T}=5000$). In this case the power spectrum (see Fig. \ref{fig:timeD}) shows a regular period spacing that, on the basis of the high frequency resolution, can be easily distinguished from the one described in Fig. \ref{fig:time}.

Even though frequency resolution may no longer be an issue in the very near future, there is a second well known factor limiting asteroseismology of SPB and $\gamma$ Dor stars: the effects of rotation on the oscillation frequency can severely complicate the high-order g-mode spectra.
Referring once more to the work by \citet{Dziembowski93}, a first requirement in order to treat rotational effects as perturbations on the oscillation frequencies is the angular rotational velocity being sufficiently smaller than the oscillation frequency ($v_{rot} \ll 2\pi\, R/P$, where $R$ is the radius of the star and $P$ the oscillation period). In the case of the models of an SPB star 
considered previously in this section, this translates into $v_{rot} \ll 100\;\rm km\;s^{-1}$ considering the modes of longest period: this is significantly larger than the average vsini ($\sim 25\; \rm km\;s^{-1}$) measured in SPB stars \citep[see e.g. ][]{Briquet07}. In the case of $\gamma$ Dor stars (see e.g. \citealt{Suarez05}) a similar estimate limits the validity of the perturbative approach to rotational velocities of $\sim 50-70\; \rm km\;s^{-1}$: for faster rotators non-perturbative approaches are needed \citep[see ][]{Dintrans00,Rieutord02}.

Even in the case of slow rotators, however, the rotational splittings may become as large as the period spacing itself. Such a large effect of rotation on the frequency spectrum can severely complicate the identification of the azimuthal order $m$ and the degree $\ell$ of the observed modes. In fact, a simple mode identification based on the regular pattern expected from high-order g-modes becomes inapplicable if the rotational splitting is of the same order as $\Delta P$ (see e.g. Fig. 13 in  \citealt{Dziembowski93}). The identification of the modes would then need to be provided by photometric and spectroscopic mode identification techniques \citep[see e.g. ][]{Balona86,Garrido90,Aerts92,Mantegazza00,Briquet03,Dupret03,Zima06}.

\section{Summary and conclusions}
\label{sec:conclusions}
In this work we investigated in detail the properties of high-order gravity in models of main-sequence stars.
 The  chemical composition gradient that develops near the outer edge of the convective core leads to a sharp variation of the \BV frequency. As we presented in Sec. \ref{sec:modetrap}, the latter is responsible for a periodic trapping of gravity modes in the region of chemical composition gradient, and it directly affects the period spacing of g modes.

In analogy with the works on white dwarfs by \citet{Brassard92} and
\citet{Montgomery03}, we show that in the case of main sequence
stars analytical approximations can be used to directly relate the
deviations from a uniform period spacing to the detailed properties
of the $\mu$-gradient region that develops near the energy
generating core. We find that a simple approximation of g-mode
periods, based on the variational principle of stellar oscillations,
is sufficient to explain the appearance of sinusoidal components in
the period spacing. This approximation (see Sec. \ref{sec:varia})
relates the periodicity of the components to the normalized buoyancy
radius of the glitch in $N$, and the amplitude of the components to
the sharpness of the feature in $N$. In particular, if the sharp
variation in $N$ is modelled as a step function, the amplitude of
such components is expected to be independent of the order of the
mode $k$; whereas if the glitch in $N$ is described with a ramp, the
amplitude of the components decreases with $k$. A more accurate
semi-analytical approximation of the period spacing, which considers
the effects of the sharp feature in $N$ on the eigenfunctions, is
also given in Sec. \ref{sec:approx2}.

We then presented a survey of the properties of high-order g modes in main sequence models of masses between 1 and 10 \msol (see Sec. \ref{sec:numeric}).
As a general result we found that, in models with a convective core, the period spacing of high-order g modes is accurately described by oscillatory components of constant amplitude, superposed to the mean period spacing predicted by the asymptotic theory of \citet{Tassoul80}.
In Sec. \ref{sec:coresms} we showed that the period spacing depends primarily on the extension and behaviour of the convective core during the main sequence and, therefore, on the mass of the star.

In models without a convective core (see Sec. \ref{sec:radiative}) the mean $\Delta P$ considerably decreases during the MS, whereas no significant deviation from a constant period spacing is present.
For an intermediate range of masses (see Sec. \ref{sec:growing}) the convective core grows during most of the MS, generating an ``unphysical'' discontinuity in $\mu$ if no mixing is added in the small semiconvective region that develops. We find that the behaviour of $\Delta P$, and in particular the appearance of periodic components, depends on the treatment of this region. It is interesting to notice that $\gamma$ Doradus stars are in the mass domain where models show a transition between growing to shrinking convective cores on the main sequence. Gravity modes could therefore represent a valuable observational test to discriminate between the different prescriptions used in stellar models \citep[see e.g.][]{Popielski05} to introduce the required mixing at the boundary of the convective core.
In models with higher masses, the convective core recedes during the main sequence (see Sec. \ref{sec:receding}): this leaves behind a $\mu$ gradient that generates clear periodic components in $\Delta P$. We found that the analytical expression derived in Sec. \ref{sec:varia} allows to accurately recover the location and sharpness of the $\mu$ gradient from the amplitude and periodicity of the components in $\Delta P$. In this mass domain, though the average period spacing does not change substantially with the age, the periodicity of the components does, and it therefore represents an indicator of the evolutionary state of the star.

In Sec. \ref{sec:extra} we showed that also extra-mixing processes
can alter the behaviour of $\Delta P$, since they affect the size
and evolution of the convective core, as well as the sharpness of
the $\mu$ gradient. We first compared models with the same $X_c$,
but computed with and without overshooting (see Sec.
\ref{sec:over0}). We found that in models with small convective
cores, or where nuclear reactions take place also in the radiative
region, the different size of the fully-mixed region changes the
periodicity of the components in $\Delta P$. In Sec.
\ref{sec:rotation} we described how chemical mixing can severely
affect the amplitude of the periodic components in $\Delta P$. In
models where turbulent mixing induced by rotation is considered, the
smoother $\mu$ profile near the core leads to a discontinuity not in
$N$ itself, but in its first derivative: as  suggested by the
analytical approximation in Sec. \ref{sec:varia}, this leads to
periodic components in $\Delta P$ whose amplitude decreases with the
order of the mode. In the case of SPB stars, in particular, we find
that the mixing induced by the typical rotation rates observed (i.e.
$\simeq 25\; \rm km\;s^{-1}$), is sufficient to alter significantly
the properties of the g-mode spectrum.

Finally in Sec.\ref{sec:realworld} we discussed the difficulties encountered in the asteroseismology of $\gamma$ Doradus and SPB stars. Even though a frequency resolution sufficient to resolve closely spaced periods will be provided by the forthcoming space-based observations, an asteroseismic inference on the internal structure will only be possible for stars with very slow rotation rates, and with reliably identified pulsation modes. Once these conditions are reached, we will be able to access the wealth of information on internal mixing which, as shown in this work, is carried by the periods of high-order gravity modes in main-sequence objects.
\section*{Acknowledgements}
A.M. and J.M. acknowledge financial support from the Prodex-ESA Contract Prodex 8 COROT (C90199). P.E. is thankful to the Swiss National Science Foundation for support.

\bibliographystyle{mn2e}
\small
\bibliography{MN-07-0975-MJ}
\label{lastpage}
\end{document}